\documentclass[%
 reprint,
 letterpaper,
superscriptaddress,
amsmath,amssymb,
aps,
pra,
]{revtex4-2}
\renewcommand{\selectlanguage}[1]{}
\usepackage{xcolor}
\usepackage[percent]{overpic}
\usepackage{graphicx}
\usepackage{dcolumn}
\usepackage{bm}
\usepackage{braket}
\usepackage{physics}
\usepackage{multirow}
\usepackage[caption=false]{subfig}

\DeclareMathOperator{\sinc}{sinc}

\newcommand{\s}[1]{\sigma_{#1}}
\newcommand{\expect}[1]{\langle #1 \rangle}

\begin{document}

\title{Dynamics of the Non-equilibrium Spin-Boson Model: A Benchmark of master equations and their validity}

\author{Gerardo Su\'arez}
\email{gerardo.suarez@phdstud.ug.edu.pl}
\affiliation{%
 International Centre for Theory of Quantum Technologies, University of Gdansk, 80-308 Gdansk, Poland\\
}%
\author{Marcin {\L}obejko}
\affiliation{Institute of Theoretical Physics and Astrophysics, Faculty of Mathematics, Physics and Informatics, University of Gda\'nsk, 80-308 Gda\'nsk, Poland}
\affiliation{%
 International Centre for Theory of Quantum Technologies, University of Gdansk, 80-308 Gdansk, Poland\\
}%

\author{Micha{\l} Horodecki}
\affiliation{%
 International Centre for Theory of Quantum Technologies, University of Gdansk, 80-308 Gdansk, Poland\\
}%

\date{\today}
\begin{abstract}

In recent years, there has been  tremendous focus on identifying whether effective descriptions of open quantum systems such as master equations,  can accurately describe the dynamics of open quantum systems. One particular question is whether they provide the correct steady state in the long time limit. Transient regime is also of interest. Description of evolution by various master equations - some of them being not complete positive -  is benchmarked against exact solutions (see e.g. Hartmann and Strunz, Phys. Rev. A 101, 012103). An important property of true evolution is its non-Markovian features, which are not captured by the simplest completely positive master equations. 

In this paper we consider a non-Markovian, yet completely positive evolution (known as refined weak coupling or cumulant equation) for the Spin-Boson model with an Overdamped Drude-Lorentz spectral density and arbitrary coupling. We bench-marked it against numerically exact solution, as well as against other master equations, for different coupling strengths and temperatures. We find the cumulant to be a better description in the weak coupling regime where it is supposed to be valid. For the examples considered it shows superiority at moderate and strong couplings in the low-temperature regime for all examples considered. In the high-temperature regime however its advantage vanishes. This indicates that the cumulant equation is a good candidate for simulations at weak to moderate coupling and low temperature. Our calculations are greatly facilitated due to our concise formulation of the cumulant equation by means of representation of the density matrix in the SU(N) basis.

\end{abstract}

\maketitle


\section{Introduction}

Open quantum systems undergo non-trivial evolution due to their coupling with their environment. One of the most popular descriptions for its dynamics namely the Gorini–Kossakowski–Lindblad-Sudarshan equation (GKLS) relies on a set of approximations, in particular, the rotating wave approximation and the Markov approximation. The first one clearly affects the accuracy of the description in both the transient and steady state regimes \cite{reconciliation, meanforcereview}, while the second is more about the transient regime \cite{breuer}. The combination of both approximations will make both coherences and populations decay ``purely exponentially" meaning they just decay without any sort of increases. On the other hand letting go of any of the approximations takes us away from this ``purely exponentially" decaying regime.

Open quantum systems will usually reach a steady state due to their interaction with the environment, when we consider the environment to be in a thermal state. The standard Markovian master equation approach indicates that our system reaches an equilibrium in a thermal state with the temperature of the bath. Depending on whether we consider Lamb-shift terms or not the equilibrium state will be according to the bare Hamiltonian, meaning the original Hamiltonian or the Lamb shifted one. 
Other approaches involving non-Markovian master equations, however, lead to steady states that are not necessarily the Gibbs state of the system Hamiltonian, a fact that has been studied extensively and for a long time in the literature \cite{reconciliation,Purkayastha2020,miyashita,Geva,romerorochin}. Recently, there was a debate surrounding consequences of this fact, the debate was due to the so called steady state coherences present in the Redfield equation,  which as their name states is a non-zero value of coherences in the energy eigenbasis of the system in the steady state. They were presented in \cite{Guarnieri}. The results of that paper however were obtained for the Redfield equation which is not completely positive, so it was argued later than this effect may come from lack of complete positivity \cite{cattaneo2021comment} rather than being a true effect. From references \cite{Flemming,tupkary,limitations} one can conclude that steady state coherences obtained from the Redfield equation are only trust worthy when coherences in the energy eigenbasis scale at most as the square of system bath couplings. In those references there are further arguments that show that there is no general way to fix the positivity problem of the Redfield equation, at least in second order, without losing accuracy in populations, coherences or both. 

In this paper we will study both transient as well as steady state regime for  a completely positive evolution 
given in a form of a dynamical map rather than a master equation - another way of modelling 
Non-Markovian open quantum system dynamics known as the cumulant equation or the refined weak coupling \cite{Rivas,Alicki89}.
  This equation is related to other equations studied in the literature, such as the coarse grained master equation \cite{Schaller_brandais,majenz_coarse_2013}. Namely, coarse grained master equations involve a time coarse graining parameter, and the cumulant equation can be thought of resulting from such master equation with the coarse graining being automatically appropriately adjusted when the time is passing \cite{lidar2020lecture}.  
  
  The cumulant equation was benchmarked against other descriptions of evolution in \cite{hartmann_accuracy_2020} for two non interacting spins coupled to a bosonic bath and a Lorentzian spectral density. Here we consider a more structured spectral density and more general coupling.
 Specifically we consider Spin Boson model with arbitrary coupling to bath, and overdamped Drude-Lorentz spectral density. 
 We will mainly be concerned with the comparison of the different master equations in the transient regime, meaning we will be mainly interested in their Non-Markovian effects. In particular, we will show that when one is interested in the dynamics of the system and not in the steady state, in particular, it is beneficial to use the refined weak coupling limit (i.e. cumulant equation) when weak coupling is considered, while the situation is different in a stronger coupling regime where the dynamics predicted via the cumulant equation might be better or worse than other more popular approaches depending on the range of temperatures considered. 

 We will also show that the cumulant equation exhibits oscillations in the coherences for long times, similar to those reported for the Redfield  equation in \cite{Guarnieri}. Notice that in \cite{Guarnieri} they are associated with Lamb-shift while for cumulant equation, they are rather related to slow convergence to the steady state. Since the equation is completely positive, when the coherences do not agree with exact solutions, this comes only from second order approximation effects, unlike in the case of the Redfield equation where they might come from the approximation, lack of positivity or a mixture of both.

We present the cumulant equation from a new perspective, exploiting the representation of the density matrix in the SU(N) basis. This allows for easy analytical calculations equation when orthogonal interactions are involved, and a way to obtain them via linear maps, we believe this formulation proves useful for analytical calculations, specially when one approximates the decay rates as proposed in \cite{winczewski2021bypassing}. This alternate formulation to solve the cumulant equation also provides with a simple way of separating the transient from the asymptotic dynamics of the system, allowing us to obtain the steady state in a simple way at least compared with previous presentations of the cumulant equation \cite{Rivas,lidar2020lecture}, which we believe may be useful for certain situations, it also allows for an straightforward way to write this equation in a master equation form.

We then show that while not oscillating for as long as those predicted by master equations those steady state coherences are present if we  solve the dynamics exactly (In the Schrodinger picture), in this paper we do this by means of the hierarchical equations of motion (HEOM) \cite{HEOM}, using the recent Qutip BoFiN implementation \cite{qutip}. This translates into the fact that the steady state is not diagonal in the basis of the system Hamiltonian, a fact that has consequences for the efficiencies of quantum machines \cite{Nazir2018}.

We also recover simpler models for the cumulant equation from our general model, for the typical Spin-Boson the treatment differs from Rivas \cite{Rivas} in that we shall present a compact, simple equation for the dynamics as opposed to the formulas provided there. We argue that this form of the cumulant equation is easier to simulate as well as more transparent, which allows one to draw conclusions from the equations more easily. For the pure dephasing model we show that the cumulant equation matches the exact solution, as does time convolutionless second order master equation but not Bloch-Redfield (BR) \cite{Doll_2008}. Code to reproduce to simulate the cumulant equation in general has been made available by the authors in \cite{repo}, the examples include the cumulant equation simulated in the form described in this paper.

Finally we discuss the role of Lamb-shift on the cumulant equation dynamics,  we discuss on the dilemma of neglecting Lamb-shift and compare the dynamics and steady state when it's present or not to the exact dynamics. We shall also expose that the discussion about Lamb-shift is more complicated in the non-Markovian regime and conclude on possible paths to adequate renormalization. It may also be argued that such Lamb-shift  that does not simply shift the energy of the system but rather has implications in its dynamics (because it does not commute with the system Hamiltonian)  appears because we shouldn't simply start for an arbitrary initial state but need to take into account the state preparation procedure, which will modify the dynamics accordingly \cite{preparation}. In this paper we don't consider such situation and study dynamics from an arbitrary initial product state.

\section{The cumulant equation for a model with composite interactions}

The system of interest in this paper is a spin with composite interactions to a bosonic bath, for  generality we will consider the most general Hamiltonian we can get for a simple Spin-Boson model which is given by:

\begin{align}\label{eq:Hamiltonian}
     H &= \underbrace{\frac{\omega_{0}}{2} \s{z}}_{H_S} + \underbrace{\sum_{k} w_{k} a_{k}^{\dagger} a_{k}}_{H_B} \nonumber \\ &+ \underbrace{\sum_{k} g_k (f_1 \s{x}+f_2 \s{y} +f_3 \s{z}) (a_{k}+a_{k}^{\dagger})}_{H_I}
\end{align}

\begin{figure}[h!]
\begin{overpic}[width=0.45\textwidth]{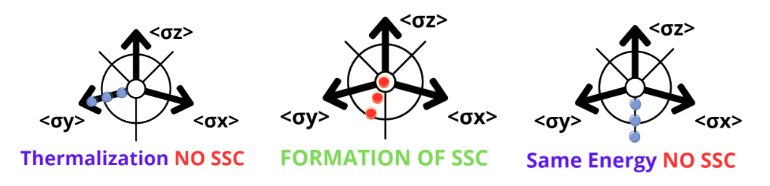}
    \put (5,20) {\Large$\displaystyle (a)$}
    \put (35,20) {\Large$\displaystyle (b)$}
    \put (70,20) {\Large$\displaystyle (c)$}
\end{overpic}
\caption{a) The interaction Hamiltonian is orthogonal to $H_{S}$, in this case the system thermalizes and no steady state coherences are generated.
b) The interaction Hamiltonian is neither orthogonal nor parallel to  $H_{S}$, in this setting steady state coherences are generated. 
c) The interaction Hamiltonian is parallel to $H_{S}$, this is the pure dephasing case.}\label{fig:old_table}
\end{figure}

The system can undergo three types of dynamics, namely thermalization, the formation of Steady State Coherences (SSC), and dephasing. The Steady-state coherences arise when composite interactions are in place, meaning there is an interaction that is both parallel and orthogonal to the system Hamiltonian $H_{S}=\frac{\omega \s{z}}{2}$ the figure above illustrates the different cases that rise in a Spin-Boson model depending on the interaction Hamiltonian $H_{I}$

\begin{itemize}
\item Figure \ref{fig:old_table}a shows the case in which we have an interaction that is orthogonal to $H_{S}$ \footnote{In this context we say an an operator is parallel to another if they commute and orthogonal if the commutator generates a different member of the SU(N) group.}, the system is driven to a Gibbs state at the temperature of the bath. No steady state coherences are generated.
\item Figure \ref{fig:old_table}b shows an interaction that is neither orthogonal nor parallel  to  $H_{S}$ but that instead is made of a composition of both orthogonal and parallel components, in this setting steady state coherences are generated 
\item Figure \ref{fig:old_table}c illustrates  $H_{S}$ parallel to the interaction. In this case, the populations of the system are not driven towards the Gibbs state at the bath's temperature, only coherences evolve and they go towards zero. 
\end{itemize}

The main model in this paper is the cumulant equation \cite{Rivas,Alicki89,winczewski2021bypassing} whose derivation is included in this manuscript  as an appendix for completeness. In general the cumulant equation is determined by the exponential of the map:

\begin{eqnarray}
 &\mathcal{K}_t [\rho_S^{(0)}]= - i  [\Lambda(t),\rho_S^{(0)}] +  \mathcal{D}(\rho)
\end{eqnarray}
where 
\begin{align}
\mathcal{D}(\rho)&=
\underset{\omega,\omega'}{\sum} \underset{\alpha,\beta}{\sum} \Gamma_{\alpha \beta}(\omega,\omega',t) \Big( A_{\alpha}(\omega') \rho_S^{(0)} A^{\dagger}_{\beta}(\omega) \nonumber  \\ &- \frac{1}{2} \{ A^{\dagger}_{\beta}(\omega) A_{\alpha}(\omega'),\rho_S^{(0)}\} \Big).
\end{align}

The usual approach of separating the interaction Hamiltonian into system and bath operators was performed meaning that we have used

\begin{align}
    H_{I} &= \sum_{k} A_{k} \otimes B_{k}  = \sum_{k,\omega} e^{-i \omega t} A_{k}(\omega) \otimes B_{k} \\ 
    A_{\alpha}(\omega) &= \sum_{\epsilon-\epsilon' =\omega} \ketbra{\epsilon} A_{\alpha} \ketbra{\epsilon'} 
\end{align}

Here
 \begin{equation}
     \Gamma_{\alpha,\beta}(w,w',t)=\int_{0}^{t} dt_1 \int_{0}^{t} dt_2 e^{i (w t_1 - w' t_2)} \mathcal{C_{\alpha,\beta}}(t_{1},t_{2})
 \end{equation} 
with
\begin{equation}
    \mathcal{C}_{\alpha,\beta}(t_{1},t_{2}) = \langle B_{\beta}(t_1) B_{\alpha}(t_2) \rangle
\end{equation}
and
\begin{eqnarray}
\Lambda(t)= \sum_{w,w'} \sum_{\alpha,\beta} \xi_{\alpha,\beta}(w,w',t) A^{\dagger}_{\alpha}(w) A_{\beta}(w')
\end{eqnarray}
with
\begin{align}
    \xi_{\alpha,\beta}(w,w',t)=   \iint_{0}^{t}\frac{dt_{1}}{2i} dt_{\!2} sgn(t_{\!1}-t_{\!2}) e^{i(\omega t_1- \omega' t_2)}   \mathcal{C}_{\beta,\alpha}(t_{1},t_{2}).
\end{align}
The dynamics of our state is then simply given by:

\begin{equation}\label{eq:linear_map}
    \rho(t)=e^{\mathcal{K}_{t}} [\rho_S^{(0)}]
\end{equation}
In general, however it can be rewritten in another way which we find more transparent and easier to work with for semi-analytical calculations \cite{Rivas,winczewski2021bypassing}. Let us choose to represent our density matrix for SU(N) following the prescription in \cite{SUN,hioe}:
\begin{equation}
    \rho = \frac{\mathbb{I}}{N} + \frac{1}{2} \sum_{j=1}^{N^{2}-1} \lambda_{j} \mu_{j}
\end{equation}
where $\lambda$'s are operators defined as follows. 
One starts by introducing the transition operators

\begin{align}
    P_{i,k} = \ket{i}\bra{k}
\end{align}
where the kets $\ket{n}$ are orthonormalized eigenstates of a linear Hermitian operator of the relevant Hilbert space. Then consider the three different types of operators

\begin{align}
  w_{j} &= -\sqrt{\frac{2} { j(j+1)}} (P_{1,1}+P_{2,2}+\dots+P_{j,j}- j  P_{ j+1 ,j+1} ) \\ 
  u_{j,k} &= P_{j,k}+P_{k,j} \\
  v_{j,k} &= i \left( P_{j,k}-P_{k,j} \right)
\end{align}
where $1 \leq l \leq N-1$ and $1 \leq j <k \leq N$. We then construct a vector made out of this operators
\begin{align}
    \vec{\lambda} &= \{\underbrace{u_{1,2},u_{1,3}, u_{1,4}, \dots}_{\frac{N(N-1)}{2} Operators} ,\underbrace{v_{1,2},v_{1,3}, v_{1,4}, \dots,}_{\frac{N(N-1)}{2} Operators} \nonumber\\ &\underbrace{w_{1},w_{2}, w_{3}, \dots,w_{N-1}}_{(N-1) Operators} \}
\end{align}
so that the total number of operators in $\vec{\lambda}$ is $N^{2}-1$, corresponding to the $N^{2}-1$ elements needed to fully identify a $N\times N$ density matrix. Notice the operators $\lambda_{i}$ satisfy

\begin{align}
    Tr\{\lambda_{i}\} &= 0 \\ 
    Tr\{\lambda_{i} \lambda_{j}\} &= 2 \delta_{i,j} 
\end{align}
which form the SU(N) Algebra \cite{hioe}. Then we can write the density operator as

\begin{align}
    \rho = \frac{\mathbb{I}}{N} + \frac{1}{2} \sum_{j=1}^{N^{2}-1} \mu_{j} \lambda_{j}
\end{align}
where
\begin{align}
\mu_{j}=    Tr\{\lambda_{i} \rho \} 
\end{align}
The vector $\vec{\mu}$ whose elements are the $\mu_{j}$ coefficients is usually called the coherence or generalized Bloch vector in the literature \cite{SUN}.

As the map eq. \eqref{eq:linear_map} is linear we can in general write its action on the density matrix as linear transformation 

\begin{align}
    \mathcal{K}_{t}[\rho] = (M \frac{\vec{\mu}}{2} + \vec{r})\cdot \vec{\lambda}
\end{align}
which can be easily exponentiated and yields (see Appendix, eq. \eqref{eq:dynamics}):

\begin{align}\label{eq:exp_linear}
    \rho(t)=\frac{\mathbb{I}}{2}+  (e^{M}-\mathbb{I}) M^{-1} \vec{r}  \cdot \vec{\lambda}+   e^{M} \frac{\vec{\mu}}{2} \cdot \vec{\lambda}
\end{align}
In fact the expression \eqref{eq:exp_linear} not only works for the exponential of a the cumulant generator but the exponential of any linear map.

\section{The Cumulant equation for a Spin-Boson model with composite interactions}

 From \eqref{eq:Hamiltonian} we may decompose the $H_{I}$ part into the bath operators and the following jump operators:

\begin{align}
    A(\omega_0)=\bar{f} \s{-}, \hspace{1.5em} A(-\omega_0)=f \s{+}, \hspace{1.5em} A(0)=f_3 \s{z} 
\end{align}
where $f=f_1-i f_2$, then from the linear coupling to the bosonic operators one obtains:

\begin{align} \label{eq:Decays}
&\Gamma(\omega,\omega',t) = t^{2}\int_{0}^{\infty} d\nu e^{i\frac{\omega-\omega'}{2} t} J(\nu) \left[ (n_{\nu}+1) \sinc\left(\frac{(\omega-\nu)t}{2}\right) \right. \nonumber \\ &\times \left. \sinc\left(\frac{(\omega'-\nu)t}{2}\right)   + n_{\nu} \sinc\left(\frac{(\omega+\nu)t}{2}\right) \sinc\left(\frac{(\omega'+\nu)t}{2}\right)   \right]
\end{align}

\begin{align} \label{eq:shifts}
    &\xi(\omega,\omega',t)=\frac{t^{2}}{2 \pi} \int_{-\infty}^{\infty} d\phi \sinc\left( \frac{\omega-\phi}{2} t\right) \sinc\left( \frac{\omega'-\phi}{2} t\right) \nonumber \\ &\times P.V \int_{0}^{\infty} d\nu J(\nu) \left[ \frac{n(\nu)+1}{\phi-\nu} + \frac{n(\nu)}{\phi+\nu} \right]
\end{align}
where $n_{\nu}$ is the Bose-Einstein distribution 

\begin{equation}
    n_{\nu}= \frac{1}{e^{\beta \nu}-1}
\end{equation}
For convenience let us introduce the following notation: from now on the functions dependence on time will not be written explicitly and their frequency dependence will be labeled as sub-indices, for example $\Gamma_{\omega,\omega'}=\Gamma(\omega,\omega',t)$ we will further simplify the notation by replacing the explicit frequencies as $\omega_{0} \xrightarrow{} -$,$-\omega_{0} \xrightarrow{} +$, $0 \xrightarrow{} z$, such that $\Gamma(\omega_{0},0,t)=\Gamma_{-z}$, using this we may finally write the cumulant equation evolution for this model as:

\begin{align}\label{eq:exp_cum}
    \rho(t)=\frac{\mathbb{I}}{2}+  (e^{M}-\mathbb{I}) M^{-1} \vec{r}  \cdot \vec{\sigma}+   e^{M} \frac{\vec{x}}{2} \cdot \vec{\sigma}
\end{align}
where

\begin{align}\label{eq:relevant_vectors}
    \vec{x}= \begin{pmatrix}
        \expect{\s{z}(0)} \\ \expect{\s{x}(0)} \\ \expect{\s{y}(0)}
    \end{pmatrix} \hspace{0.5em} 
    \vec{r}=\begin{pmatrix}
        |f|^{2} \frac{\Gamma_{++}-\Gamma_{--}}{2} \\ -\Im(\Gamma_{z}^{-}) \\ \Re(\Gamma_{z}^{-}) 
    \end{pmatrix} \\
\end{align}
Where M is given by \eqref{eq:relevant_matrix}

\onecolumngrid

\begin{align}\label{eq:relevant_matrix}
   M= \begin{pmatrix}-\abs{f}^{2}\Gamma &  - 2 f_{1} \xi_{z} -\Im(\Gamma_{z}^{+}) &  2 f_{2} \xi_{z} + \Re(\Gamma_{z}^{+})  \\ -\Im(\Gamma_{z}^{+}) + 2 f_{1} \xi_{z} & - \left(\frac{\abs{f}^{2}}{2} \Gamma + 2 f_{3}^{2} \Gamma_{zz}+ \Re(f^{2} \Gamma_{-+}) \right) & \abs{f}^{2} \xi -\Im(f^{2} \Gamma_{-+})\\  \Re(\Gamma_{z}^{+})-2 f_{2} \xi_{z} & - \left( \abs{f}^{2} \xi + \Im(f^{2} \Gamma_{-+}) \right) & -\frac{\abs{f}^{2}}{2} \Gamma - 2 f_{3}^{2} \Gamma_{zz}+ \Re(f^{2} \Gamma_{-+})\end{pmatrix}
\end{align}

\twocolumngrid

\section{Dynamics of the system   when Lamb-shift is neglected}

In this section, we compare our model which was computed using the cumulant equation with other more established approaches in the literature. This section serves as bench-mark of our Non-Markovian equation in different scenarios. Since Lamb-shift itself is usually neglected in the approaches considered in literature, this benchmark should be relevant. In a later section we will see how the Lamb-shift correction affects the accuracy of the cumulant equation. A similar benchmark can be found in \cite{hartmann_accuracy_2020} where the focus is on the regime of validity of the several equations studied in that paper, here we focus both on the regime of validity and on which equation is better at which regime, while the conclusions found in that paper are correct here we indicate that there is more to the story as we see that the cumulant equation or refined weak coupling is typically more faithful to the exact dynamics in its regime of validity. This fact combined with the cumulant equation being a CPTP map make it a good alternative to simulate open quantum systems in the weak coupling regime, specially with respect to the several efforts to make the Bloch-Redfield equation completely positive \cite{Cattaneo_2019,Potts_2021,delapradilla2024taming}

From this section forward all simulations shown in plots have 

\begin{align}
    \rho(0)=\begin{pmatrix}
0.5 & 0.5 \\
0.5 & 0.5
\end{pmatrix}
\end{align}
as the initial state and

\begin{equation}
    J(\omega) = \frac{2 \lambda \omega \gamma}{\gamma^{2} +\omega^{2}}
\end{equation}
as the spectral density.  As a figure of merit we will use quantum fidelity \cite{Nielsen}, which for mixed states is given by:

\begin{equation}
    \mathcal{F}(\rho,\sigma) = \left(Tr[ \sqrt{\sqrt{\rho}\sqrt{\sigma}\sqrt{\rho}}]\right)^{2}
\end{equation}

Throughout the manuscript, the fidelity will be taken with respect to the HEOM approach, we will use the short hand notation

\begin{equation}
    \mathcal{F}= \mathcal{F}(\rho(t),\rho_{HEOM}(t))
\end{equation}

\subsection{Pure Dephasing and the conundrum of reproducing exact results at second order}

If we want to recover a pure dephasing model, it suffices to set $f=0$ for which our linear transformation becomes 

\begin{align}
    \begin{pmatrix}a' \\ b' \\ c' \end{pmatrix} &= \underbrace{\begin{pmatrix} 0 &  0 &  0  \\0 & -  2 f_{3}^{2} \Gamma_{zz} & 0\\  0 & 0 &  - 2 f_{3}^{2} \Gamma_{zz} \end{pmatrix}   }_{M} \begin{pmatrix}a \\ b \\ c \end{pmatrix} 
\end{align}
\normalsize
In this case, the dynamics are easily solvable as Eq. \eqref{eq:dynamics} becomes

\begin{align}
  \rho(t) &= \frac{\mathbb{I}}{2} +  \vec{\sigma}^{T} e^{M} \vec{a} = \left(
\begin{array}{cc}
 a+\frac{1}{2} & e^{- 2 f_{3}^{2} \Gamma_{zz}} (b-i c) \\
 e^{- 2 f_{3}^{2} \Gamma_{zz}} (b+i c) & \frac{1}{2}-a \\
\end{array}
\right)
\end{align}
From this, we can see the dynamics doesn't induce any change on the diagonal, namely there are no energy exchanges between the system an environment,the interaction simply kills coherences for this sort of evolution as  it is expected. For longer times $\Gamma_{zz}$ increases with $t$ then the steady state is simply given by

\begin{align}
  \rho_{steady} & \left(
\begin{array}{cc}
 a+\frac{1}{2} & 0 \\
 0 & \frac{1}{2}-a \\
\end{array}
\right)
\end{align}

\begin{figure}[h!]
\begin{overpic}[width=0.45\textwidth]{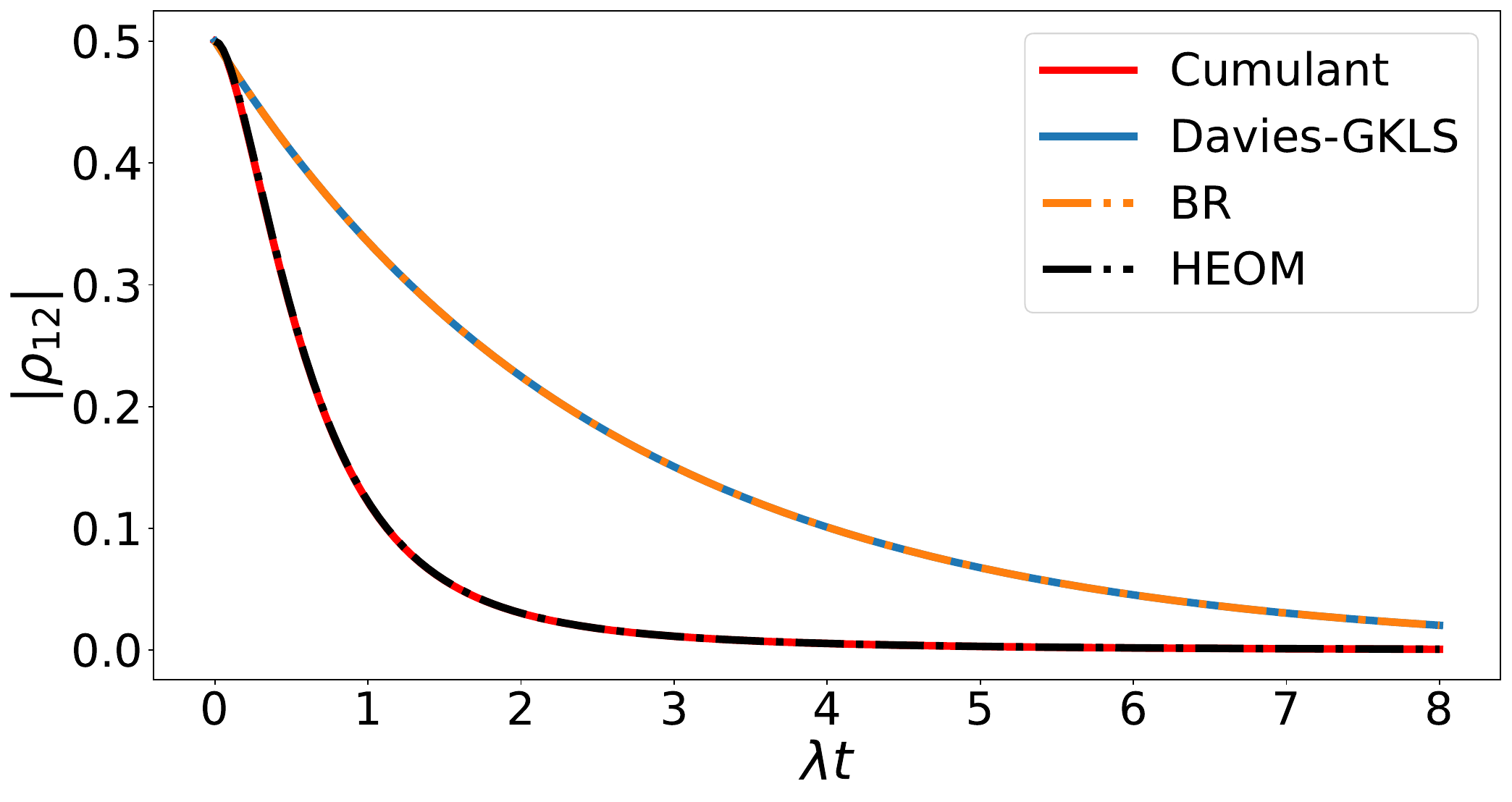}
        \put (-6,45) {\Large$\displaystyle (a)$}
\end{overpic}

\begin{overpic}[width=0.45\textwidth]{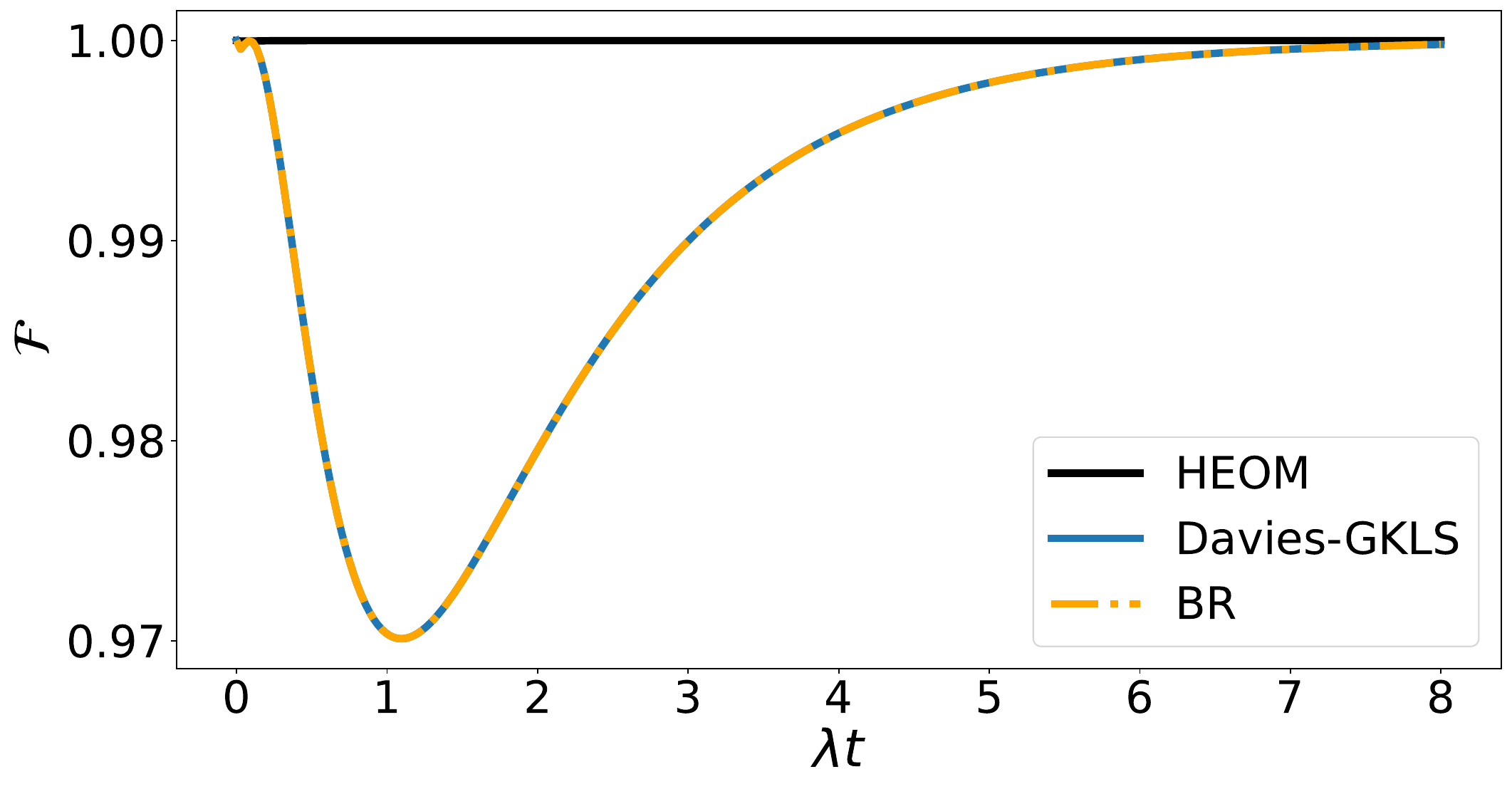}
        \put (-6,45) {\Large$\displaystyle (b)$}
\end{overpic}

\begin{overpic}[width=0.45\textwidth]{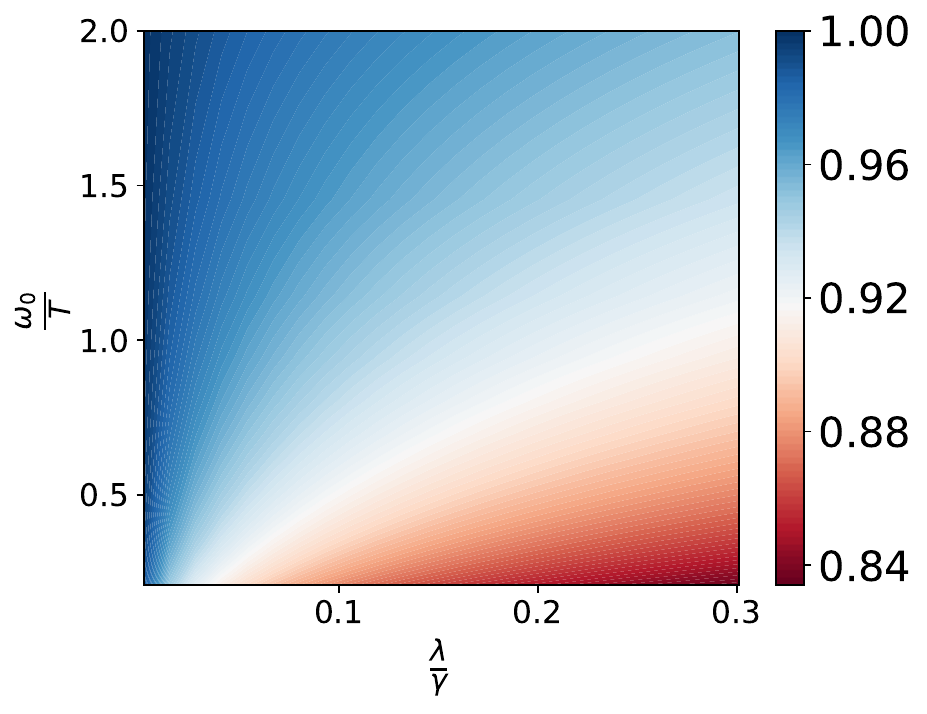}
        \put (-6,65) {\Large$\displaystyle (c)$}
\end{overpic}

\caption{(a) Shows the absolute value of the coherence in pure dephasing evolution (b) Shows the fidelity to the exact solution for the different approaches under pure dephasing evolution, cumulant is excluded as it matches the exact solution (c) Shows the Fidelity of the Davies-GKLS and Bloch-Redfield equation for a a wide range of parameters. The dynamics for the pure dephasing model are shown, it corresponds to $f_{1}=f_{2}=0$ and $f_{3}=1$. For (a) and (b)  the parameters used for the simulation were $\frac{\omega_{0}}{T} = 4$,$\frac{\lambda}{\gamma}=\frac{1}{4}$, $\gamma= 5 \omega_{0}$. We can see HEOM has converged to the analytical exact solution, it should be mentioned that it required $N_{k}=8$ Matsubara exponents, and  8 ADOS, thus the computational effort is significant when compared with the cumulant equation, furthermore at lower temperatures, more exponents may be needed, and the ADOS may not be enough, thus resulting in higher computational effort while the computational effort for the cumulant equation remains constant.}\label{fig:validity}
\end{figure} 
Notice we would have the exact same solution if we had considered Lamb-shift, from Eq. \eqref{eq:general_ls} we can see that Lamb-shift plays no role in this scenario as it is proportional to the identity. As a matter of fact, the cumulant equation for this model is equivalent to the exact solution. Let us consider the exact solution for a pure dephasing model as described by \cite{lidar2020lecture,Ekert}  the coherences decay according to

\begin{align}
    \rho_{12}(t) = e^{- \gamma(t) } \rho_{12}(0)
\end{align}
where 
\begin{align}
    \gamma(t)= 4  \int_{0}^{\infty} d \omega J(\omega) \coth(\frac{\beta \omega}{2}) \left( \frac{1-\cos(\omega t)}{\omega^2}\right)
\end{align}
For comparison let us quickly rewrite this. Note that $\cos(\omega t)=1-2 \sin(\frac{\omega t}{2})^{2}$ such that we may write:
\begin{align}
    \gamma(t)=  2 t^{2} \int_{0}^{\infty} d \omega J(\omega) \coth(\frac{\beta \omega}{2})  \sinc \left(\frac{\omega t}{2} \right)^{2}
\end{align}
While substituting directly in Eq. \eqref{eq:Decays} results in 
\begin{align}
    \Gamma_{zz} &=t^{2}\int_{0}^{\infty} d\nu  J(\nu)  \sinc\left(\frac{\nu}{2}t\right)^{2}  (2 n_{\nu}+1)   \\
    \Gamma_{zz} &=t^{2}\int_{0}^{\infty} d\nu  J(\nu)  \sinc\left(\frac{\nu}{2}t\right)^{2}  \coth{\left(\frac{\beta \nu}{2}\right)} 
\end{align}

By simple comparison we see that $\gamma(t)=2\Gamma_{zz}$ and that coherences are decaying in the same way, indicating that the cumulant equation is exact for pure dephasing. The cumulant is not the only equation that has this feat the time convolutionless master equation (TCL) at second order (Redfield \cite{Schaller_notes,REDFIELD19651}) is also exact for this model \cite{Doll_2008} but the commonly used Bloch-Redfield master equation \cite{bloch,qutip,Schaller_notes} which we compare to cumulant through this paper does not, this example shows a clear, simple scenario where we can see the superiority of the cumulant equation against similar approaches like Bloch-Redfield. On the other hand, it is a map, rather than an integral-differential equation like the TCL beating it in simplicity but still being able to produce exact solutions. This is remarkable given that this is independent of the coupling and it is indeed a conundrum that even for arbitrary strong coupling we are able to produce the exact dynamics with just a second-order expansion.

In Fig. \ref{fig:validity} we can see a contour plot that shows the minimum fidelity the BR equation has for a given set of parameters, we can see that the area where it works with high fidelity is rather restricted, which is understandable given that it is only meant to operate in the weak coupling limit, still the cumulant equation which was derived in the weak coupling limit holds, showing a clear advantage to study dephasing scenarios.

In this setting even the simplest equation, namely the Davies-Gorini–Kossakowski–Sudarshan–Lindblad (Davies-GKLS) master equation is in good agreement with the exact solution, the TCL and cumulant equation exhibit an advantage as they match the exact solution whereas Bloch-Redfield does not.

\subsection{The standard Spin-Boson Model}

The standard Spin-Boson model \cite{Rivas,Schaller_notes} can be recovered from \eqref{eq:Hamiltonian} by making make $f_{3}=f_{2}=0$, The linear transformation giving rise to the cumulant equation dynamics is then given by:

\begin{align}\label{eq:M_sx}
    M &= \begin{pmatrix}- \Gamma &  0 & 0 \\ 0 & -\frac{\Gamma}{2} + \Re( \Gamma_{-+}) & -\left( \Im(\Gamma_{-+})+\xi \right)  \\ 0 &  \xi -  \Im(\Gamma_{-+}) & -\frac{\Gamma}{2} - \Re( \Gamma_{-+})\end{pmatrix}  \\ \vec{r}&=\begin{pmatrix}\frac{(\Gamma_{++}-\Gamma_{--})}{2} \\ 0 \\ 0 \end{pmatrix}
\end{align}
The evolution of the system is obtained from \eqref{eq:exp_cum} easily, in particular one remarkable fact is that since the matrix $M$ rank is $2$ this allows us to obtain analytical expressions for its exponential easily, providing a way to get analytical answers for the Spin-Boson model out of this Non-Markovian description. 

Another advantage of this form of the evolution is that we may separate the asymptotic dynamics from the memory dependent one,  as we know that $e^{M}\xrightarrow[]{} 0$ as $t \xrightarrow[]{} \infty$ for this model as the real part of its eigenvalues are always negative

Therefore, if we only aim at knowing the steady state of the system then it suffices to look at the non-vanishing terms of \eqref{eq:exp_cum} in the long time limit, which results in :

\begin{align}
    \rho_{steady} &=\frac{\mathbb{I}}{2}- \vec{\sigma} \cdot M^{-1} \vec{r}=\vec{\sigma} \cdot \begin{pmatrix}
    \frac{\Gamma_{++}-\Gamma_{--}}{2 \Gamma} \\ 0 \\ 0 
    \end{pmatrix} \\
 &= \frac{\mathbb{I}}{2} +\frac{\Gamma_{++}-\Gamma_{--}}{2 \Gamma} \s{z}.
\end{align}

\begin{figure}[htb!]
\begin{overpic}[width=0.45\textwidth]{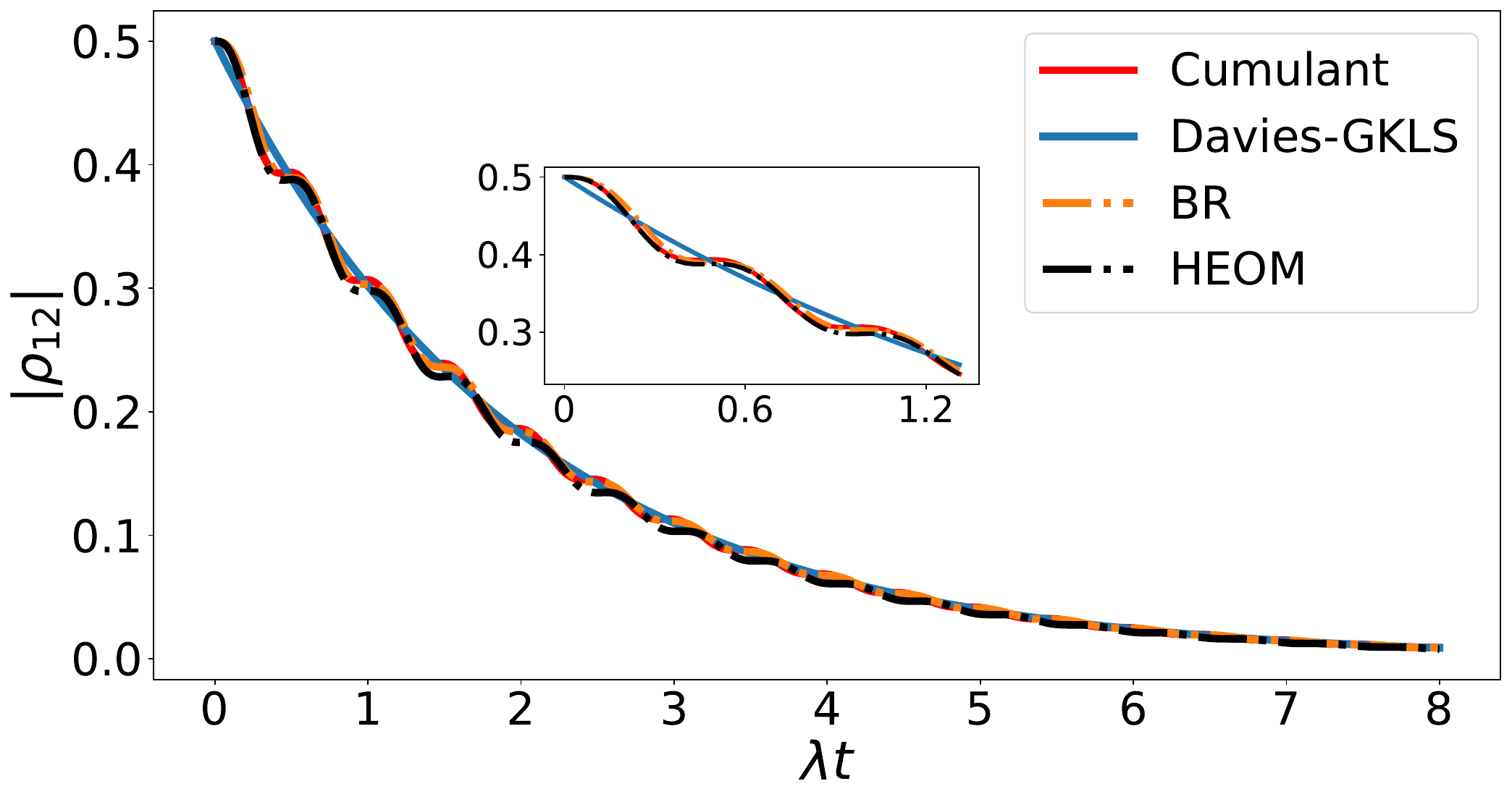}
        \put (-6,45) {\Large$\displaystyle (a)$}
\end{overpic}
\begin{overpic}[width=0.45\textwidth]{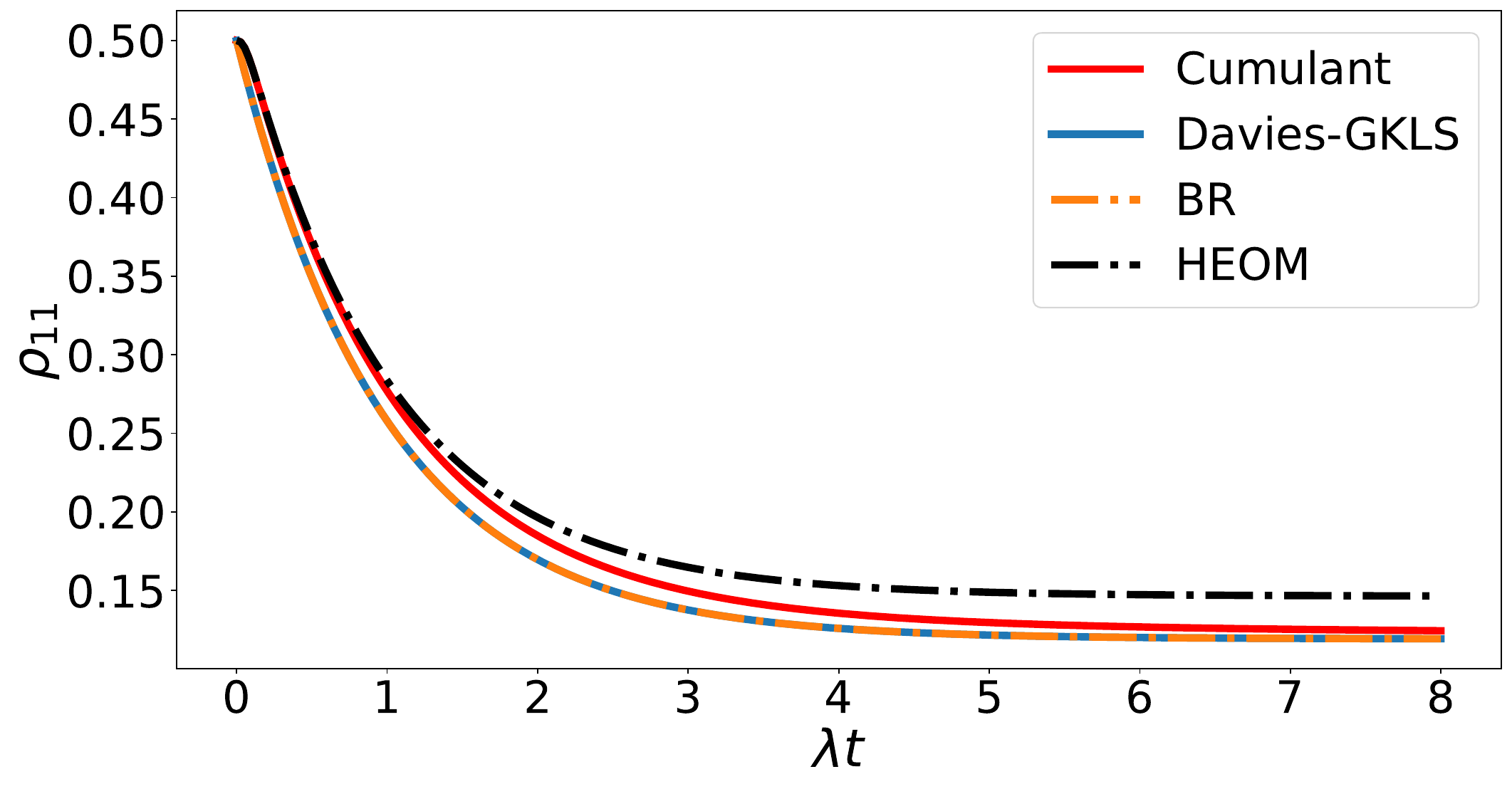}
        \put (-6,45) {\Large$\displaystyle (b)$}
\end{overpic}
\begin{overpic}[width=0.45\textwidth]{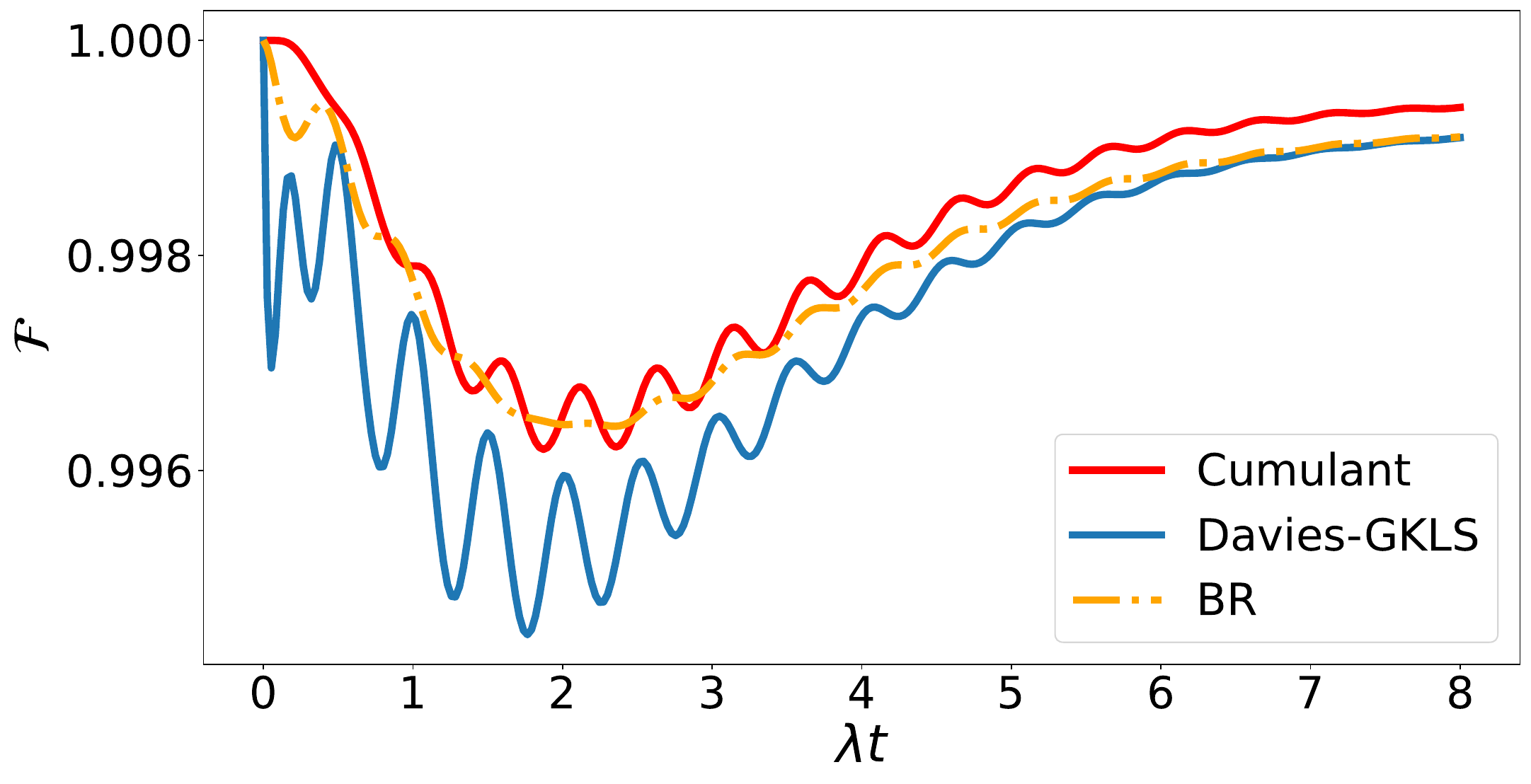}
        \put (-6,45) {\Large$\displaystyle (c)$}
\end{overpic}
\caption{(a) Shows the absolute value of coherence for the different approaches, an inset is added so that one can clearly discern the advantages of the cumulant and Bloch-Redfield methods when describing coherences (b) Shows the evolution of the population for each of the different approaches (c) Shows the fidelity with respect to the HEOM method. The dynamics for the Spin-Boson shown corresponds to $f_{3}=f_{2}=0$ and $f_{1}=1$, The parameters used were $\frac{\omega_{0}}{T} = 2$,$\frac{\lambda}{\gamma}=0.01$, $\gamma=5 \omega_{0}$. }\label{fig:sx}
\end{figure} 

This can obtain analytically in the long time limit, substituting the adequate time dependent coefficients into Eq. \eqref{eq:Decays}:

\begin{align}
\Gamma_{--}+\Gamma_{++} &=  t^{2} \int_{0}^{\infty} d\omega J(\omega) \coth\left( \frac{\omega}{2T}\right)  \nonumber \\ &\left( \sinc^{2}\left(  \frac{(\omega-\omega_0)t}{2} \right) + \sinc^{2}\left(  \frac{(\omega+\omega_0)t}{2} \right) \right) \\
\Gamma_{++}-\Gamma_{--} &= t^{2} \int_{0}^{\infty} d\omega J(w)  \nonumber \\ &\left( \sinc^{2}\left(  \frac{(\omega-\omega_0)t}{2}  \right)- \sinc^{2}\left(  \frac{(\omega+\omega_0)t}{2} \right) \right) \label{eq:Integrable}
\end{align}
Since we only care about the steady state of the system at this point we may take the limit of $t\xrightarrow{}\infty$ which greatly simplifies the analysis as the limit of the time-dependent function as :

\begin{align}
    \lim_{t\xrightarrow{}\infty} t^{2} \sinc^{2} \left( \frac{(\omega-\omega_{0})t}{2}\right) = 2 \pi t \delta(\omega - \omega_0)
\end{align}
Thus we have
\begin{align} \label{eq:steady_state}
    \rho_{steady} &- \frac{\mathbb{I}}{2} = -\frac{\Gamma_{--}-\Gamma_{++}}{ 2 \Gamma}  \s{z} \\&= \frac{ \int_{0}^{\infty} d\omega J(\omega) \left( \delta(\omega-\omega_0) - \delta(\omega+\omega_{0})\right) \s{z}}{ 2  \int_{0}^{\infty} d\omega J(\omega) \coth\left(\frac{\omega}{2T}\right) \left( \delta(\omega-\omega_{0}) + \delta(\omega+\omega_{0})\right) }   \\ &= \frac{1}{2} \left( \frac{J(\omega_{0})- J(-\omega_{0})}{\left( J(\omega_{0}) - J(-\omega_{0}) \right) \coth\left(\frac{\beta \omega_{0}}{2}\right)}\right) \s{z}
\end{align}
Which can finally be written down as:

\begin{align}
    \rho_{steady}=\frac{1}{2} \left( \mathbb{I} + \tanh\left( \frac{\beta \omega_0}{2}\right) \sigma_z \right)
\end{align}

Thus the steady state is just Gibbs state to bare Hamiltonian, in agreement to what is expected for a second order description \cite{reconciliation,Purkayastha2020} as off-diagonal corrections vanish for orthogonal couplings in the Spin-Boson model. Regarding the true evolution, we know that 
indeed there are no steady state coherences \cite{winczewski2023renormalization,Purkayastha2020,reconciliation} in this scenario, but there are corrections to populations that are a consequence of $\mathcal{O}(\lambda^{4})$ effects, so we see that in this case the cumulant equation does not reach the correct 
the stationary state, as opposed to what was expected \cite{reconciliation}. 
Let us however remark here that corrections for populations can be derived from so called quasi-steady state computed perturbatively from cumulant equation (see Ref. \cite{reconciliation}) and that other master equations (Davies-GKLS and Bloch-Redfield) fail in the same way. As we see in Fig. \ref{fig:sx} initially the populations 
are better described by cumulant than the other master equations, which can be considered as a trace of 
the mentioned corrections of \cite{reconciliation}.


Since in our linear map the dynamics of populations and the dynamics of coherences are decoupled this allows one to exponentiate the matrix from Eq. \eqref{eq:M_sx} without making any sort of assumption, which in turn allows us to write the evolution of the population and coherences compactly, since we are neglecting the Non-Markovian Lambshift in this section, the analytical solution becomes even simpler, we obtain:

\begin{align}
    \expect{\sigma_z(t)} &= e^{-\Gamma(t)} \expect{\sigma_{z}(0)} + \frac{(e^{-\Gamma(t)}-1) (\Gamma_{++}(t)-\Gamma_{--}(t))}{ \Gamma(t)} \\
    \expect{\s{+}(t)} &= \frac{e^{-|\Gamma_{-+}|-\frac{\Gamma}{2}}}{2 |\Gamma_{-+}|} \Bigg( \Gamma_{-+} \expect{\sigma_{-}(0)} (e^{2 |\Gamma_{-+}|} - 1)  \nonumber \\  &+ |\Gamma_{-+}| \expect{\sigma_{+}(0)} (e^{2 |\Gamma_{-+}|} + 1) \Bigg)\\ 
   \expect{\s{-}(t)} &= \frac{e^{-|\Gamma_{-+}|-\frac{\Gamma}{2}}}{2 |\Gamma_{-+}|} \Bigg( \Gamma_{-+} \expect{\sigma_{+}(0)} (e^{2 |\Gamma_{-+}|} - 1)  \nonumber \\  &+ |\Gamma_{-+}| \expect{\sigma_{-}(0)} (e^{2 |\Gamma_{-+}|} + 1) \Bigg)
\end{align}

In terms of the dynamics one can see that the cumulant equation is the most favorable approximate description for the parameters chosen in Fig. \ref{fig:sx}. As we are in the weak coupling regime, we can see a pretty good agreement with the exact dynamics even for the simplest equation, namely, the Davies-GLKS. Still the cumulant equation is a better description in early and long times when compared to the other master equations presented here, it is also as good as Bloch-Redfield in intermediate times, this indicates higher orders of the cumulant equation might be worth pursuing in the same way the TCL equation is often written down.

\begin{figure}[h!]
\begin{overpic}[width=\textwidth]{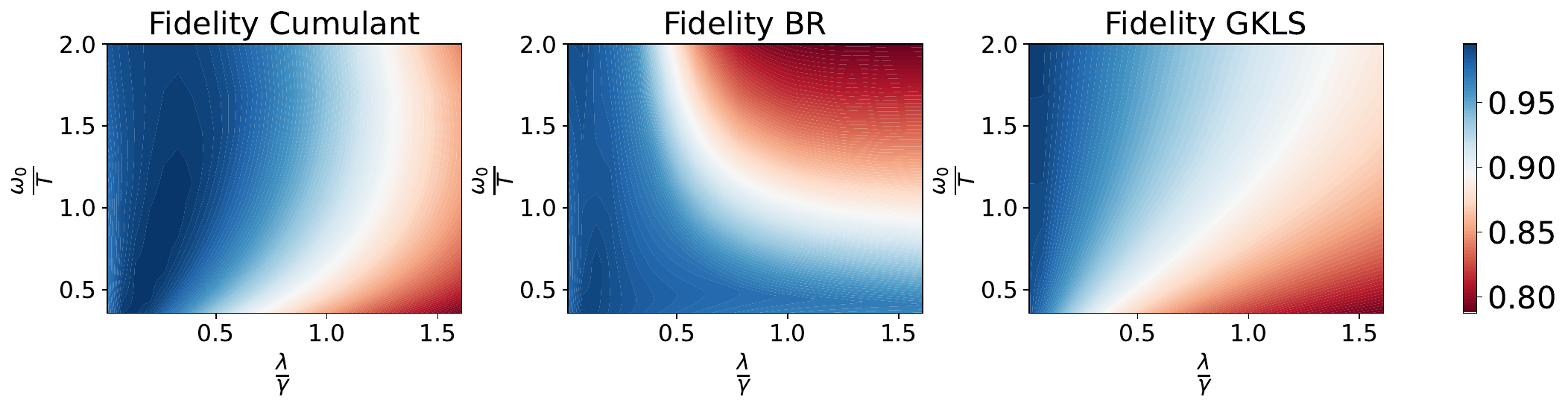}
        \put (0,22) {\Large$\displaystyle (a)$}
\end{overpic}
\centering
\begin{overpic}[width=0.35\textwidth]{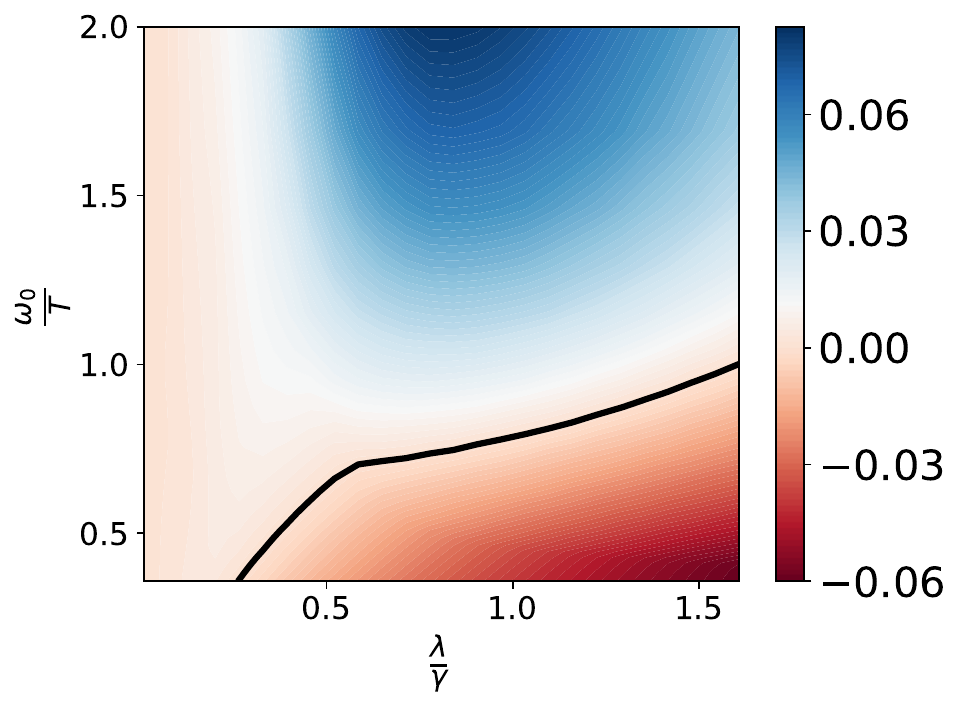}
        \put (-6,70) {\Large$\displaystyle (b)$}
\end{overpic} \quad\quad\quad
\begin{overpic}[width=0.35\textwidth]{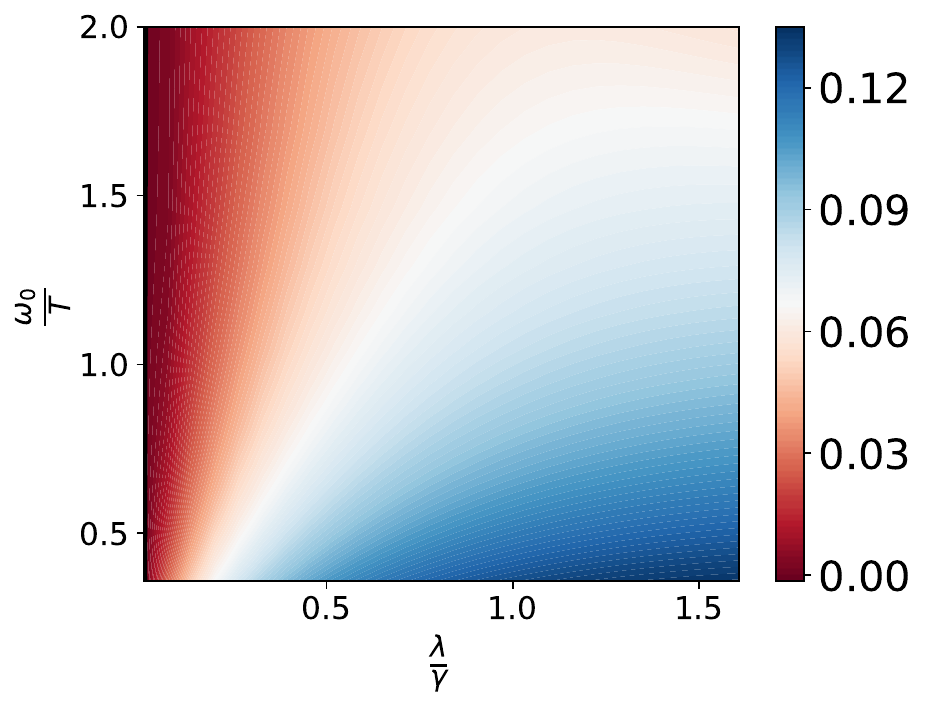}
        \put (-6,70) {\Large$\displaystyle (c)$}
\end{overpic}
\caption{a) Shows the minimum fidelity of each of the three approximate descriptions considered with respect to HEOM for a wide range of parameters b)Shows the difference in minimum fidelity between the cumulant equation and the Bloch Redfield equation, the black line divides the region where the difference is positive (cumulant has higher minimum fidelity) and where it is negative. c) Shows the difference in minimum fidelity between the cumulant equation and the Davies-GKLS equation, the cumulant equation is a better description for all chosen parameters}\label{fig:fid_sx}
\end{figure}

In figure \ref{fig:fid_sx} we see that the cumulant equation is not only better for our chosen parameters, but that it is a better description for a wide range of parameters. Particularly for the example considered, it is only a worse description in the regime of high ``$\frac{\lambda}{\gamma}$" and high temperature, a regime where we expect second order descriptions to fail.   Bare in mind that the  black line that divides the regime where cumulant is better that Bloch-Redfield may change shape for different values of $\gamma$ with respect to $\omega_{0}$ and different spectral densities, here we have chosen $\gamma=\omega_{0}$. However in general we see that the  cumulant equation is a better description at small $\frac{\lambda}{\gamma}$ and low temperature, while Bloch Redfield is better in the opposite regime.

\onecolumngrid

\vfill
\break
\twocolumngrid
\subsection{The Composite interaction case}

In the previous two sections we recovered models for which the cumulant equation has been studied in the literature, the difference so far  has been on how to solve the cumulant dynamics \cite{Rivas} and the comparison to other master equations, however we believe our way of solving might prove useful for analytical calculations, specially when approximation of the integrals will be used in the same spirit as in \cite{winczewski2021bypassing,approx_plenio,3lvl}. 

\begin{figure}[htb!]

\begin{overpic}[width=0.45\textwidth]{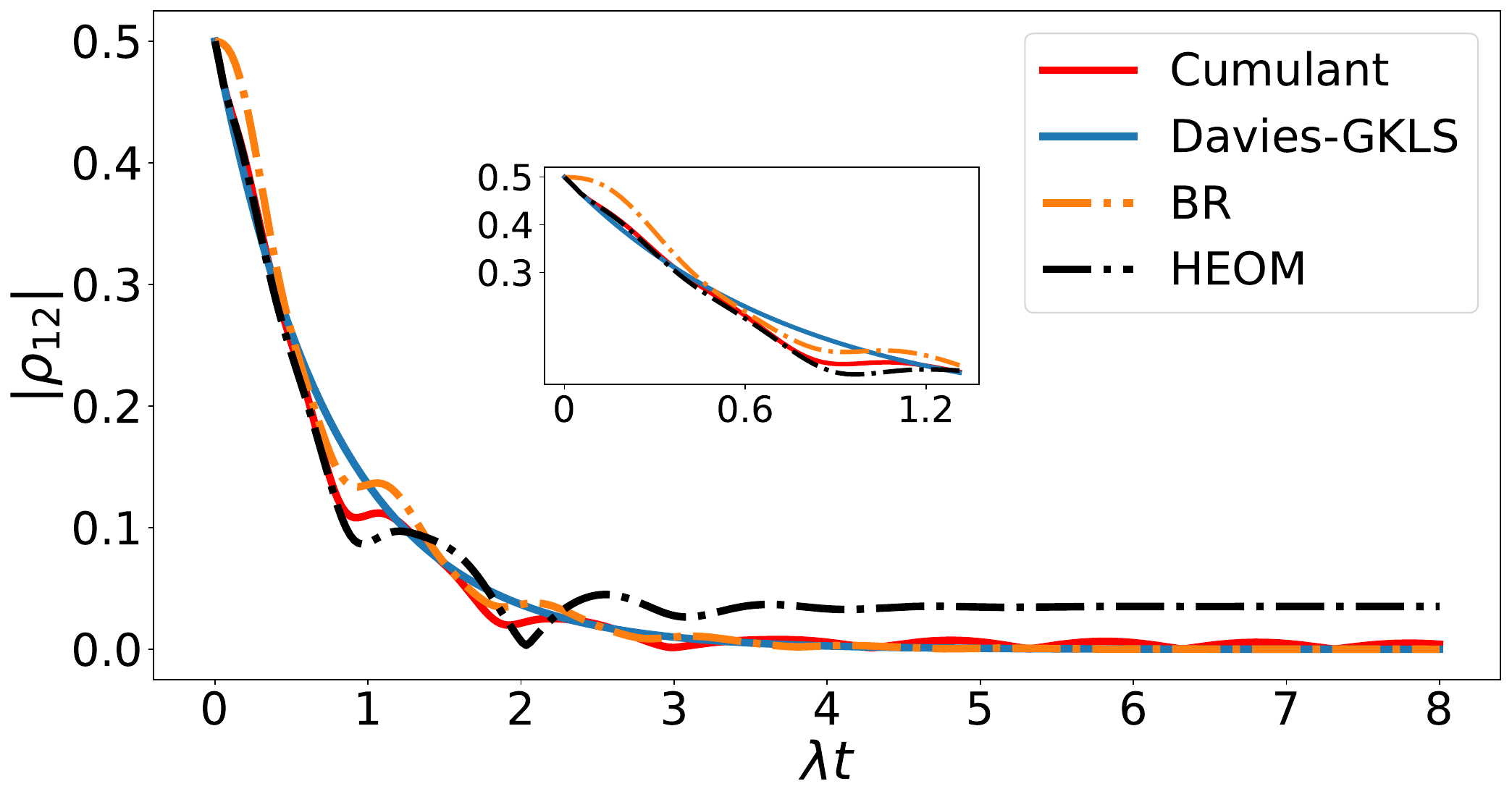}
        \put (-6,45) {\Large$\displaystyle (a)$}
\end{overpic}
\begin{overpic}[width=0.45\textwidth]{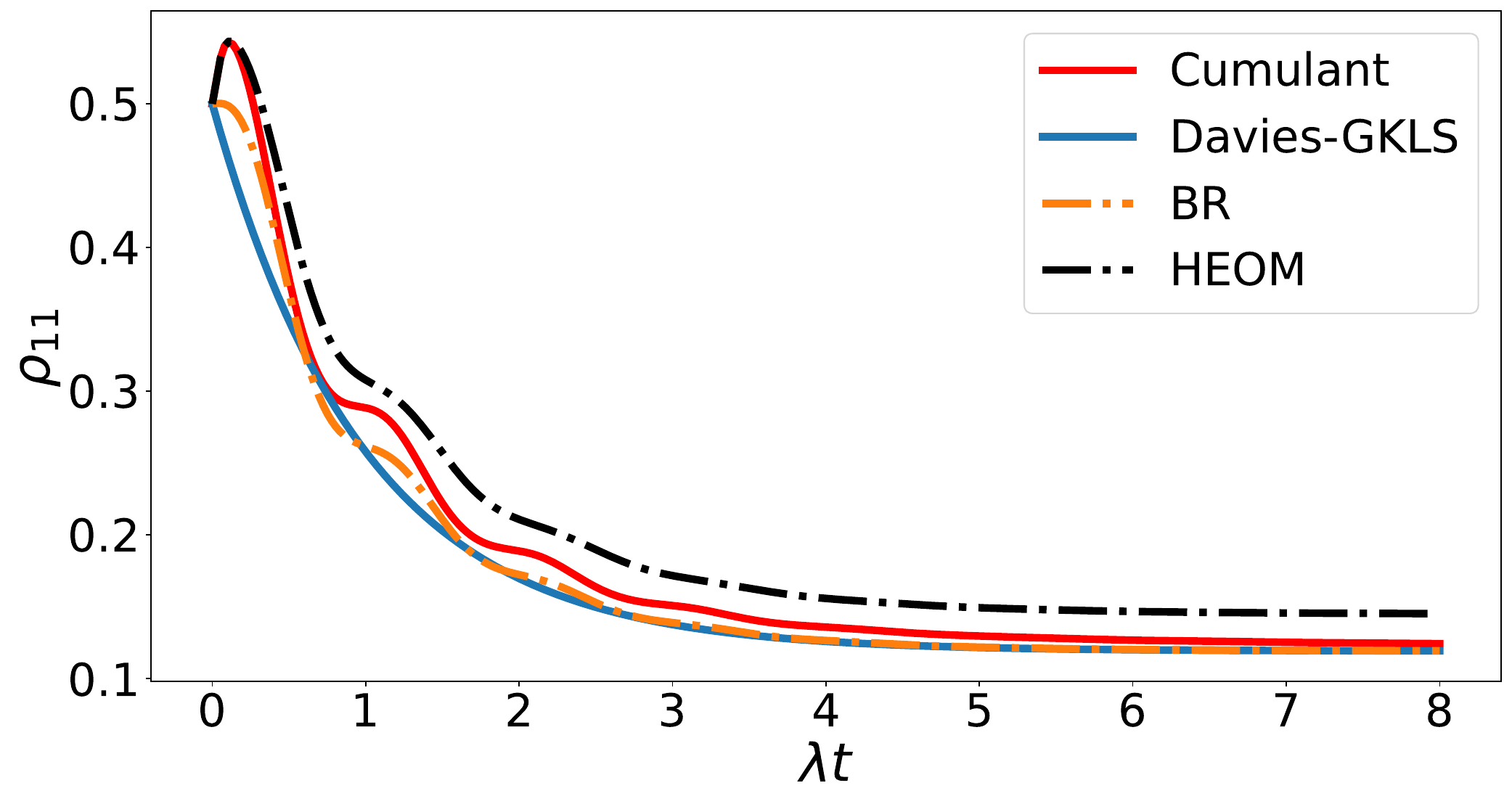}
        \put (-6,45) {\Large$\displaystyle (b)$}
\end{overpic}
\begin{overpic}[width=0.45\textwidth]{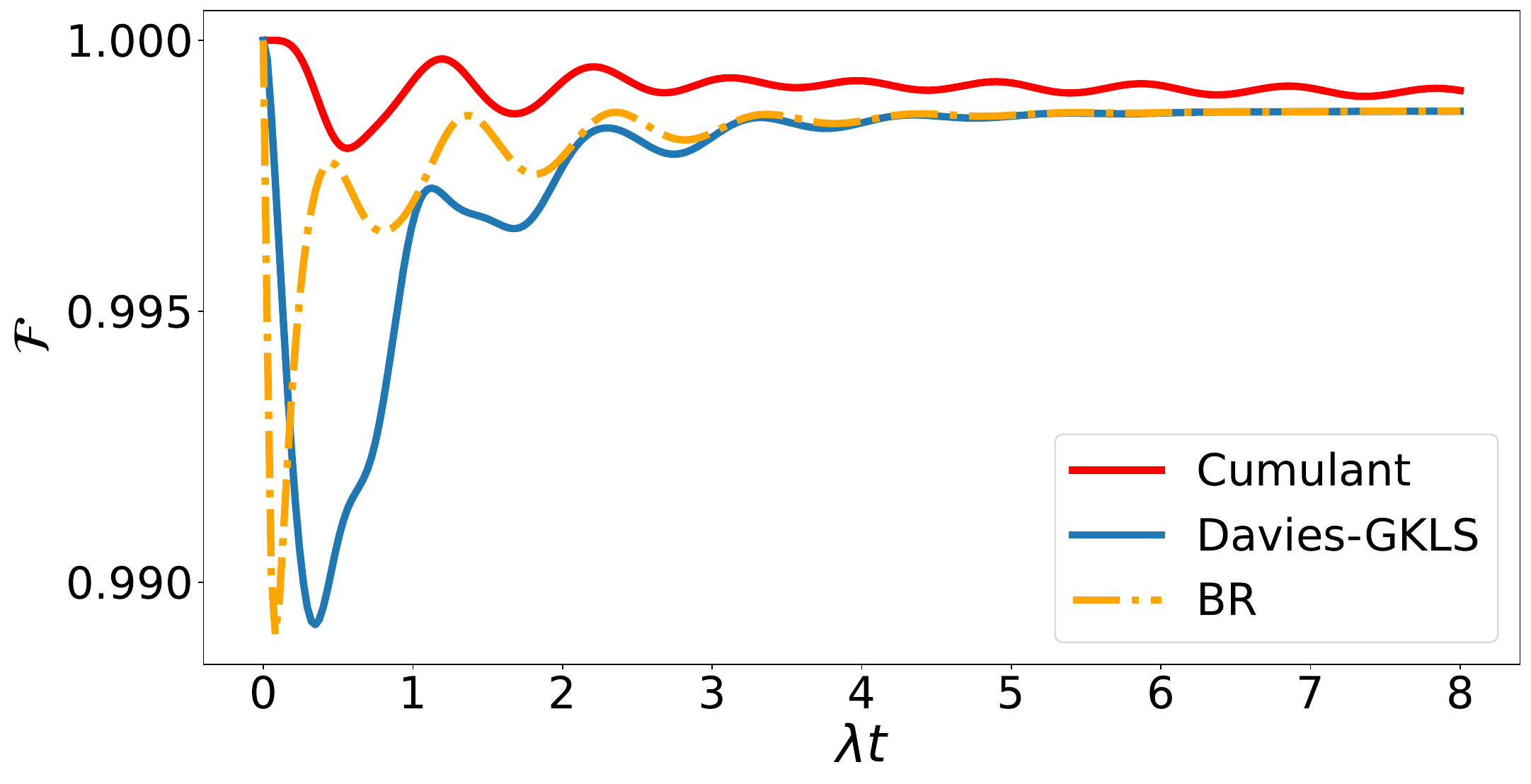}
        \put (-6,45) {\Large$\displaystyle (c)$}
\end{overpic}
    \caption{(a) Shows the absolute value of the coherence for the different approaches considered (b) Shows the evolution of the population for the different approaches considered (c) Shows the fidelity with respect to HEOM for each of the methods considered. The plots corresponds to $f_{2}=0$ and $f_{3}=f_{1}=1$, for the plots shown $\frac{\omega_{0}}{T} = 2$,$\frac{\lambda}{\gamma}=0.01$. }\label{fig:sx_sy}
\end{figure} 
A first attempt in this direction was put forward in \cite{winczewski2021bypassing}, our results are now bench-marked with respect to other approaches, showing agreement and even advantage.In this section we extend these models by including additional couplings, which is sometimes called the Non-equilibrium spin-Boson model \cite{Xu2016}, see appendix \ref{ap:qubit_model} to see how Hamiltonian \eqref{eq:Hamiltonian} is related to this model. In the case of composite interactions the matrix is rank 3, which makes analytical expressions complicated as they depend on a cubic root, and therefore it's hard to get Eq. \eqref{eq:exp_cum} in a compact form unless approximations are done on the integrals. 
The result of the dynamics can be observed in the Fig. \ref{fig:sx_sy}.  
\begin{figure}[h!]
\centering
\begin{overpic}[width=0.45\textwidth]{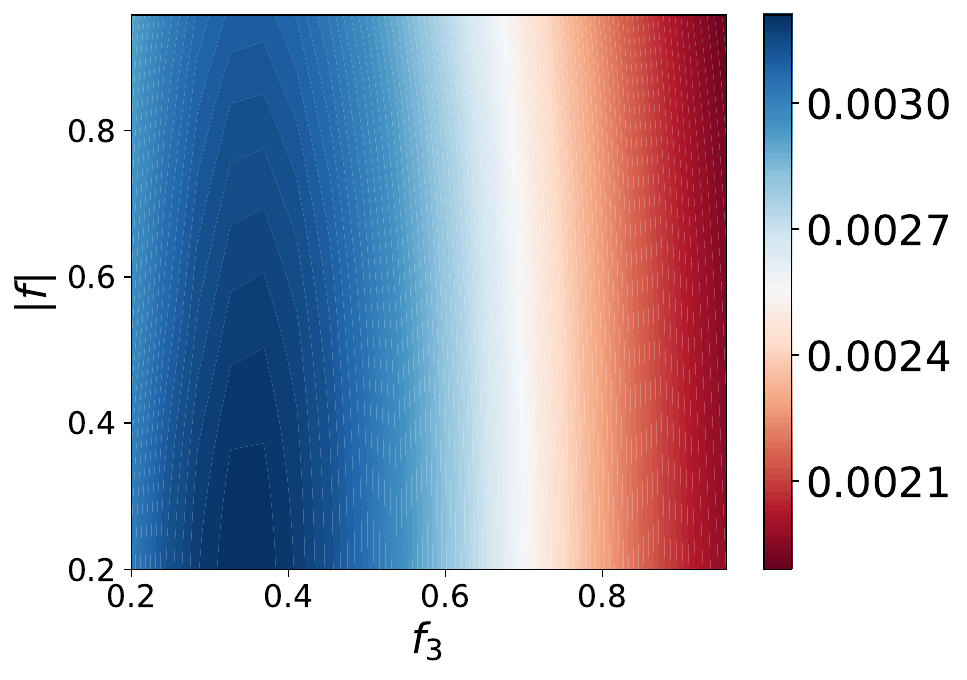}
        \put (-3,65) {\Large$\displaystyle (a)$}
\end{overpic}
\begin{overpic}[width=0.45\textwidth]{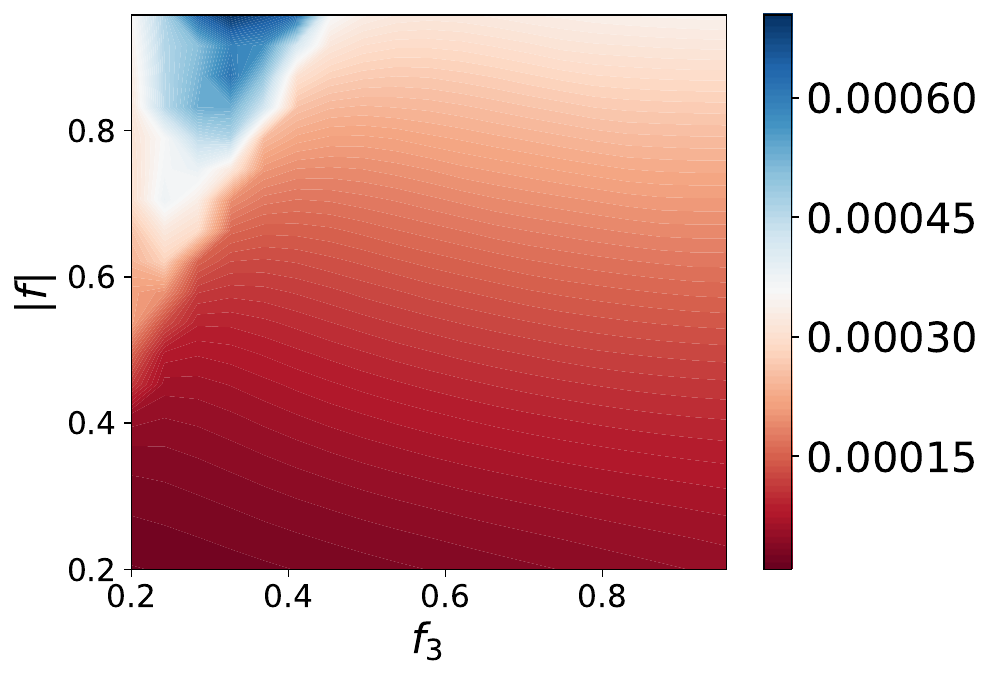}
        \put (-3,65) {\Large$\displaystyle (b)$}
\end{overpic}
  \caption{(a) The minimum fidelity of the cumulant compared to the Davies-GKLS equation (b) The minimum fidelity of the cumulant compared to the Bloch-Redfield equation. These plots show the different in minimum fidelity between the standard master equations and the cumulant equation, when the couplings between the parallel and orthogonal components of the Hamiltonian are imbalanced. For the simulations we fixed $\frac{\omega_{0}}{T}=2,\frac{\lambda}{\gamma} =0.01$}\label{fig:imbalanced}
\end{figure}
We can see that in this case the cumulant equation has an advantage over the more standard Bloch-Redfield equation while the advantage is small. The fact that the cumulant equation can reproduce better the early time populations is the main source of it's advantage with respect to the Bloch-Redfield equation. While it looses this advantage in time, as it is closer in populations while it slowly goes to the bare Gibbs state it has a better fidelity until reaching the steady state. 

Notice that small oscillations in coherence remain in the cumulant equation while the other master equations have already reached a fixed value. This is due to the slow decay of the cumulant equation to the Davies-GKLS steady state, as we have mentioned this gives a reminiscence of the steady state coherences as they are long lived.

One may wonder if the parameters chosen above namely equal $f_{1}=f_{3}$ played a role in the superiority of the cumulant equation, Fig. \ref{fig:imbalanced} shows that this is not the case by showing the difference in minimum fidelities over a range of different $|f|$ and $f_{3}$.

\begin{figure}[h!]

\begin{overpic}[width=\textwidth]{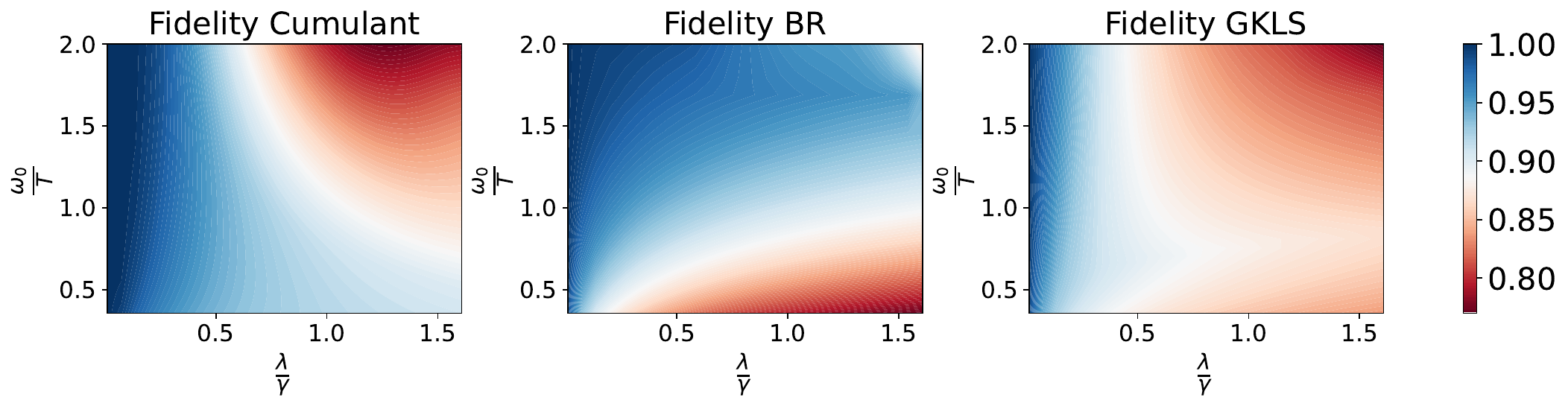}
        \put (-3,22) {\Large$\displaystyle (a)$}
\end{overpic}
\centering

\begin{overpic}[width=.35\textwidth]{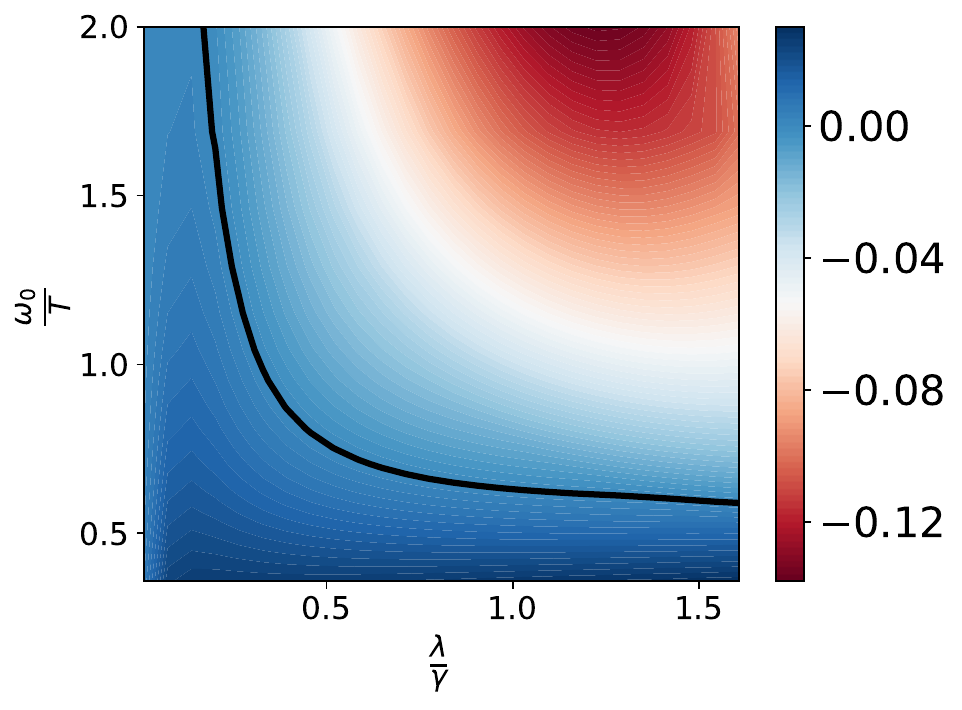}
        \put (-6,65) {\Large$\displaystyle (b)$}
\end{overpic}
 \quad \quad \quad
\begin{overpic}[width=.35\textwidth]{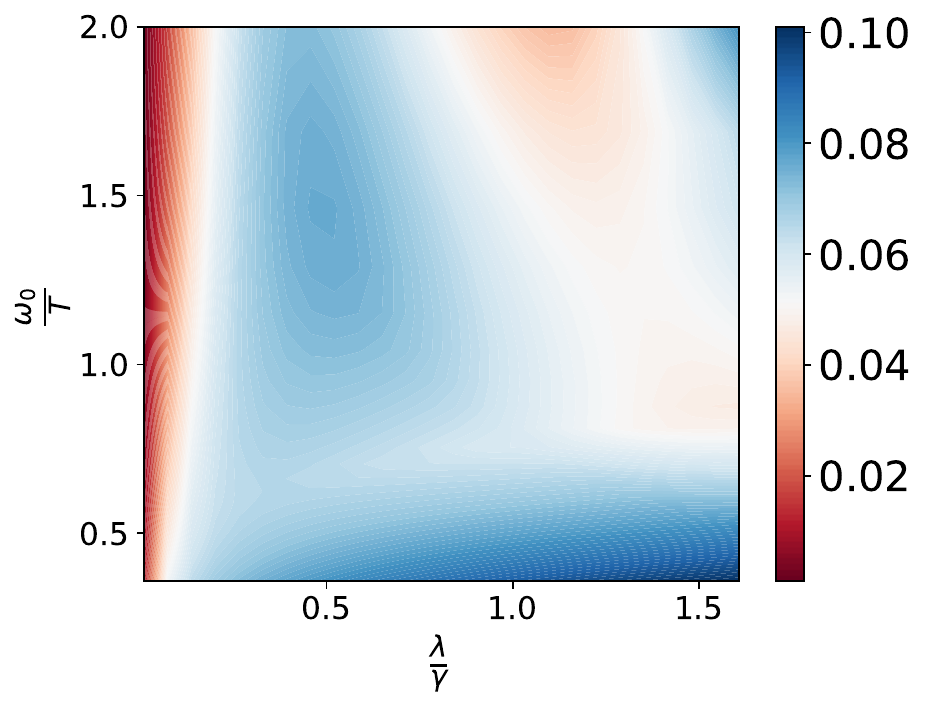}
        \put (-6,65) {\Large$\displaystyle (c)$}
\end{overpic}

\caption{a) Shows the minimum fidelity of all master equations against HEOM. Both plot (b) and (c) show Difference in minimum fidelity between the cumulant equation and (b)the Bloch Redfield equation, the black line divides the region where the difference is positive (cumulant has higher minimum fidelity) and where it is negative. (c) the Davies-GKLS equation, the cumulant equation is a better description for all chosen parameters.}\label{fig:fid_sx_sz}
\end{figure}

In this section we observed that the Bloch-Redfield equation regime of validity seems to be larger than that of the cumulant equation/refined weak coupling as mentioned in \cite{hartmann_accuracy_2020}. The cumulant equation is a more faithful description when $\frac{\lambda}{\gamma} \ll 1$ which corresponds to the weak coupling regime where master equations are valid descriptions.  As Figure \ref{fig:fid_sx_sz} shows the cumulant description is more faithful in the regime of of small ``$\frac{\lambda}{\gamma}$" and low temperature, just as in figure \ref{fig:fid_sx}, while for stronger coupling which one is a better description seems model dependent, from these examples we can conclude that the cumulant equation is a better description at low temperature in the weak coupling regime 

\onecolumngrid

\section{The effect on Lamb-shift on the cumulant equation}

Most work on open quantum systems that has been done with master equations neglects Lamb-shift  corrections \cite{albert2018lindbladians,Jeske,cattaneo_local_2019,lidar2020lecture,vinjanampathy_quantum_2016,rivascritical}. One of the main reasons people give for neglecting it, is that it is usually small (or that it's effect is small which follows from this). 
 However, this does not make sense as this is a cut-off dependent quantity and might diverge in general. Another, stronger argument is that if the shift occurs in the Hamiltonian part it must also occur in the dissipative part, otherwise there is some kind of inconsistency in the equation. People who advocate for the latter argument say that one should take the physical (already dressed/Lamb-shifted ) Hamiltonian and derive the master equation in its interaction picture where no Lamb-shift exists, this would amount to having a master equation where Lamb-shift is neglected, however its interpretation is different, for a more in-depth discussion see \cite{winczewski2023renormalization}. This argument is at least solid and justifies neglecting Lamb-shift when it commuted with the original system Hamiltonian or when it is time independent as it is a constant shift so the rotation maps a time independent problem to a time independent problem.  In the opposite case such a rotation would map a time independent problem into a time dependent one, so taking a time independent Hamiltonian in this renormalized picture does not make sense or at the very least needs further justification.

\begin{figure}[h!]
\begin{overpic}[width=.33\textwidth]{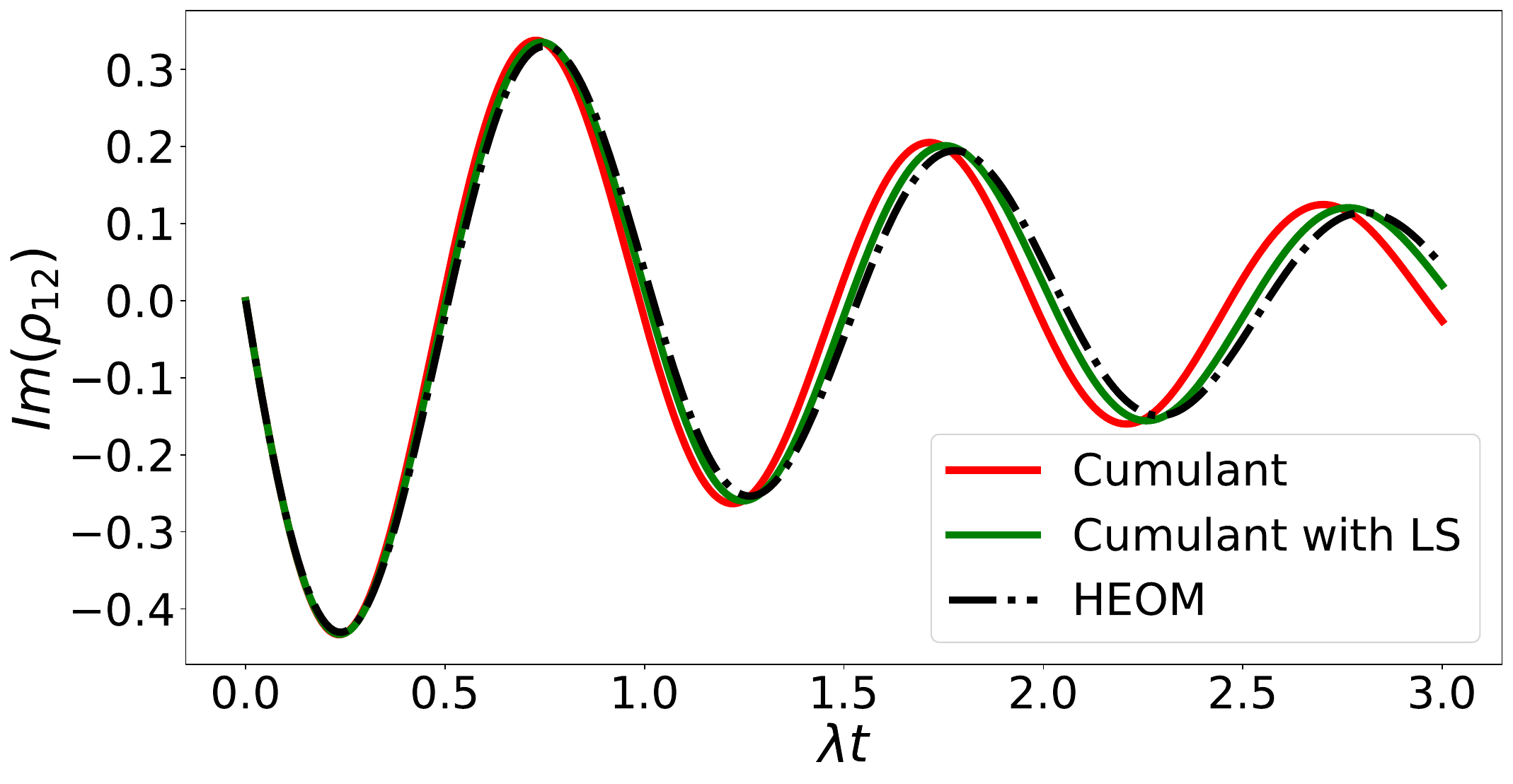}
        \put (14,42) {\Large$\displaystyle (a)$}
\end{overpic}
\begin{overpic}[width=.33\textwidth]{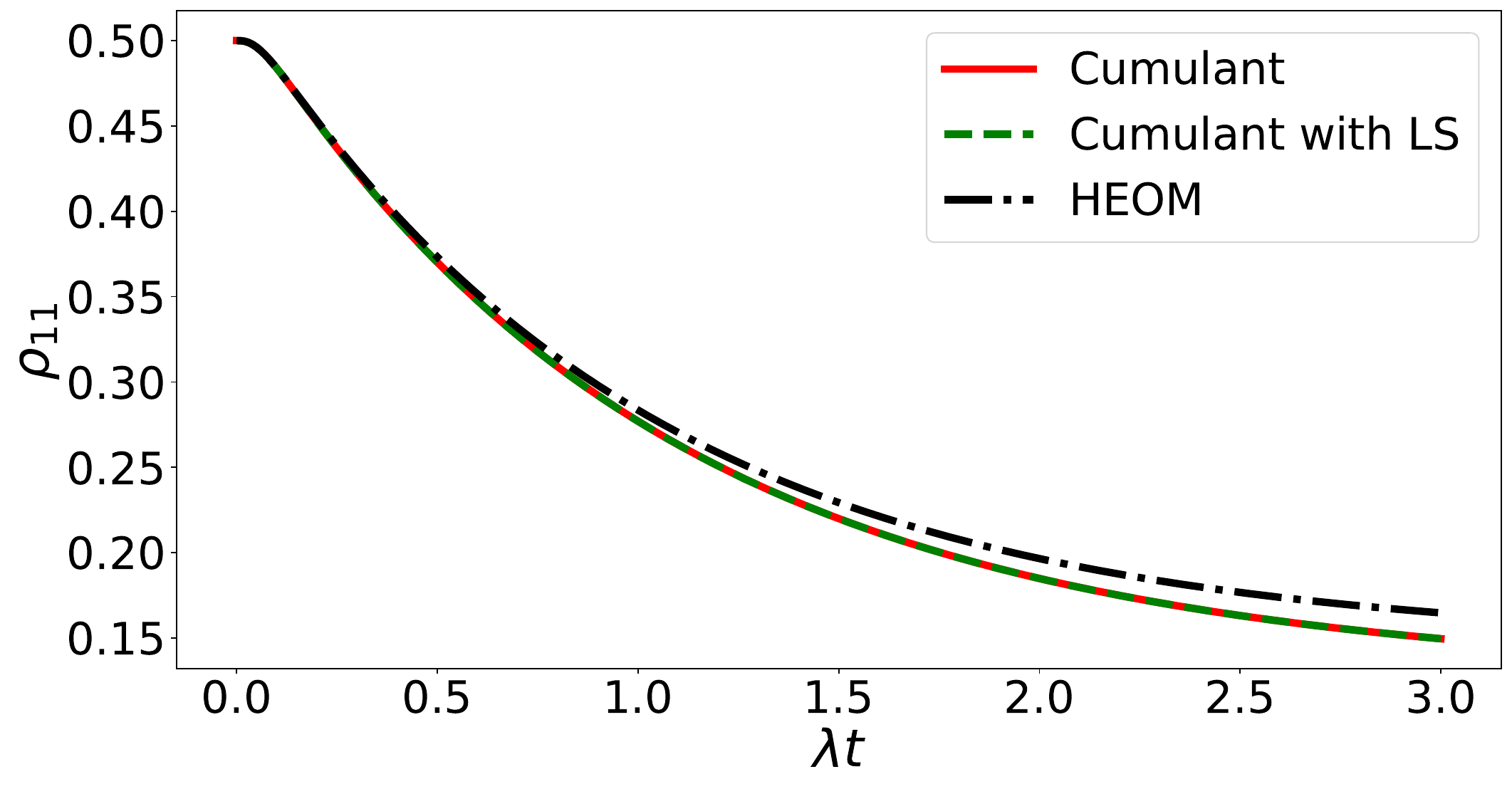}
        \put (14,34) {\Large$\displaystyle (b)$}
\end{overpic}
\begin{overpic}[width=.33\textwidth]{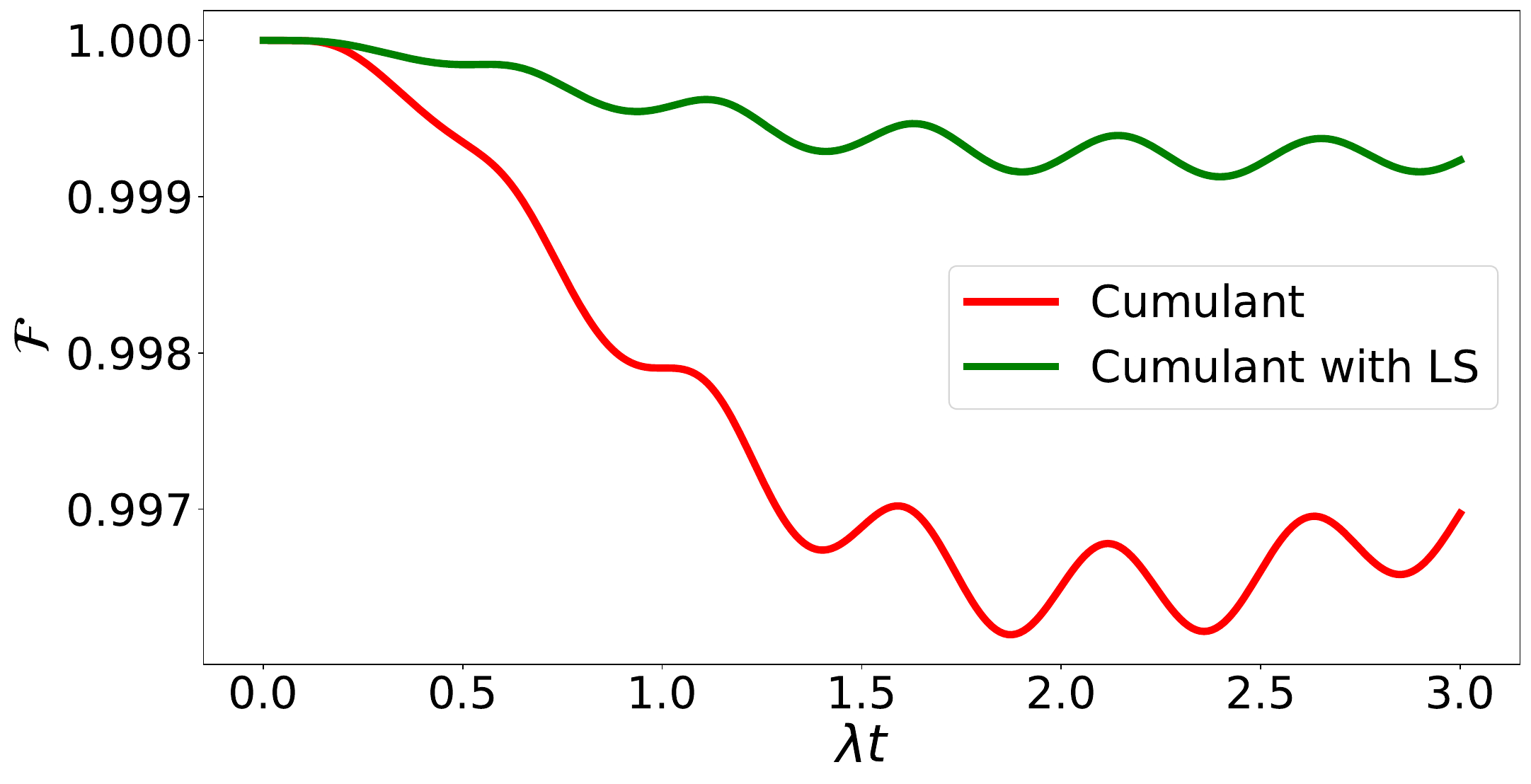}
        \put (14,35) {\Large$\displaystyle (c)$}
\end{overpic}
\begin{overpic}[width=.33\textwidth]{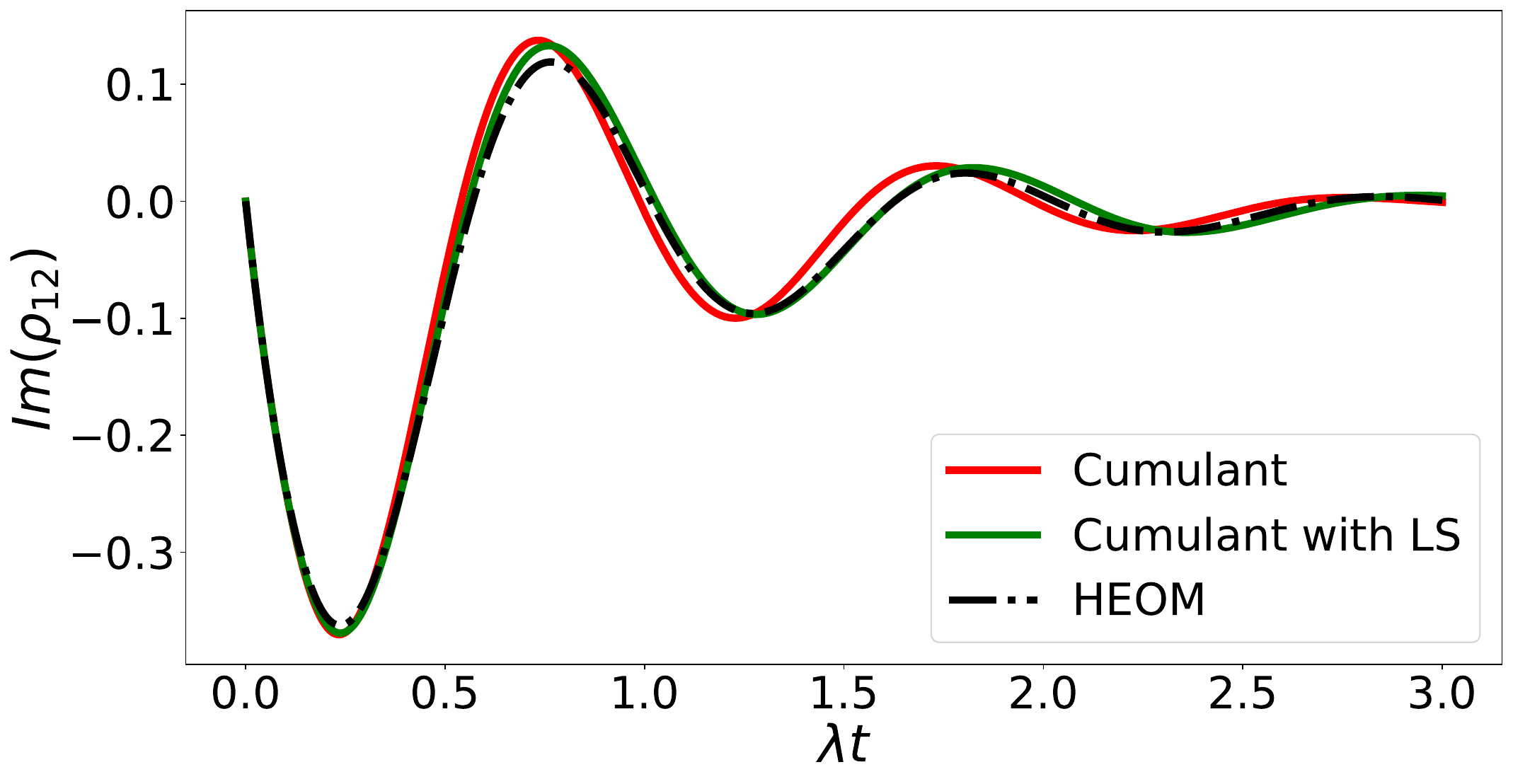}
        \put (12,42) {\Large$\displaystyle (d)$}
\end{overpic}
\begin{overpic}[width=.33\textwidth]{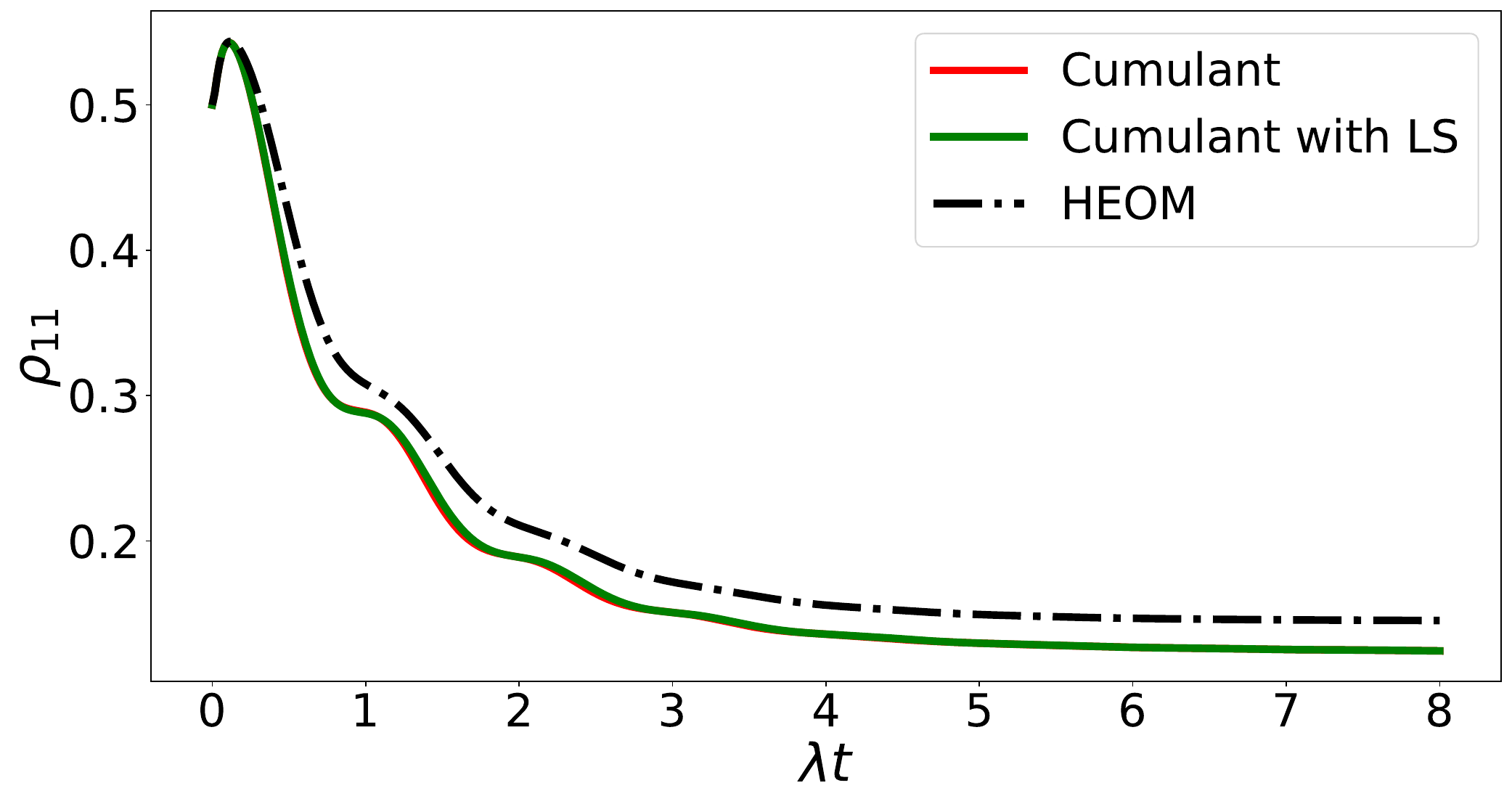}
        \put (14,14) {\Large$\displaystyle (e)$}
\end{overpic}
\begin{overpic}[width=.33\textwidth]{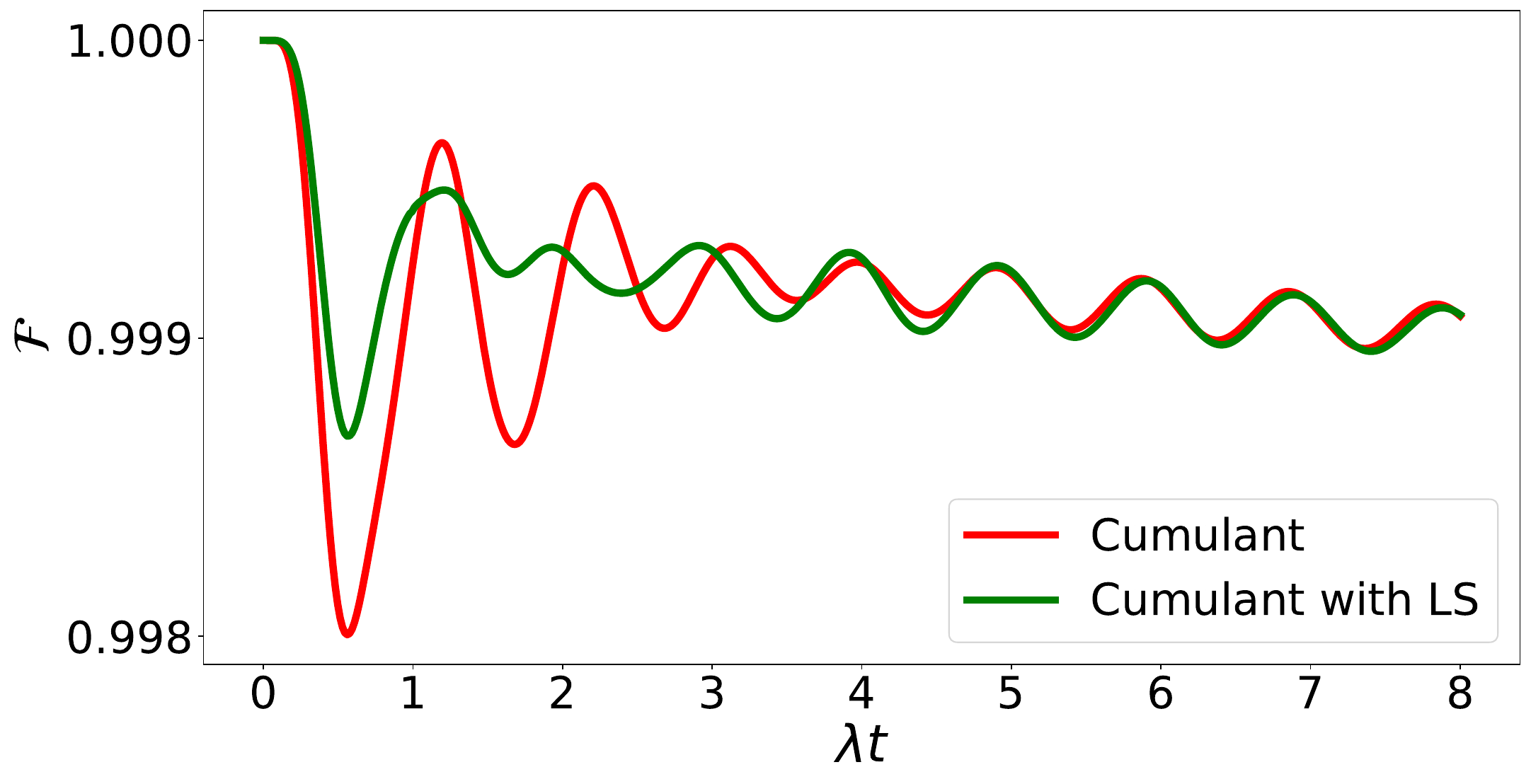}
        \put (34,42) {\Large$\displaystyle (f)$}
\end{overpic}
    \caption{The dynamics for the Spin-Boson model are shown. For plots (a)-(c) it corresponds to $f_{3}=f_{2}=0$ and $f_{1}=1$, while for plots (d)-(f) it corresponds to $f_{2}=0$ and $f_{3}=f_{1}=1$.  The parameters used for the simulations were $\frac{\omega_{0}}{T} = 2$,$\frac{\lambda}{\gamma}=0.01$  and $\gamma=5\omega_{0}$. (a)The imaginary part of coherence when  $H_{I} \propto \s{x}$. (b) Population when $H_{I} \propto \s{x}$ (c) Fidelity with respect to HEOM when $H_{I}\propto \s{x}$ (d)The imaginary part of coherence when  $H_{I} \propto \s{x}+\s{z}$. (e) Population when $H_{I} \propto \s{x}+\s{z}$ (f) Fidelity with respect to HEOM when $H_{I}\propto \s{x}+\s{z}$ }\label{fig:sx_sy_ls}
\end{figure} 

While Lamb shift can always be seen as a constant drift in the case for Markovian master equations it may not always be true when dealing with Non-Markovian master equations, as it may not always commute with the Hamiltonian of the system. In particular, for this simple model it is not the case as was shown in \cite{Guarnieri,reconciliation}. This happens when we have composite interactions and it may be the reason master for which equations show the so-called steady state coherences. The fact that the the numerically exact solution (HEOM) exhibits such a steady state coherence may point to the direction that Lamb-shift does indeed make the master equation more faithful and therefore one should indeed shift. 

When it comes to the cumulant equation Lamb-shift, as far as we are aware of its effect on dynamics has not been studied elsewhere (in \cite{Rivas} Lamb-shift is calculated but its effects on dynamics are not observed or discussed  ). Here we see, that it may be beneficial to use the time dependent Lamb-shift as one gets a higher fidelity at the expense of more computation time. However, we may notice that the main effect of the Lamb-shift is to have a better description of the coherences, as the Lamb-shift correction makes the coherence be in sync with the numerically exact solution. It's effects on populations is negligible though, so when looking at longer time scales using the Lamb-shift correction may not be worth it, as the cumulant equation will slowly converge to the steady state solution of the Davies-GKLS equation.  On short and intermediate time scales we can see an improvement by using the Lamb-shift correction on average, however at certain times our fidelity is slightly reduced due to amplitude mismatch in some cases, as it can be seen from the Fig. \ref{fig:sx_sy_ls}. The findings of this paper seem to indicate that Lamb-shift is indeed beneficial to the descriptions of dynamics, however the improvement it obtains for the system is small, it remains an open question whether it can have a big impact on other systems.

\section{Conclusions}

We have revisited the study of the Spin-Boson model in the cumulant/refined weak coupling approach and proposed an alternative approach to solve it by using the fact that the super operator acts as a linear map. This allows us to come up with a simple analytical answer when contrasted with the previous studies in the literature.  We have also  extended the literature by studying the so called non-equilibrium Spin-Boson model, and shown that in the weak coupling regime the cumulant equation corresponds better to the true dynamics than other popular approaches such as the Bloch-Redfield equation, showing that the cumulant equation is a good candidate to study dynamics in this regime. This regime is the most important one for this model as it corresponds to a charge qubit (see Appendix \ref{ap:qubit_model}).

One big part of the advantage of the cumulant equation is that it picks up a correction to the diagonal of the density matrix in finite time which slowly vanishes.
Therefore it is overall quite good for a pretty long time. 

On the other hand, cumulant equation exhibit slowly dying oscillations of coherences. 
In \cite{cattaneo2021comment,Guarnieri}
such oscillations were also observed under Bloch-Redfield master equation, yet they were not dying. They looked like steady ones, but this was the effect of second order expansion. 
For longer times this is not anymore valid, and the oscillations have been also slowly dying. It was argued in \cite{cattaneo2021comment} that the presence of these is due to violation of complete positivity. However we now see that the completely positive cumulant equation also exhibit such oscillations, and therefore we believe that it may be instead a feature of the master equations based on Born approximation. 
We thus believe that higher orders of the cumulant expansion might cure this. 


Overall, the cumulant equation provides a useful tool for the study of open systems in the weak coupling regime, especially when time dynamics rather than steady state is considered.  Our form of its solution allows one to separate the transient dynamics from the steady state dynamics in a convenient fashion, that may prove useful in the study of quantum heat engines.

\begin{acknowledgments} We thank Luis Cort-Barrada, Marek Winczewski, Neill Lambert and Paul Menczel for extensive discussions and valuable suggestions. This work was supported by the International Research Agendas Programme (IRAP) of the Foundation for Polish Science (FNP), with structural funds from the European Union (EU). GS acknkowledges the support by the Polish National Science Centre grant OPUS-21 (No: 2021/41/B/ST2/03207). MH is supported by the QuantERA II Programme (No 2021/03/Y/ST2/00178, acronym ExTRaQT) that has received funding from the European Union’s Horizon 2020.\end{acknowledgments}

\appendix
\onecolumngrid
\section{Derivation of The cumulant equation for the non equilibrium Spin-Boson model}

Let us first find the Lamb-shift which is given by:

\begin{equation*}
\Lambda(t)= \sum_{\omega,\omega'} \xi(\omega,\omega',t) A^{\dagger}(\omega)A(\omega')
\end{equation*}

After expanding the sum this becomes
\begin{align}\label{eq:general_ls}
    \Lambda(t)=\frac{\abs{f}^{2}}{2} \left( \xi_{++}+\xi_{--}\right)+ \frac{\abs{f}^{2}}{2} \left( \xi_{++}-\xi_{--}\right) \s{z} +\bar{f} f_{3} (\xi_{-z}-\xi_{z+}) \s{-}+f f_{3} (\xi_{z-}-\xi_{+z}) \s{+} +f_{3}^{2} \xi_{zz}
\end{align}

As anything commutes with the identity one may drop that term, as well as rewrite this in terms of $\s{x},\s{y}$,but first:

\begin{align}
    \xi(\omega,\omega',t)=\frac{t^{2}}{4 \pi} \int_{-\infty}^{\infty} d\phi \sinc\left( \frac{\omega-\phi}{2} t\right) \sinc\left( \frac{\omega'-\phi}{2} t\right) P.V \int_{0}^{\infty} d\nu \left[ \frac{\gamma(\nu)}{\phi-\nu} + \frac{\gamma(-\nu)}{\phi+\nu} \right]
\end{align}

From there we can see that the terms where one omega is zero are the same, namely $\xi(\omega,0,t)=\xi(0,\omega',t)$ using that:

\begin{align}
   \Lambda(t) &= \frac{\abs{f}^{2}}{2} \left( \xi_{++}-\xi_{--}\right) \s{z} +\bar{f} f_{3} (\xi_{-z}-\xi_{z+}) \s{-}+f f_{3} (\xi_{z-}-\xi_{+z}) \s{+}  \\ 
   &= \frac{\abs{f}^{2}}{2} \left( \xi_{++}-\xi_{--}\right) \s{z} + f_{3} (\xi_{z-}-\xi_{z+})(f_{1} \s{x} + f_{2} \s{y})
\end{align}

Now we define:

\begin{align}
    R[O] &= \sum_{\omega,\omega'} R_{\omega,\omega'}[O] \\
    R_{\omega,\omega'}[O] &= \Gamma(\omega,\omega',t) \left( A(\omega') O A^{\dagger}(\omega) - \frac{\{ A^{\dagger}(\omega) A(\omega ') , O \}}{2} \right)
\end{align}

we will now use sub-indices to refer to omegas $-=\omega,+=-\omega,z=0$ using that notation we may write \footnote{This convention has been chosen to match existing literature, when the reduced model is considered \cite{Rivas}}:

\begin{align}
   \mathcal{Z}[\rho] =-i \left(\frac{\abs{f}^{2}}{2} \left( \xi_{--}-\xi_{++}\right) [\s{z},\rho]+ f_{3} (\xi_{z+}-\xi_{z-})(f_{1} [\s{x},\rho] + f_{2} [\s{y},\rho]) \right) + R[\rho]
\end{align}

However we may write a qubit density matrix as:

\begin{equation*}
    \rho=\frac{1}{2} +b \s{x} + c \s{y}+d \s{z}
\end{equation*}

Since that map acts linearly one may find it on each operator separately, to this quickly let us compute the action of $R_{\omega,\omega'}$ on each operator:

- The identity:
\begin{align}
\begin{matrix}
      R_{++}[I]=- \Gamma_{--} \abs{f}^{2} \s{z} & R_{+-}[I]=0 & R_{z+}[I]= 2 \Gamma_{z-} \bar{f} f_3 \s{-} \\
    R_{-+}[I]= 0  & R_{--}[I]= \Gamma_{++} \abs{f}^{2} \s{z}  & R_{z-}[I] = - 2 \Gamma_{z+} f f_3 \s{+} \\ R_{+z}[I]= 2 \Gamma_{-z}  f_3 f \s{+} & R_{-z}[I]= -2 \Gamma_{+z} f_3 \bar{f} \s{-} & R_{zz}[I]=0  
\end{matrix}
\end{align}


- $\s{z}$
\begin{align}
\begin{matrix}
    R_{++}[\s{z}]=- \Gamma_{--} \abs{f}^{2} \s{z}  & R_{+-}[\s{z}]=0 & R_{z+}[\s{z}]= \Gamma_{z-} \bar{f} f_3 \s{-} \\
    R_{-+}[\s{z}]=0 & R_{--}[\s{z}]=- \Gamma_{++}\abs{f}^{2} \s{z} & R_{z-}[\s{z}]=\Gamma_{z+}f f_3 \s{+} \\R_{+z}[\s{z}]=\Gamma_{-z} f f_3 \s{+} & R_{-z}[\s{z}]=\Gamma_{+z} \bar{f} f_3 \s{-} & R_{zz}[\s{z}]=0
\end{matrix}
\end{align}



- $\s{x}$

\begin{align}
\begin{matrix}
R_{++}[\s{x}]=- \Gamma_{--} \abs{f}^{2} \frac{\s{x}}{2} & R_{-+}[\s{x}]=\Gamma_{+-}\bar{f}^{2} \s{-} & R_{z+}[\s{x}]= \Gamma_{-z} \bar{f} f_3 \frac{\s{z}}{2} \\
    R_{+-}[\s{x}]=\Gamma_{-+} f^{2} \s{+} & R_{--}[\s{x}]=-\Gamma_{++} \abs{f}^{2}\frac{\s{x}}{2} & R_{z-}[\s{x}]= \Gamma_{z+}  f f_3 \frac{\s{z}}{2} \\R_{+z}[\s{x}]= \Gamma_{-z} f_3 f \frac{\s{z}}{2} & R_{-z}[\s{x}]=\Gamma_{+z} f_3 \bar{f} \frac{\s{z}}{2} & R_{zz}[\s{x}]=-\Gamma_{zz}  2 f_{3}^{2} \s{x}
\end{matrix}
\end{align}


- $\s{y}$

\begin{align}
\begin{matrix}
    R_{++}[\s{y}]=- \abs{f}^{2} \frac{\s{y}}{2} & R_{-+}[\s{y}]=- i \bar{f}^{2}  \s{-} & R_{z+}[\s{y}]= \bar{f} f_{3} \frac{-i \s{z}}{2} \\ 
    R_{+-}[\s{y}]=i f^{2} \s{+} & R_{--}[\s{y}]=-\abs{f}^{2} \frac{\s{y}}{2} & R_{-z}[\s{y}]=\bar{f}f_3 (\frac{-i \s{z}}{2}) \\ R_{+z}[\s{y}]=f_3 f (\frac{i \s{z}}{2}) & R_{z-}[\s{y}]=f_3 f (\frac{i \s{z}}{2}) & R_{zz}[\s{y}]=- 2 f_{3}^{2} \s{y}
\end{matrix}
\end{align}


Using those we can compute the action of R, so we finally have:

\begin{align}
R[I] =& \left|{f}\right|^{2} \left( \Gamma_{++}- \Gamma_{--} \right) \s{z} + \left(\Gamma_{-z} f f_{3} - \Gamma_{+z} f_{3} \overline{f} + \Gamma_{z-} f_{3} \overline{f} - \Gamma_{z+} f f_{3}\right) \s{x} + i \left(\Gamma_{-z} f f_{3} + \Gamma_{+z} f_{3} \overline{f} - \Gamma_{z-} f_{3} \overline{f} - \Gamma_{z+} f f_{3}\right) \s{y}  \\
R[\s{z}] =& -\left|{f}\right|^{2} \left( \Gamma_{--} + \Gamma_{++} \right) \s{z} + i f_{3} \left(\frac{\Gamma_{-z} f }{2} - \frac{\Gamma_{+z}  \overline{f}}{2} - \frac{\Gamma_{z-}  \overline{f}}{2} + \frac{\Gamma_{z+} f }{2}\right) \s{y} + f_{3} \left(\frac{\Gamma_{-z} f }{2} + \frac{\Gamma_{+z}  \overline{f}}{2} + \frac{\Gamma_{z-}  \overline{f}}{2} + \frac{\Gamma_{z+} f }{2}\right) \s{x}  \\
R[\s{x}] =& f_{3} \left(\frac{\Gamma_{-z} f}{2} + \frac{\Gamma_{+z} \overline{f}}{2} + \frac{\Gamma_{z-} \overline{f}}{2} + \frac{\Gamma_{z+} f}{2}\right) \s{z} + i \left(\frac{\Gamma_{-+} f^{2}}{2} - \frac{\Gamma_{+-} \overline{f}^{2}}{2}\right) \s{y}  \\ &+ \left(\frac{\Gamma_{-+} f^{2}}{2} + \frac{\Gamma_{+-} \overline{f}^{2}}{2} - 2 \Gamma_{zz} f_{3}^{2} - \left( \frac{\Gamma_{--}}{2} + \frac{\Gamma_{++}}{2}\right) \left|{f}\right|^{2}\right) \s{x} \\
R[\s{y}] =& \left(- \frac{\Gamma_{-+} f^{2}}{2} - \frac{\Gamma_{+-} \overline{f}^{2}}{2} - 2 \Gamma_{zz} f_{3}^{2} + \left(- \frac{\Gamma_{--}}{2} - \frac{\Gamma_{++}}{2}\right) \left|{f}\right|^{2}\right) \s{y} \\ &+ i \left(f_{3} \left(\frac{\Gamma_{-z} f}{2} - \frac{\Gamma_{+z} \overline{f}}{2} - \frac{\Gamma_{z-} \overline{f}}{2} + \frac{\Gamma_{z+} f}{2}\right) \s{z} + \left(\frac{\Gamma_{-+} f^{2}}{2} - \frac{\Gamma_{+-} \overline{f}^{2}}{2}\right) \s{x}\right)
\end{align}

However, we can see from the definition that $\Gamma_{zw}=\overline{\Gamma_{wz}}$ so the expressions can be written as:

\begin{equation}
R[I] = \left|{f}\right|^{2} \left( \Gamma_{++}- \Gamma_{--} \right) \sigma_{z} + 2  \Re\left(f_{3} f (\Gamma_{-z}-\Gamma_{z+})\right) \sigma_{x} - 2  \Im\left(f_{3} f (\Gamma_{-z}-\Gamma_{z+})\right) \sigma_{y} 
\end{equation}
\begin{equation}
R[\s{z}] = -\left|{f}\right|^{2} \left( \Gamma_{--} + \Gamma_{++} \right) \sigma_{z} -\Im\left(f_{3} f (\Gamma_{-z}+\Gamma_{z+})\right)  \sigma_{y} + \Re\left(f_{3} f (\Gamma_{-z}+\Gamma_{z+})\right) \sigma_{x}  \\
\end{equation}

\begin{equation}
R[\s{x}] =  \left(\Re\left( \Gamma_{-+}f^{2}\right) - 2 \Gamma_{zz} f_{3}^{2} - \left( \frac{\Gamma_{--}}{2} + \frac{\Gamma_{++}}{2}\right) \left|{f}\right|^{2}\right) \sigma_{x}+\Re\left( f_{3} f (\Gamma_{-z}+\Gamma_{z+})\right) \sigma_{z} -\Im\left( \Gamma_{-+}f^{2}\right) \sigma_{y} 
\end{equation}

\begin{equation}
R[\s{y}] = -\left( \Re\left( \Gamma_{-+}f^{2}\right) + 2 \Gamma_{zz} f_{3}^{2} + \left( \frac{\Gamma_{--}}{2} + \frac{\Gamma_{++}}{2}\right) \left|{f}\right|^{2}\right) \sigma_{y} +   \Im\left( f_{3} f (\Gamma_{-z}+\Gamma_{z+})\right) \sigma_{z} - \Im\left( \Gamma_{-+}f^{2}\right) \sigma_{x}
\end{equation}

Using this one may finally obtain the action of the super operator on each of the Pauli matrices. Since everything commutes with the identity there is no contribution from the Hamiltonian part then:

\begin{equation}
    \mathcal{Z}[I]= \left|{f}\right|^{2} \left( \Gamma_{++}- \Gamma_{--} \right) \sigma_{z} + 2  \Re\left(f_{3} f (\Gamma_{-z}-\Gamma_{z+})\right) \sigma_{x} - 2  \Im\left(f_{3} f (\Gamma_{-z}-\Gamma_{z+})\right) \sigma_{y}
\end{equation}

For the Pauli matrices we have non-zero Hamiltonian contribution given by:

\begin{align}
[\Lambda(t),\s{i}]=\left(\frac{\abs{f}^{2}}{2} \left( \xi_{++}-\xi_{--}\right) [\s{z},\s{i}]+ f_{3} (\xi_{z-}-\xi_{z+})(f_{1} [\s{x},\s{i}] + f_{2} [\s{y},\s{i}]) \right)
\end{align}

where $\s{i}$ can be any of the Pauli matrices, using the equation above one finds for $\s{x}$:

\begin{align}
[\Lambda(t),\s{x}]=& \left(\frac{\abs{f}^{2}}{2} \left( \xi_{++}-\xi_{--}\right) (2 i \s{y}) + f_{2} f_{3} (\xi_{z-}-\xi_{z+}) (-2 i \s{z}) \right) \\
=& i \left(\abs{f}^{2} \left( \xi_{++}-\xi_{--}\right) \s{y} - 2   f_{2} f_{3} (\xi_{z-}-\xi_{z+}) \s{z}  \right)
\end{align}

for $\s{y}$

\begin{align}
[\Lambda(t),\s{y}] =& \left(\frac{\abs{f}^{2}}{2} \left( \xi_{++}-\xi_{--}\right) (-2 i \s{x})+ f_{3} (\xi_{z-}-\xi_{z+})f_{1} (2 i \s{z}) \right) \\
=& i \left(2 f_{1} f_{3} (\xi_{z-}-\xi_{z+})  \s{z}-\abs{f}^{2} \left( \xi_{++}-\xi_{--}\right)  \s{x} \right)
\end{align}

and finally for $\s{z}$:

\begin{align}
    [\Lambda(t),\s{z}]=&  f_{3} (\xi_{z-}-\xi_{z+})(f_{1} (- 2 i \s{y}) + f_{2} (2 i \s{x}) ) \\
    =& 2 i f_{3} (\xi_{z-}-\xi_{z+}) (f_{2} \s{x}-f_{1} \s{y})
\end{align}

Remember that 

\begin{align}
    \mathcal{Z}[x]= - i [\Lambda(t),x]+ R(x)
\end{align}

With all of the equations above we may now write the action of the super operator on each of the Pauli matrices, we start with $\s{z}$:

\begin{align}
      \mathcal{Z}[\s{z}] &=  2 f_{3} (\xi_{z-}-\xi_{z+}) (f_{2} \s{x}-f_{1} \s{y})  -\left|{f}\right|^{2} \left( \Gamma_{--} + \Gamma_{++} \right) \sigma_{z} - \Im\left( f f_{3}(\Gamma_{-z}+\Gamma_{z+})\right) \sigma_{y} +  \Re\left( f_{3} f(\Gamma_{z-}+\Gamma_{z+})\right) \sigma_{x}
\end{align}

Then for $\s{x}$:

\begin{align}
          \mathcal{Z}[\s{x}] &=   |f|^{2} (\xi_{++}-\xi_{--}) \s{y} - 2 f_{2} f_{3} (\xi_{z-}-\xi_{z+}) \s{z} - \Im(f^{2} \Gamma_{-+}) \s{y} + \Re(f f_{3}(\Gamma_{-z}+\Gamma{z-}))\s{z} \nonumber \\ &- \left( \frac{\abs{f}^{2}}{2} (\Gamma_{--}+\Gamma_{++})+ 2 f_{3}^{2} \Gamma_{zz}- \Re(f^{2} \Gamma_{-+}) \right)\s{x}
\end{align}

Finally for $\s{y}$

\begin{align}
          \mathcal{Z}[\s{y}] &= - \abs{f}^{2} (\xi_{++}-\xi_{--}) \s{x} + 2 f_{3} f_{1} (\xi_{z-}-\xi_{z+}) \s{z} - \left( \frac{\abs{f}^{2}}{2} (\Gamma_{--}+\Gamma_{++}) + 2 f_{3}^{2} \Gamma_{zz} + \Re(f^{2} \Gamma_{-+})\right) \s{y} \nonumber \\ &- \Im(f^{2} \Gamma_{-+}) \s{x} - \Im(f f_{3}(\Gamma_{z-}+\Gamma_{z+})) \s{z}
\end{align}

To find the evolution of the density matrix, one can then use the fact that any qubit density matrix can be written as 

\begin{equation}
    \rho =\frac{\mathcal{I}+\vec{r} \cdot \vec{\sigma}}{2}  = \frac{\mathcal{I}}{2} + c \s{x} +b \s{y} +a \s{z} 
\end{equation}

Before moving forward let us introduce the following notation:

\begin{align}
    \xi = \xi_{++}- \xi_{--} \\ 
    \Gamma_{z}^{-} = f_{3} f (\Gamma_{-z}-\Gamma_{z+})\\
    \Gamma_{z}^{+} = f_{3} f (\Gamma_{z-}+ \Gamma_{z+}) \\ 
    \Gamma = \Gamma_{--} + \Gamma_{++} \\ 
    \xi_{z} = f_{3} (\xi_{z+}-\xi_{z-})
\end{align}

Using the previous results one finds that, and the notation defined above one obtains:

\begin{align}
    \mathcal{Z}[\rho] &=  \mathcal{Z}\left[\frac{\mathcal{I}}{2}+ a \s{z} + b \s{x} +c \s{y}\right]=\frac{\mathcal{Z}[\mathcal{I}]}{2}+ a \mathcal{Z}[\s{z}] + b \mathcal{Z}[\s{x}]+ c \mathcal{Z}[\s{y}] \\
    &= \frac{I}{2}  \left( \abs{f}^{2} \left( \Gamma_{++}-\Gamma_{--}\right) \s{z} + 2  \Re(\Gamma_{z}^{-}) \s{x} - 2  \Im(\Gamma_{z}^{-}) \s{y} \right) \nonumber \\
    &+a   \left( -\abs{f}^{2} \Gamma \s{z} + \left(\Re(\Gamma_{z}^{+}) - 2 f_{2} \xi_z \right) \s{x} -  \left(  \Im(\Gamma_{z}^{+}) - 2 f_{1} \xi_z\right) \s{y}\right) \nonumber \\
    &- c \left( \left( \frac{\abs{f}}{2} \Gamma  +2 f_{3}^{2} \Gamma_{zz} + \Re(f^{2}\Gamma_{-+})\right) \s{y} +\Im(f^{2} \Gamma_{-+}) \s{x} + \Im(\Gamma_{z}^{+}) \s{z} + \abs{f}^{2} \xi \s{x} + 2 f_{1} \xi_{z} \s{z} \right)\nonumber \\
    &+ b \left(\abs{f}^{2} \xi \s{y} + 2 f_{2} \xi_{z} \s{z} + \Re(\Gamma_{z}^{+}) \s{z}  - \Im(f^{2} \Gamma_{-+}) \s{y} -\left(\frac{\abs{f}^{2}}{2} \Gamma + 2 f_{3}^{2} \Gamma_{zz} - \Re(f^{2} \Gamma_{-+}) \right) \s{x}\right)
\end{align}

One then collects terms proportional to each of the Pauli matrices so that one has:

\begin{align}
    \mathcal{Z}[\rho] &= \left( \frac{\abs{f}^{2} (\Gamma_{++}-\Gamma{++})}{2} - a \abs{f}^{2} \Gamma -c (\Im(\Gamma_{z}^{+})+2 f_1 \xi_z) + b  (2 f_{2} \xi_{z}+ \Re(\Gamma_{z}^{+})) \right) \s{z} \nonumber \\ 
    &+ \left( - \Im(\Gamma_{z}^{-}) + a 
    \left(  2  f_{1} \xi_{z}-\Im(\Gamma_{z}^{+})  \right) - b \left( \Im(f^{2}\Gamma_{-+}) -  \abs{f}^{2} \xi \right)  -c \left( \frac{\abs{f}^{2}}{2} \Gamma + 2 f_{3}^{2} \Gamma_{zz} + \Re(f^{2}\Gamma_{-+}) \right) \right) \s{y} \nonumber \\
    &+ \left( \Re(\Gamma_{z}^{+})-c(\abs{f}^{2} \xi+\Im(f^{2}\Gamma_{-+})) + a(\Re(\Gamma_{z}^{+}) - 2\xi_{z} f_{2})-b \left( \frac{\abs{f}^{2}}{2} \Gamma +2 f_{3}^{2} \Gamma_{zz} - \Re(f^{2} \Gamma_{-+})\right) \right) \s{x}
\end{align}

This may be written as a linear transformation of the coefficients:

\begin{align}
    \begin{pmatrix}a' \\ b' \\ c' \end{pmatrix} &= \underbrace{\begin{pmatrix}-\abs{f}^{2}\Gamma &  2 f_{2} \xi_{z} + \Re(\Gamma_{z}^{+}) &  - 2 f_{1} \xi_{z} -\Im(\Gamma_{z}^{+})   \\ \Re(\Gamma_{z}^{+}) - 2 f_{2} \xi_{z} & - \left(\frac{\abs{f}^{2}}{2} \Gamma + 2 f_{3}^{2} \Gamma_{zz} - \Re(f^{2} \Gamma_{-+}) \right) & -\left(\abs{f}^{2} \xi + \Im(f^{2} \Gamma_{-+}) \right)\\  2 f_{1} \xi_{z}-\Im(\Gamma_{z}^{+}) &  \left( \abs{f}^{2} \xi - \Im(f^{2} \Gamma_{-+}) \right) & - \left(\frac{\abs{f}^{2}}{2} \Gamma + 2 f_{3}^{2} \Gamma_{zz} + \Re(f^{2} \Gamma_{-+})\right)\end{pmatrix}   }_{M} \begin{pmatrix}a \\ b \\ c \end{pmatrix} \\ &+\underbrace{\begin{pmatrix}\abs{f}^{2}\frac{(\Gamma_{++}-\Gamma_{--})}{2} \\  \Re(\Gamma_{z}^{-}) \\ - \Im(\Gamma_{z}^{-}) \end{pmatrix}}_{\vec{r}}
\end{align}
\normalsize

We may now rewrite the action of the super operator as a linear transformation and obtain an form for the evolution based on the previous map:

\begin{align}
    \mathcal{K}_{2}[\rho]= (M \vec{a} + \vec{r}) \cdot \vec{\sigma}
\end{align}

Where 

\begin{equation}
    \vec{\sigma}=\begin{pmatrix}
    \s{z} \\ \s{x} \\ \s{y}
    \end{pmatrix}
\end{equation}

Now applying the map again will have the same form, it will just be applied on $\rho=a' \s{z}+b'\s{x}+c\s{y}$, so this time the term proportional to the identity ($\vec{r}$) does not come into play so applying the map once again results in:

\begin{align}
    \mathcal{Z}^{2}[\rho] &= \left( M(M \vec{a} + \vec{r}) \right) \cdot \vec{\sigma}
\end{align}
which again has the form $\rho=a' \s{z}+b'\s{x}+c\s{y}$. Using this fact we can find that applying the super operator $n$ times results in 

\begin{align}
    \mathcal{Z}^{n}[\rho] &= \left( M^{n-1}(M \vec{a} + \vec{r}) \right) \cdot \vec{\sigma}
\end{align}

finally the exponential can be computed as :

\begin{align} 
    e^{\mathcal{Z}}[\rho]&=\rho+ \sum_{n=1}^{\infty}\frac{M^{n-1}(M \vec{a} + \vec{r})}{n!} \cdot \vec{\sigma} \\ 
    &= \rho - (\vec{a}+M^{-1}\vec{r}) \cdot \vec{\sigma} + \sum_{n=0}^{\infty}\frac{M^{n} \vec{a} }{n!} \cdot \vec{\sigma} + \sum_{n=0}^{\infty}\frac{M^{n-1} \vec{r} }{n!} \cdot \vec{\sigma} \\ 
    &= \rho -  (\vec{a}+M^{-1}\vec{r})\cdot \vec{\sigma} +  e^{M} \vec{a}\cdot \vec{\sigma} +  e^{M} M^{-1} \vec{r} \cdot \vec{\sigma}\\
   e^{\mathcal{Z}}[\rho]  &=\frac{\mathbb{I}}{2}+  (e^{M}-\mathbb{I}) M^{-1} \vec{r}\cdot \vec{\sigma} +   e^{M} \vec{a} \cdot \vec{\sigma}\label{eq:dynamics}
\end{align} 

\section{Relationship with the Charge Qubit}\label{ap:qubit_model}

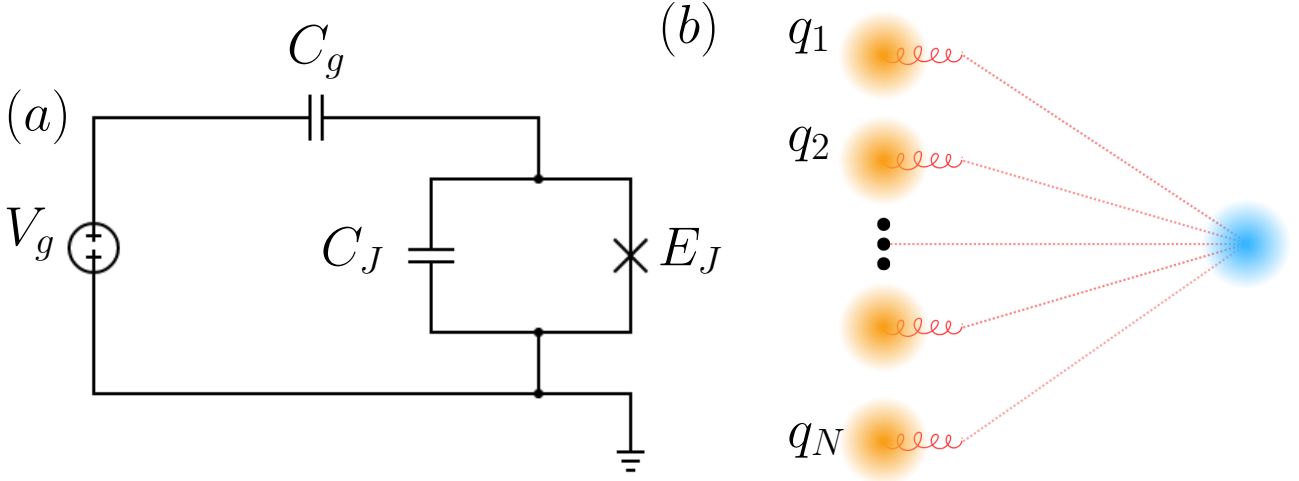
\begin{figure}[htb]
\centering
\subfloat[\label{fig:circuit}]{%
 \centering
    \begin{overpic}[width=0.5\textwidth]{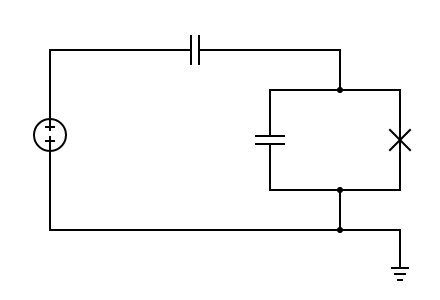}
        \put (-2,38) {\huge$\displaystyle V_{g}$}
        \put (40,65) {\huge$\displaystyle C_{g}$}
        \put (45,35) {\huge$\displaystyle C_{J}$}
        \put (95,35) {\huge$\displaystyle E_{J}$}
        \put (-2,55) {\huge$\displaystyle (a)$}
    \end{overpic}
} \hspace{1.2 cm}
\subfloat[\label{fig:qubit}]{%
    \centering
    \begin{overpic}[width=0.4\textwidth]{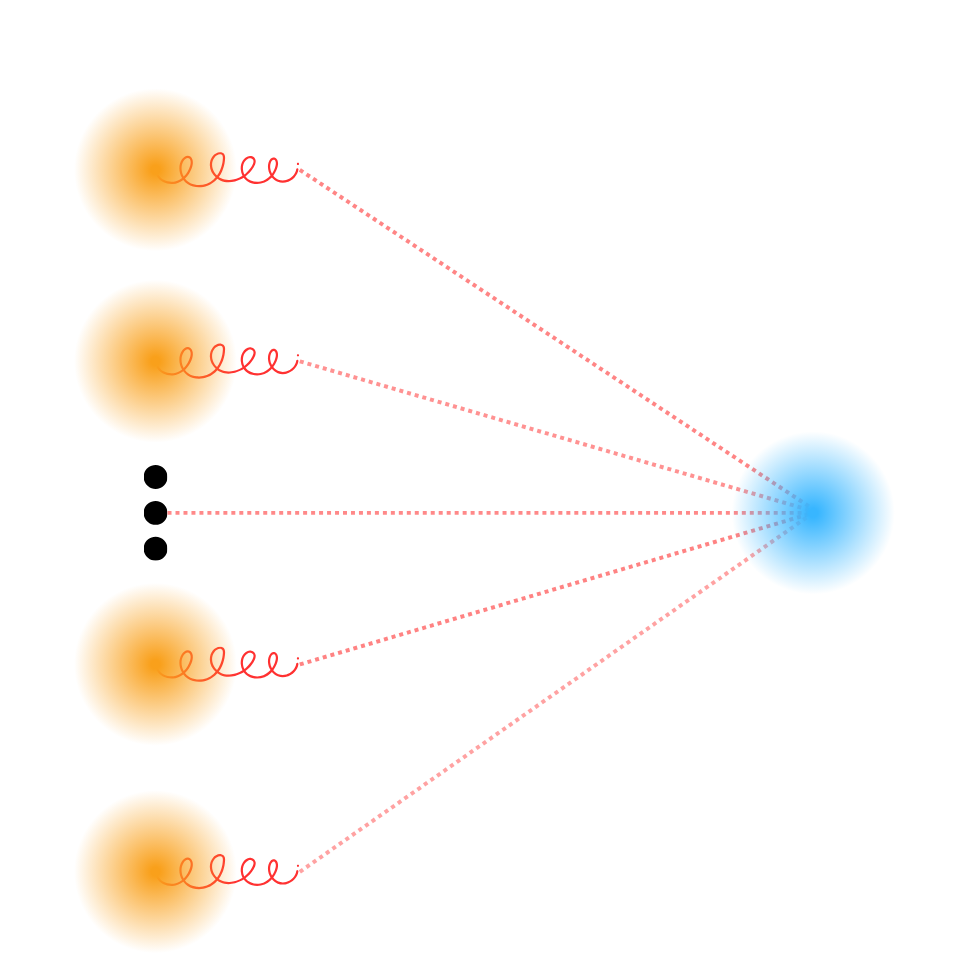}
        \put (-2,85) {\huge$\displaystyle q_{1}$}
        \put (-2,65) {\huge$\displaystyle q_{2}$}
        \put (-2,10) {\huge$\displaystyle q_{N}$}
        \put (-26,85) {\huge$\displaystyle (b)$}
    \end{overpic}
}

\caption{(a) Depicts a typical Charge Qubit schematic. (b) A graphical description of the Spin-Boson model. This figure shows the two possible representations of the model, on the one hand it can be thought of as a charge qubit where we are only interested in the first two levels and that is in the weakly coupled regime Fig. \ref{fig:circuit} on the other hand it can be thought of as a spin particle coupled to a Bosonic bath Fig. \ref{fig:qubit} where the coupling to the bath is not orthogonal or parallel to the system Hamiltonian as illustrated in the Table 1}
\end{figure}

One of the simplest elements in superconducting circuits is the Cooper par box (CPB) or charge qubit \cite{Girvin},  which is a superconducting candidate for a qubit whose basis states are charge state which represent the presence of cooper pairs in the superconducting island, the simplified schematics of a charge qubit are shown on Fig. \ref{fig:qubit}, The line with the cross is used to denote a tunnel junction, $C_{J}$ denotes the junction's capacitance while $E_{J}$ is the Josephson energy, $C_{g}$ and $V_{g}$ are the the gate voltage and capacitance used to control the superconducting junction. The Hamiltonian of such a super conducting circuit in standard circuit QED is given by

\begin{equation} \label{eq:charge_qubit}
    H = E_{C} (\hat{N}-N_{g})^{2} - E_{J} \cos(\hat{\phi})
\end{equation}

where $E_{C}$ is the junctions charging energy, $\hat{N}$  and $\hat{\phi}$ correspond to the canonical operators that quantize charge and flux. If we rewrite \eqref{eq:charge_qubit} in the Cooper pair number basis we obtain:

\begin{equation}
    H= E_{C} \sum_{N} (N-N_{g})^{2} | N \rangle \langle  N |  -\frac{E_{J}}{2} \sum_{N} | N + 1 \rangle \langle  N | + | N + 1 \rangle \langle  N | 
\end{equation}

When the Josephson Energy is much bigger than the capacitive energy $E_{J} \gg E_{C}$, we find ourselves in the small coupling regime. In this regime $N_{g}=N+\frac{1}{2}+\Delta_{g}$, if we then now restrict ourselves to the two lowest energy levels we obtain

\begin{equation}
    H = \frac{w_{0}}{2} \sigma_{z}  - \frac{E_{J}}{2} \sigma_{x}
\end{equation}

where $w_{0}= 2 E_{C} \Delta_{g}$, while the coupling from \eqref{eq:Hamiltonian}  may look a bit unnatural to the reader, it comes from the CPB model we just briefly reviewed 
 by following the procedure in \cite{Purkayastha2020,Schaller_notes,Xu2016}  we will follow the arguments given in those references. Without loss of generality let us consider 

\begin{equation}\label{eq:noneq-spin}
     H = \underbrace{\frac{\omega_{0}}{2} \s{z} + \frac{\Delta}{2} \s{x} + \frac{\delta}{2} \s{y} }_{H_S} + \underbrace{\sum_{k} w_{k} a_{k}^{\dagger} a_{k}}_{H_B} + \underbrace{\sum_{k} g_k \s{z} (a_{k}+a_{k}^{\dagger})}_{H_I}
\end{equation}

which corresponds to the CPB qubit hamiltionian subjected to dephasing by a bosonic bath when $\delta=0$. The first step towards deriving a master equation is to diagonalize the system,  to obtain the eigenbasis $H_{s} \ket{\psi} = E_{n} \ket{\psi}$, for this two level system we will obtain:

\begin{eqnarray}\label{eq:eigenbasis}
    E_{\pm} &=& \pm \frac{\Omega}{2}  \\
    \ket{+} &=&\sqrt{\frac{\delta^{2} +\Delta^{2}}{\delta^{2} +\Delta^{2} + (\omega_{0}+\Omega)^{2}} } \left( \ket{1} + \frac{\omega_{0}+\Omega}{\Delta + i \delta } \ket{0}\right) \\ 
     \ket{-} &=& \frac{\Omega +\omega_{0}}{\sqrt{\Delta^{2}+\delta^{2}+(\Omega+\omega_{0})}} \left( \ket{1}- \frac{(\Delta-i \delta)}{\omega_{0}+\Omega} \ket{0}\right)
\end{eqnarray}

Where $\Omega=\sqrt{\delta^{2} +\Delta^{2} + \omega_{0}^{2}}$. Then we need to rewrite $\s{z}$ in this basis which is given by:
\begin{eqnarray}
    \s{z}^{r} &=& \sum_{i,j=+,-}\bra{j} \sigma_{z} \ket{i} \ket{i}\bra{j} 
\end{eqnarray}

So in this basis we have:

\begin{eqnarray}
\s{z}^{r}=\begin{pmatrix}
        \bra{+} \s{z} \ket{+} & \bra{+} \s{z} \ket{-}  \\ \bra{-} \s{z} \ket{+} & \bra{-} \s{z} \ket{-} 
\end{pmatrix}
\end{eqnarray}

We can then decompose this matrix in terms of the pauli matrices in this bases given by:

\begin{eqnarray}
    \Tilde{\sigma}_{z} &= \ket{+}\bra{+} - \ket{-}\bra{-} \\     \Tilde{\sigma}_{x} &= \ket{+}\bra{-} + \ket{-}\bra{+} \\
    \Tilde{\sigma}_{y} &= -i \left(\ket{+}\bra{+} - \ket{-}\bra{-} \right)  
\end{eqnarray}

We may find the decomposition due to the fact that $Tr[\sigma_{i}]=0$ and  $\{ \mathbb{I},\Tilde{\sigma}_{x},\Tilde{\sigma}_{y},\Tilde{\sigma}_{z}\}$ form a complete basis, so that:

\begin{eqnarray}
\s{z}^{r} &=& c_{x} \Tilde{\sigma}_{x}+ c_{y} \Tilde{\sigma}_{y} +c_{z}\Tilde{\sigma}_{z} \\
Tr[\s{z}^{r} \Tilde{\sigma}_{i}] &=& c_{i}
\end{eqnarray}

Where we skipped the identity because our original matrix is trace-less. We then find:

\begin{eqnarray}
    c_{z}&=&\frac{2 \left( \omega_{0}^{2} -\Delta \sqrt{\Delta^{2}+\delta^{2}}\right)}{\Omega^{2}}   \\ 
    c_{y}&=& \frac{-2 \delta}{\Omega} \\
    c_{x}&=& -\frac{2 \omega_{0} \left( \Delta+\sqrt{\delta^{2} + \Delta^{2}}\right)}{\Omega^{2}}
\end{eqnarray}

Then the Hamiltonian in this rotated picture can be written as:

\begin{equation}\label{eq:noneq-spin_rotated}
     \Tilde{H} = \underbrace{\frac{\Omega}{2} \Tilde{\sigma}_{z} }_{H_S} + \underbrace{\sum_{k} w_{k} a_{k}^{\dagger} a_{k}}_{H_B} + \underbrace{\sum_{k} g_k \left(c_{z} \Tilde{\sigma}_{z}+ c_{x} \Tilde{\sigma}_{x}+ c_{y} \Tilde{\sigma}_{y} \right) (a_{k}+a_{k}^{\dagger})}_{H_I}
\end{equation}

Which is our model of interest, and the one whose steady state coherences have been debated \cite{cattaneo2021comment,Guarnieri,guarnieriErratum}, as we showed in this section it simply corresponds to diagonalizing the CPB qubit Hamiltonian 

\section{Non-Markovianity of the model}

In previous studies the Non-Markovianity of the Spin-Boson model under evolution described by the cumulant equation has been shown 
for $x$-coupling with bath
\cite{winczewski2021bypassing,Rivas}. In this section we present results for more general coupling with bath, and compare them with the Bloch-Redfield equation as well as with the HEOM. This serves as a confirmation that the Non-Markovianity described by the cumulant equation is not an approximation effect. The measure of Non-Marovianity used is the trace distance measure

\begin{equation}
    \frac12 || \rho(t) - \sigma(t) || = \frac12 \Tr{\sqrt{(\rho(t) - \sigma(t))^\dagger (\rho(t) - \sigma(t))}}
\end{equation}

\begin{figure}[htb]
\begin{overpic}[width=0.45\textwidth]{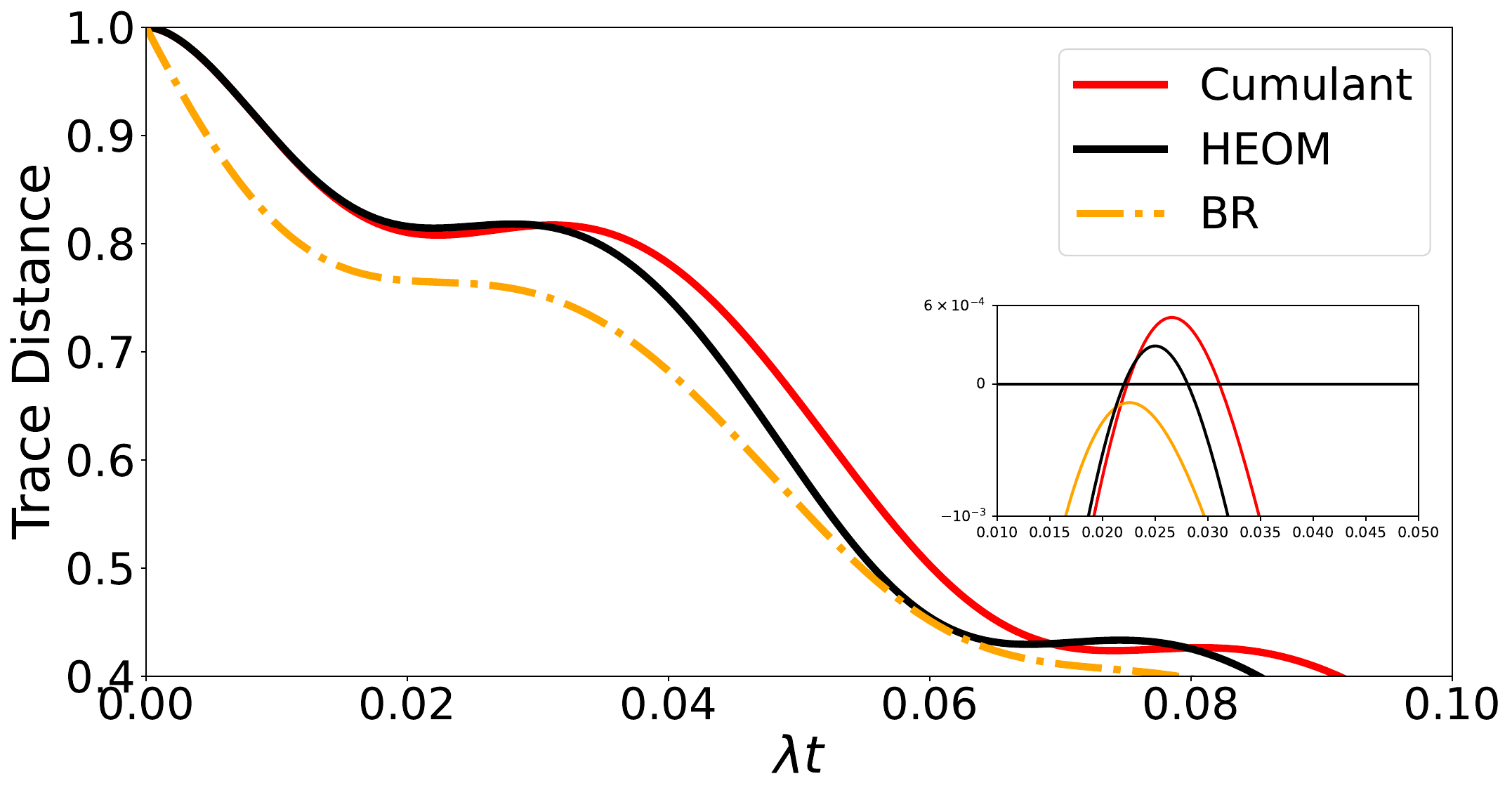}
        \put (-6,45) {\Large$\displaystyle (a)$}
\end{overpic} \quad \quad \quad
\begin{overpic}[width=0.45\textwidth]{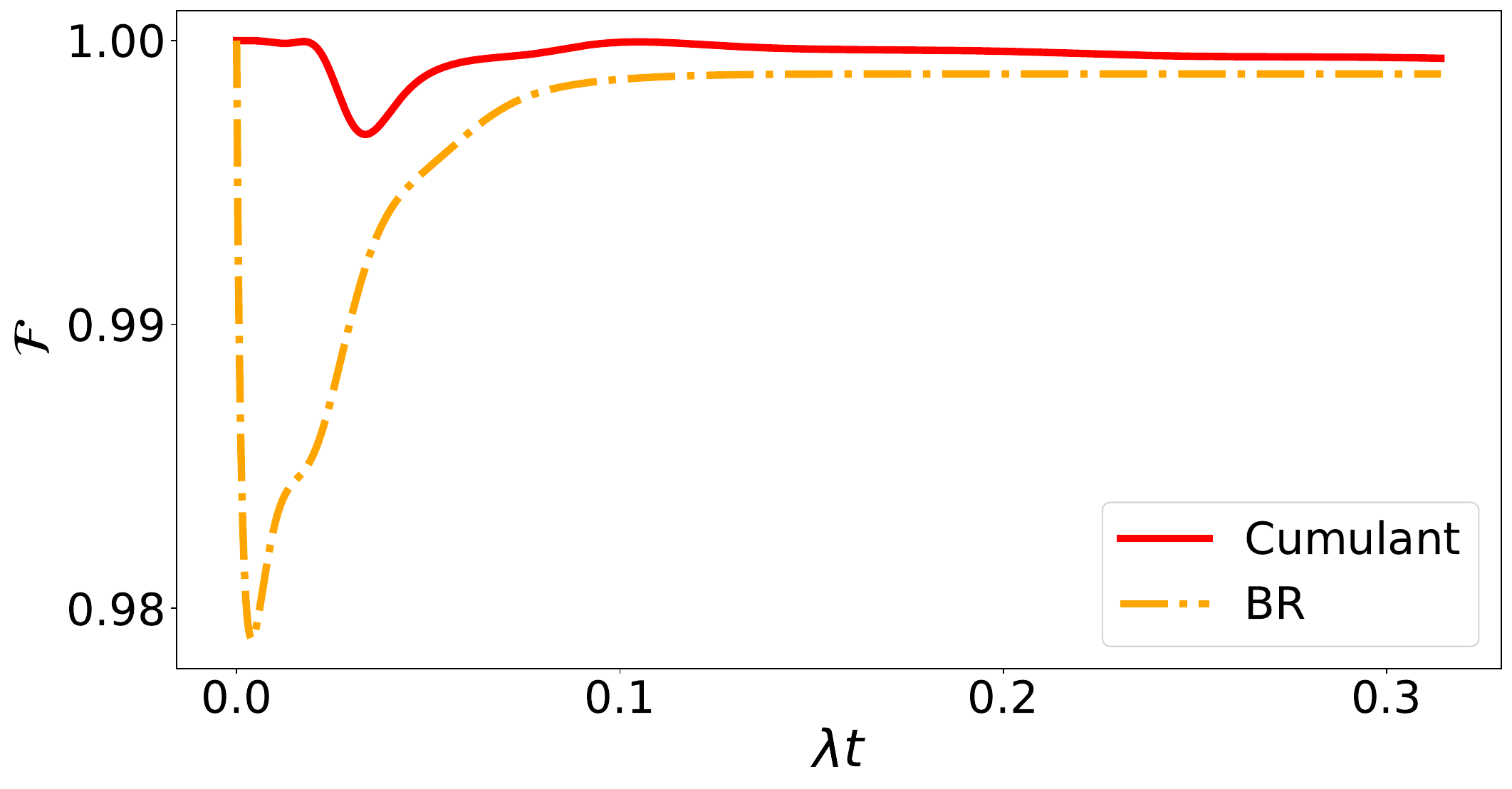}
        \put (-6,45) {\Large$\displaystyle (b)$}
\end{overpic}
  \caption{(a)Trace distance measure of the the different approaches for the Non-equilibrium Spin-Boson model. An inset zooming in is provided so that increase in trace distance can be observed clearly. Notice the cumulant equation correctly shows the Non-Markovianity of the model even when overestimating the increase of the trace distance (b) Fidelity to HEOM. The parameters used for both plots are   $f_{1}=5$,$f_{3}=1$ for $\frac{\lambda}{\gamma}=0.01$,$\frac{\omega_{0}}{T} =1$.}\label{fig:NM}
\end{figure}

The initial states for the two evolutions were chosen to be orthogonal specifically we chose
\begin{equation}
\rho(0)=\begin{pmatrix}
0.5 & 0.5i \\
-0.5i & 0.5
\end{pmatrix}=\overline{\sigma(0)}.
\end{equation}
For Markovian evolutions the trace distance between the evolutions decreases monotonically, while for Non-Markovian evolutions it does not, usually the increase in the measure is small \cite{measures}. To visualize the increase in the trace distance better better, an inset has been placed in Fig. \ref{fig:NM} the inset contains the derivative of the trace distance and it is shown in logarithmic scale, a positive value indicates increase of the trace distance measure and therefore Non-Markovianity

Indeed, the cumulant equation 
correctly points out non-Markovian regimes, although it overestimates the magnitude of the Non-Markovianity, serving therefore for a qualitative indicator, a fact that was not discussed in previous studies. 

For example when considering the parameters shown in the Fig. \ref{fig:NM}, the Non-Markovianity is clearly overestimated. Yet the cumulant equation still shows Non-Markovianity while the Bloch-Redfield (also known as Markovian Redfield) doesn't. 

At the same time this section works as a word of caution against taking this measures calculated from master equations too seriously, as in some regimes it may overestimate the Non-Markovianity of the situation, while still being a more faithful description of the exact dynamics. or may show Non-Markovianity where the exact solution shows none, we have observed that the cumulant equation presents deviations from the numerically exact solution in the latter case. While this may be seen as at disadvantage of the cumulant equation we would like to remark that the cumulant equation still shows Non-Markovianity while the Bloch-Redfield (also known as Markovian Redfield) doesn't, and at the same time exhibits higher fidelity at least in the setups considered. 

\section{Comparison with the time dependent Redfield equation}

Throughout the manuscript we compared the cumulant equation to the Bloch-Redfield equation. One may wonder if the cumulant equation is still a better description in the weak coupling limit when we consider the full Redfield equation. 

The Redfield equation in the interaction picture can be written as \cite{reconciliation}

\begin{eqnarray} 
\frac{d}{dt} \tilde \rho(t) = \tilde{\mathcal{L}}_t [\tilde \rho(t)]  &=&  i [\tilde \rho(t), \sum_{\omega,\omega'} \sum_{\alpha, \beta} \tilde{\mathcal{S}}_{\alpha \beta}(\omega,\omega',t) A^\dag_\alpha (\omega) A_\beta (\omega')]\\
 &+&  \sum_{\omega,\omega'} \sum_{\alpha, \beta}  \tilde \gamma_{\alpha \beta}(\omega, \omega', t) \left( A_\beta (\omega') \tilde \rho(t) A^\dag_\alpha(\omega) -\frac{1}{2} \{A^\dag_\alpha (\omega) A_\beta (\omega'), \tilde \rho(t) \} \right) 
\end{eqnarray}

where $\tilde{\mathcal{S}}_{\alpha \beta}(\omega,\omega',t)$ is the Lamb-shift term which we will neglect in this comparison, and $\tilde \gamma_{\alpha \beta}(\omega, \omega', t)$ is given by:

\begin{align}
    \tilde \gamma_{\alpha \beta}(\omega, \omega', t) &= e^{i(\omega-\omega')t} \gamma_{\alpha \beta}(\omega, \omega',t) \\
    \gamma_{\alpha \beta}(\omega, \omega', t) &= \Gamma_{\alpha \beta}(\omega', t) + \Gamma^*_{\beta \alpha}(\omega, t), \label{gamma_omega_omega}\\
    \Gamma_{\alpha \beta}(\omega, t) &= \int_0^{t} ds \ e^{i \omega s } \langle B_\alpha (s) B_\beta (0) \rangle. 
\end{align}

One thing to notice is that to get the super-operator that generates the dynamics requires a similar effort to the one from the cumulant equation, however solving the integro-differential equation will require more evaluations, which turns this method slower than the cumulant equation. 

\begin{figure}[htb]
\begin{overpic}[width=0.45\textwidth]{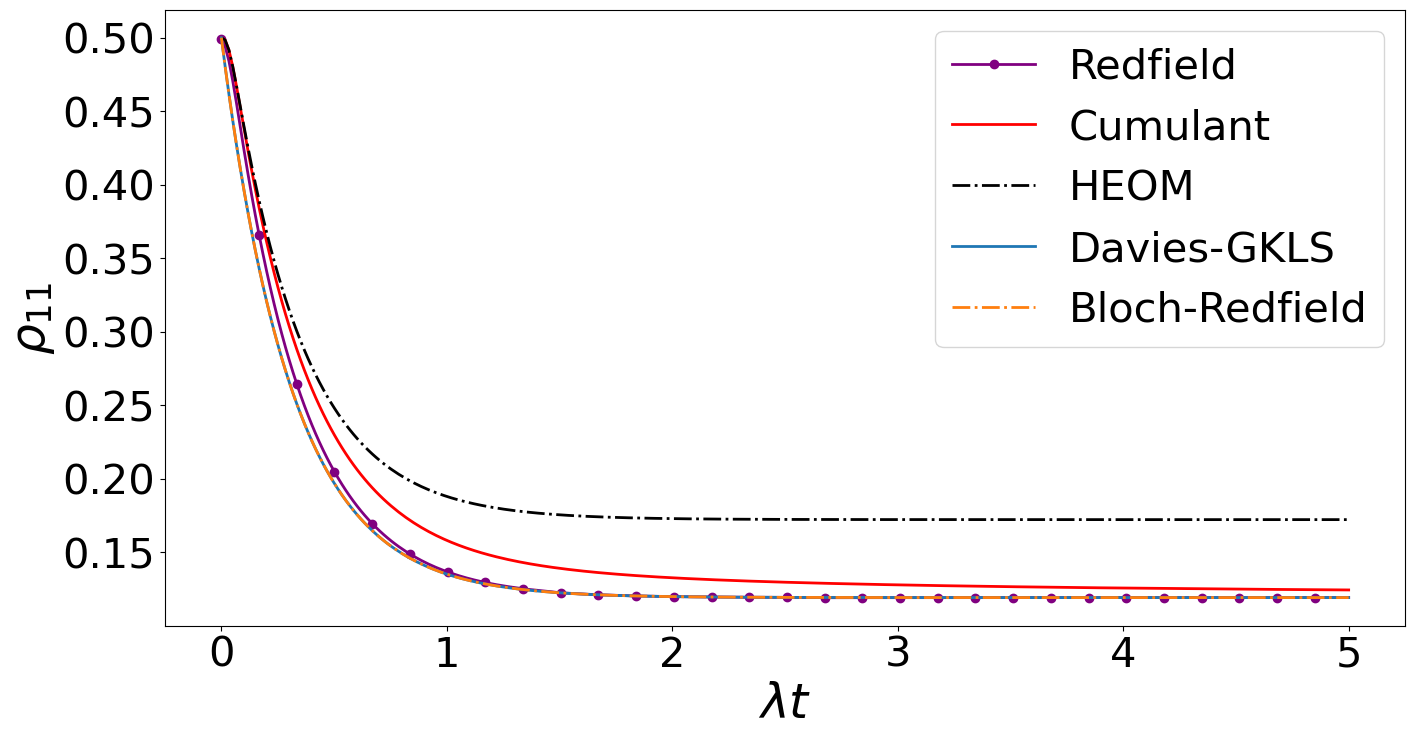}
        \put (-6,45) {\Large$\displaystyle (a)$}
\end{overpic} \quad \quad \quad
\begin{overpic}[width=0.45\textwidth]{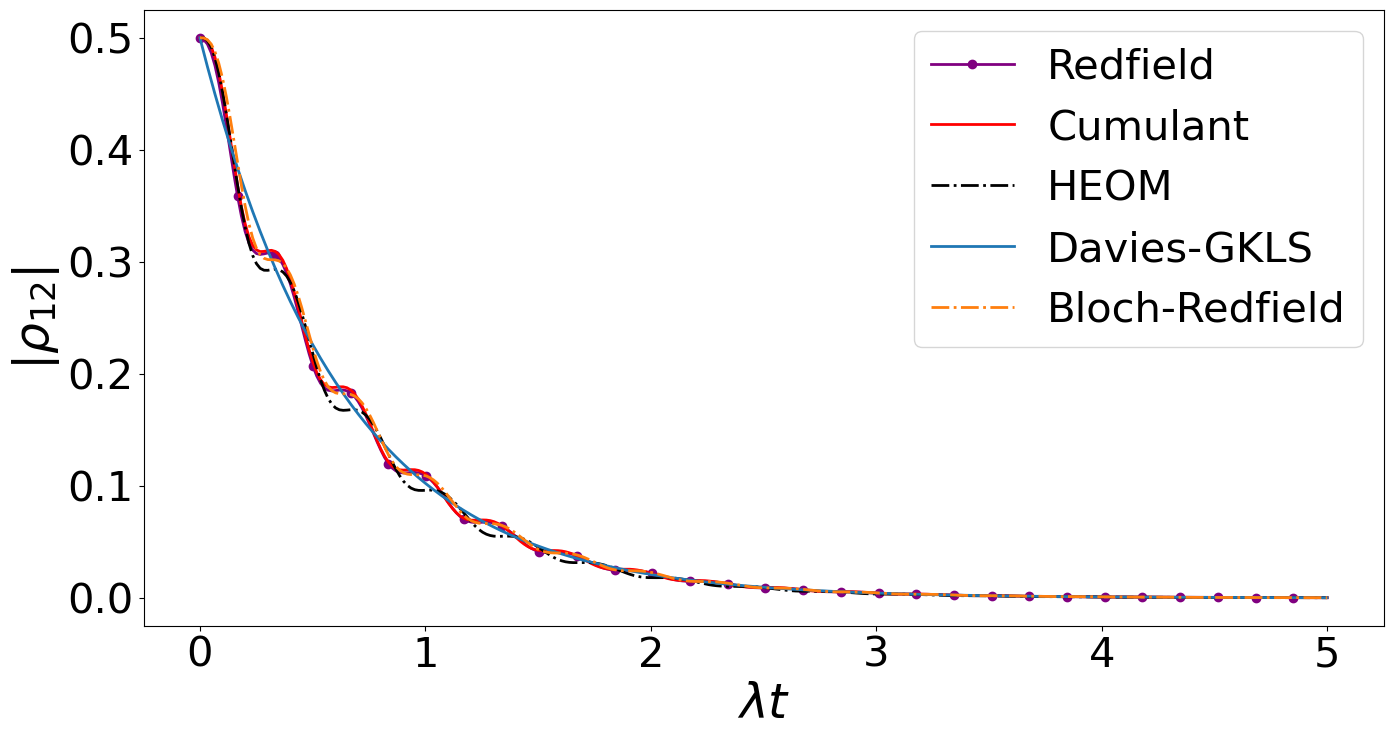}
        \put (-6,45) {\Large$\displaystyle (b)$}
\end{overpic} \quad \quad \quad
\begin{overpic}[width=0.45\textwidth]{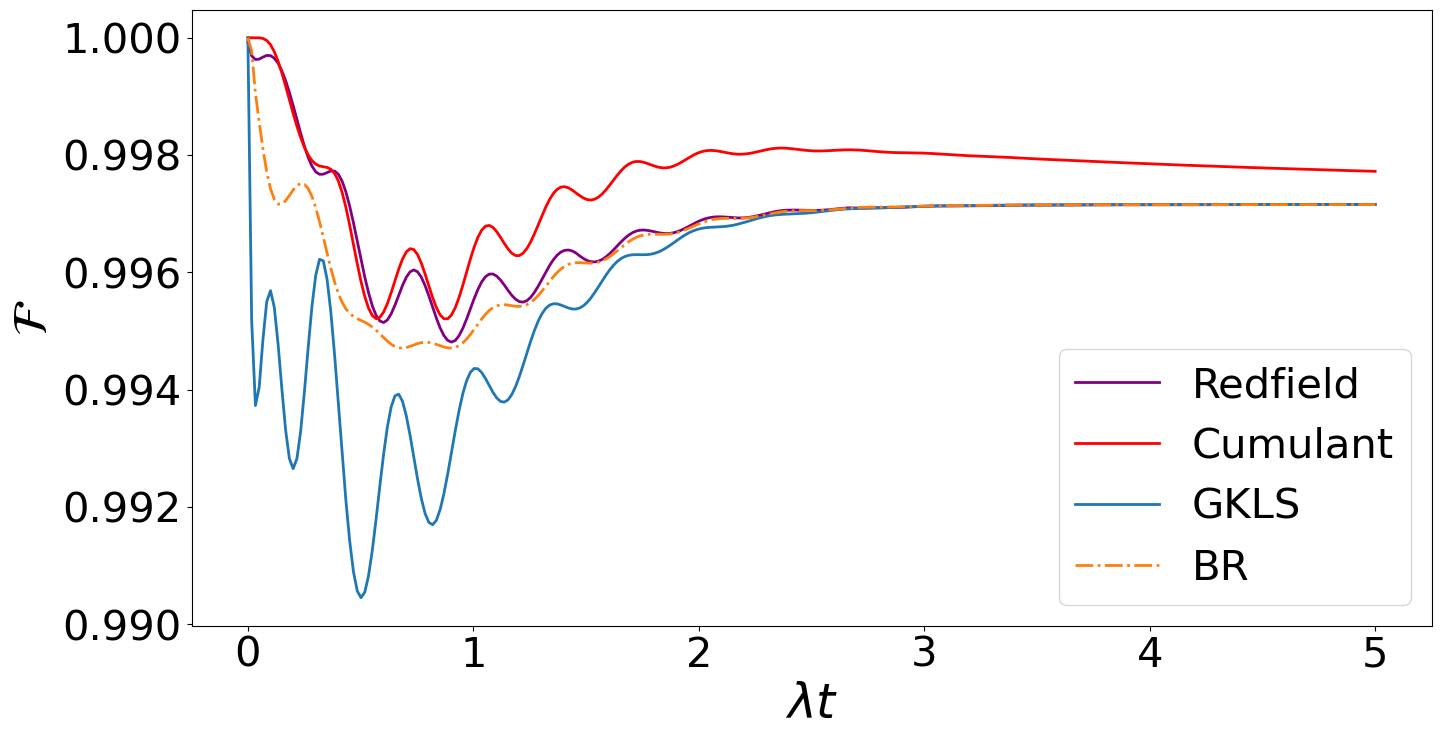}
        \put (-6,45) {\Large$\displaystyle (c)$}
\end{overpic} \quad \quad \quad
  \caption{(a) Shows the population evolution for each of the different methods, (b) Shows the absolute value of the coherence, (c) Shows the fidelity of the approximate methods to the exact solution. The plots shows Comparison of the dynamics computed with the different approaches for the Non-equilibrium Spin-Boson model with $f_{1}=1$,$f_{3}=f_{2}=0$ for $\frac{\lambda}{\gamma}=0.1$,$\frac{\omega_{0}}{T} =2$ }\label{fig:redfield}
\end{figure} 

We see that the advantage of the cumulant equation over Redfield is quite similar to the Bloch-Redfield equation, this can be visualized in Fig. \ref{fig:redfield}, conversely on other simulations we see that the disadvantages of the cumulant are around the same, so even when comparing to the time dependent Redfield equation the conclusions from the text hold, where we see the cumulant equation is a better description in the weak coupling limit at low temperature, while  in the strong coupling and high temperature regimes, which model is better seems to be model dependent. In those strong coupling regimes even Numerically exact solutions may have a hard time due to numerical instabilities and bad approximations to correlation functions \cite{instabilities,thoss_review}, on top of that exact solutions increase the size of the Hilbert space of study making simulation of higher dimensional systems inaccessible \cite{Lambert2019,menczel2024nonhermitian,Strathearn_2017,DAMPF}. On the other hand several transformations can be applied to reduce the effective coupling to the environment, these procedures also ``renormalize" the master equation making the state closer to that of the exact solution \cite{Xu2016,becker} in combination with these techniques the cumulant equation could be a prime candidate for the simulation of open systems at low temperature

\bibliography{apssamp}

\providecommand{\noopsort}[1]{}\providecommand{\singleletter}[1]{#1}%
\begin{thebibliography}{56}%
\makeatletter
\providecommand \@ifxundefined [1]{%
 \@ifx{#1\undefined}
}%
\providecommand \@ifnum [1]{%
 \ifnum #1\expandafter \@firstoftwo
 \else \expandafter \@secondoftwo
 \fi
}%
\providecommand \@ifx [1]{%
 \ifx #1\expandafter \@firstoftwo
 \else \expandafter \@secondoftwo
 \fi
}%
\providecommand \natexlab [1]{#1}%
\providecommand \enquote  [1]{``#1''}%
\providecommand \bibnamefont  [1]{#1}%
\providecommand \bibfnamefont [1]{#1}%
\providecommand \citenamefont [1]{#1}%
\providecommand \href@noop [0]{\@secondoftwo}%
\providecommand \href [0]{\begingroup \@sanitize@url \@href}%
\providecommand \@href[1]{\@@startlink{#1}\@@href}%
\providecommand \@@href[1]{\endgroup#1\@@endlink}%
\providecommand \@sanitize@url [0]{\catcode `\\12\catcode `\$12\catcode `\&12\catcode `\#12\catcode `\^12\catcode `\_12\catcode `\%12\relax}%
\providecommand \@@startlink[1]{}%
\providecommand \@@endlink[0]{}%
\providecommand \url  [0]{\begingroup\@sanitize@url \@url }%
\providecommand \@url [1]{\endgroup\@href {#1}{\urlprefix }}%
\providecommand \urlprefix  [0]{URL }%
\providecommand \Eprint [0]{\href }%
\providecommand \doibase [0]{https://doi.org/}%
\providecommand \selectlanguage [0]{\@gobble}%
\providecommand \bibinfo  [0]{\@secondoftwo}%
\providecommand \bibfield  [0]{\@secondoftwo}%
\providecommand \translation [1]{[#1]}%
\providecommand \BibitemOpen [0]{}%
\providecommand \bibitemStop [0]{}%
\providecommand \bibitemNoStop [0]{.\EOS\space}%
\providecommand \EOS [0]{\spacefactor3000\relax}%
\providecommand \BibitemShut  [1]{\csname bibitem#1\endcsname}%
\let\auto@bib@innerbib\@empty
\bibitem [{\citenamefont {\L{}obejko}\ \emph {et~al.}(2024)\citenamefont {\L{}obejko}, \citenamefont {Winczewski}, \citenamefont {Su\'arez}, \citenamefont {Alicki},\ and\ \citenamefont {Horodecki}}]{reconciliation}%
  \BibitemOpen
  \bibfield  {author} {\bibinfo {author} {\bibfnamefont {M.}~\bibnamefont {\L{}obejko}}, \bibinfo {author} {\bibfnamefont {M.}~\bibnamefont {Winczewski}}, \bibinfo {author} {\bibfnamefont {G.}~\bibnamefont {Su\'arez}}, \bibinfo {author} {\bibfnamefont {R.}~\bibnamefont {Alicki}},\ and\ \bibinfo {author} {\bibfnamefont {M.}~\bibnamefont {Horodecki}},\ }\bibfield  {title} {\bibinfo {title} {Corrections to the hamiltonian induced by finite-strength coupling to the environment},\ }\href {https://doi.org/10.1103/PhysRevE.110.014144} {\bibfield  {journal} {\bibinfo  {journal} {Phys. Rev. E}\ }\textbf {\bibinfo {volume} {110}},\ \bibinfo {pages} {014144} (\bibinfo {year} {2024})}\BibitemShut {NoStop}%
\bibitem [{\citenamefont {Trushechkin}\ \emph {et~al.}(2022)\citenamefont {Trushechkin}, \citenamefont {Merkli}, \citenamefont {Cresser},\ and\ \citenamefont {Anders}}]{meanforcereview}%
  \BibitemOpen
  \bibfield  {author} {\bibinfo {author} {\bibfnamefont {A.~S.}\ \bibnamefont {Trushechkin}}, \bibinfo {author} {\bibfnamefont {M.}~\bibnamefont {Merkli}}, \bibinfo {author} {\bibfnamefont {J.~D.}\ \bibnamefont {Cresser}},\ and\ \bibinfo {author} {\bibfnamefont {J.}~\bibnamefont {Anders}},\ }\bibfield  {title} {\bibinfo {title} {{Open quantum system dynamics and the mean force Gibbs state}},\ }\bibfield  {journal} {\bibinfo  {journal} {AVS Quantum Science}\ }\textbf {\bibinfo {volume} {4}},\ \href {https://doi.org/10.1116/5.0073853} {10.1116/5.0073853} (\bibinfo {year} {2022}),\ \bibinfo {note} {012301},\ \Eprint {https://arxiv.org/abs/https://pubs.aip.org/avs/aqs/article-pdf/doi/10.1116/5.0073853/16493686/012301\_1\_online.pdf} {https://pubs.aip.org/avs/aqs/article-pdf/doi/10.1116/5.0073853/16493686/012301\_1\_online.pdf} \BibitemShut {NoStop}%
\bibitem [{\citenamefont {Breuer}\ and\ \citenamefont {Petruccione}(2002)}]{breuer}%
  \BibitemOpen
  \bibfield  {author} {\bibinfo {author} {\bibfnamefont {H.~P.}\ \bibnamefont {Breuer}}\ and\ \bibinfo {author} {\bibfnamefont {F.}~\bibnamefont {Petruccione}},\ }\href@noop {} {\emph {\bibinfo {title} {The theory of open quantum systems}}}\ (\bibinfo  {publisher} {Oxford University Press},\ \bibinfo {address} {Great Clarendon Street},\ \bibinfo {year} {2002})\BibitemShut {NoStop}%
\bibitem [{\citenamefont {Purkayastha}\ \emph {et~al.}(2020)\citenamefont {Purkayastha}, \citenamefont {Guarnieri}, \citenamefont {Mitchison}, \citenamefont {Filip},\ and\ \citenamefont {Goold}}]{Purkayastha2020}%
  \BibitemOpen
  \bibfield  {author} {\bibinfo {author} {\bibfnamefont {A.}~\bibnamefont {Purkayastha}}, \bibinfo {author} {\bibfnamefont {G.}~\bibnamefont {Guarnieri}}, \bibinfo {author} {\bibfnamefont {M.~T.}\ \bibnamefont {Mitchison}}, \bibinfo {author} {\bibfnamefont {R.}~\bibnamefont {Filip}},\ and\ \bibinfo {author} {\bibfnamefont {J.}~\bibnamefont {Goold}},\ }\bibfield  {title} {\bibinfo {title} {Tunable phonon-induced steady-state coherence in a double-quantum-dot charge qubit},\ }\href {https://doi.org/10.1038/s41534-020-0256-6} {\bibfield  {journal} {\bibinfo  {journal} {npj Quantum Information}\ }\textbf {\bibinfo {volume} {6}},\ \bibinfo {pages} {27} (\bibinfo {year} {2020})}\BibitemShut {NoStop}%
\bibitem [{\citenamefont {Mori}\ and\ \citenamefont {Miyashita}(2008)}]{miyashita}%
  \BibitemOpen
  \bibfield  {author} {\bibinfo {author} {\bibfnamefont {T.}~\bibnamefont {Mori}}\ and\ \bibinfo {author} {\bibfnamefont {S.}~\bibnamefont {Miyashita}},\ }\bibfield  {title} {\bibinfo {title} {Dynamics of the density matrix in contact with a thermal bath and the quantum master equation},\ }\href {https://doi.org/10.1143/JPSJ.77.124005} {\bibfield  {journal} {\bibinfo  {journal} {Journal of the Physical Society of Japan}\ }\textbf {\bibinfo {volume} {77}},\ \bibinfo {pages} {124005} (\bibinfo {year} {2008})},\ \Eprint {https://arxiv.org/abs/https://doi.org/10.1143/JPSJ.77.124005} {https://doi.org/10.1143/JPSJ.77.124005} \BibitemShut {NoStop}%
\bibitem [{\citenamefont {Geva}\ \emph {et~al.}(2000)\citenamefont {Geva}, \citenamefont {Rosenman},\ and\ \citenamefont {Tannor}}]{Geva}%
  \BibitemOpen
  \bibfield  {author} {\bibinfo {author} {\bibfnamefont {E.}~\bibnamefont {Geva}}, \bibinfo {author} {\bibfnamefont {E.}~\bibnamefont {Rosenman}},\ and\ \bibinfo {author} {\bibfnamefont {D.}~\bibnamefont {Tannor}},\ }\bibfield  {title} {\bibinfo {title} {{On the second-order corrections to the quantum canonical equilibrium density matrix}},\ }\href {https://doi.org/10.1063/1.481928} {\bibfield  {journal} {\bibinfo  {journal} {The Journal of Chemical Physics}\ }\textbf {\bibinfo {volume} {113}},\ \bibinfo {pages} {1380} (\bibinfo {year} {2000})},\ \Eprint {https://arxiv.org/abs/https://pubs.aip.org/aip/jcp/article-pdf/113/4/1380/10829106/1380\_1\_online.pdf} {https://pubs.aip.org/aip/jcp/article-pdf/113/4/1380/10829106/1380\_1\_online.pdf} \BibitemShut {NoStop}%
\bibitem [{\citenamefont {Romero-Rochin}\ and\ \citenamefont {Oppenheim}(1989)}]{romerorochin}%
  \BibitemOpen
  \bibfield  {author} {\bibinfo {author} {\bibfnamefont {V.}~\bibnamefont {Romero-Rochin}}\ and\ \bibinfo {author} {\bibfnamefont {I.}~\bibnamefont {Oppenheim}},\ }\bibfield  {title} {\bibinfo {title} {Relaxation properties of two-level systems in condensed phases},\ }\href {https://doi.org/https://doi.org/10.1016/0378-4371(89)90051-4} {\bibfield  {journal} {\bibinfo  {journal} {Physica A: Statistical Mechanics and its Applications}\ }\textbf {\bibinfo {volume} {155}},\ \bibinfo {pages} {52} (\bibinfo {year} {1989})}\BibitemShut {NoStop}%
\bibitem [{\citenamefont {Guarnieri}\ \emph {et~al.}(2018)\citenamefont {Guarnieri}, \citenamefont {Kol\'a\ifmmode~\check{r}\else \v{r}\fi{}},\ and\ \citenamefont {Filip}}]{Guarnieri}%
  \BibitemOpen
  \bibfield  {author} {\bibinfo {author} {\bibfnamefont {G.}~\bibnamefont {Guarnieri}}, \bibinfo {author} {\bibfnamefont {M.}~\bibnamefont {Kol\'a\ifmmode~\check{r}\else \v{r}\fi{}}},\ and\ \bibinfo {author} {\bibfnamefont {R.}~\bibnamefont {Filip}},\ }\bibfield  {title} {\bibinfo {title} {Steady-state coherences by composite system-bath interactions},\ }\href {https://doi.org/10.1103/PhysRevLett.121.070401} {\bibfield  {journal} {\bibinfo  {journal} {Phys. Rev. Lett.}\ }\textbf {\bibinfo {volume} {121}},\ \bibinfo {pages} {070401} (\bibinfo {year} {2018})}\BibitemShut {NoStop}%
\bibitem [{\citenamefont {Cattaneo}\ and\ \citenamefont {Manzano}(2021)}]{cattaneo2021comment}%
  \BibitemOpen
  \bibfield  {author} {\bibinfo {author} {\bibfnamefont {M.}~\bibnamefont {Cattaneo}}\ and\ \bibinfo {author} {\bibfnamefont {G.}~\bibnamefont {Manzano}},\ }\href@noop {} {\bibinfo {title} {Comment on "steady-state coherences by composite system-bath interactions"}} (\bibinfo {year} {2021}),\ \Eprint {https://arxiv.org/abs/2106.09138} {arXiv:2106.09138 [quant-ph]} \BibitemShut {NoStop}%
\bibitem [{\citenamefont {Fleming}\ and\ \citenamefont {Cummings}(2011)}]{Flemming}%
  \BibitemOpen
  \bibfield  {author} {\bibinfo {author} {\bibfnamefont {C.~H.}\ \bibnamefont {Fleming}}\ and\ \bibinfo {author} {\bibfnamefont {N.~I.}\ \bibnamefont {Cummings}},\ }\bibfield  {title} {\bibinfo {title} {Accuracy of perturbative master equations},\ }\href {https://doi.org/10.1103/PhysRevE.83.031117} {\bibfield  {journal} {\bibinfo  {journal} {Phys. Rev. E}\ }\textbf {\bibinfo {volume} {83}},\ \bibinfo {pages} {031117} (\bibinfo {year} {2011})}\BibitemShut {NoStop}%
\bibitem [{\citenamefont {Tupkary}\ \emph {et~al.}(2022{\natexlab{a}})\citenamefont {Tupkary}, \citenamefont {Dhar}, \citenamefont {Kulkarni},\ and\ \citenamefont {Purkayastha}}]{tupkary}%
  \BibitemOpen
  \bibfield  {author} {\bibinfo {author} {\bibfnamefont {D.}~\bibnamefont {Tupkary}}, \bibinfo {author} {\bibfnamefont {A.}~\bibnamefont {Dhar}}, \bibinfo {author} {\bibfnamefont {M.}~\bibnamefont {Kulkarni}},\ and\ \bibinfo {author} {\bibfnamefont {A.}~\bibnamefont {Purkayastha}},\ }\bibfield  {title} {\bibinfo {title} {Fundamental limitations in lindblad descriptions of systems weakly coupled to baths},\ }\href {https://doi.org/10.1103/PhysRevA.105.032208} {\bibfield  {journal} {\bibinfo  {journal} {Phys. Rev. A}\ }\textbf {\bibinfo {volume} {105}},\ \bibinfo {pages} {032208} (\bibinfo {year} {2022}{\natexlab{a}})}\BibitemShut {NoStop}%
\bibitem [{\citenamefont {Tupkary}\ \emph {et~al.}(2022{\natexlab{b}})\citenamefont {Tupkary}, \citenamefont {Dhar}, \citenamefont {Kulkarni},\ and\ \citenamefont {Purkayastha}}]{limitations}%
  \BibitemOpen
  \bibfield  {author} {\bibinfo {author} {\bibfnamefont {D.}~\bibnamefont {Tupkary}}, \bibinfo {author} {\bibfnamefont {A.}~\bibnamefont {Dhar}}, \bibinfo {author} {\bibfnamefont {M.}~\bibnamefont {Kulkarni}},\ and\ \bibinfo {author} {\bibfnamefont {A.}~\bibnamefont {Purkayastha}},\ }\bibfield  {title} {\bibinfo {title} {Fundamental limitations in lindblad descriptions of systems weakly coupled to baths},\ }\href {https://doi.org/10.1103/PhysRevA.105.032208} {\bibfield  {journal} {\bibinfo  {journal} {Phys. Rev. A}\ }\textbf {\bibinfo {volume} {105}},\ \bibinfo {pages} {032208} (\bibinfo {year} {2022}{\natexlab{b}})}\BibitemShut {NoStop}%
\bibitem [{\citenamefont {Rivas}(2017)}]{Rivas}%
  \BibitemOpen
  \bibfield  {author} {\bibinfo {author} {\bibfnamefont {A.}~\bibnamefont {Rivas}},\ }\bibfield  {title} {\bibinfo {title} {Refined weak-coupling limit: Coherence, entanglement, and non-markovianity},\ }\href {https://doi.org/10.1103/PhysRevA.95.042104} {\bibfield  {journal} {\bibinfo  {journal} {Phys. Rev. A}\ }\textbf {\bibinfo {volume} {95}},\ \bibinfo {pages} {042104} (\bibinfo {year} {2017})}\BibitemShut {NoStop}%
\bibitem [{\citenamefont {Alicki}(1989)}]{Alicki89}%
  \BibitemOpen
  \bibfield  {author} {\bibinfo {author} {\bibfnamefont {R.}~\bibnamefont {Alicki}},\ }\bibfield  {title} {\bibinfo {title} {Master equations for a damped nonlinear oscillator and the validity of the markovian approximation},\ }\href {https://doi.org/10.1103/PhysRevA.40.4077} {\bibfield  {journal} {\bibinfo  {journal} {Phys. Rev. A}\ }\textbf {\bibinfo {volume} {40}},\ \bibinfo {pages} {4077} (\bibinfo {year} {1989})}\BibitemShut {NoStop}%
\bibitem [{\citenamefont {Schaller}\ and\ \citenamefont {Brandes}(2008)}]{Schaller_brandais}%
  \BibitemOpen
  \bibfield  {author} {\bibinfo {author} {\bibfnamefont {G.}~\bibnamefont {Schaller}}\ and\ \bibinfo {author} {\bibfnamefont {T.}~\bibnamefont {Brandes}},\ }\bibfield  {title} {\bibinfo {title} {Preservation of positivity by dynamical coarse graining},\ }\href {https://doi.org/10.1103/PhysRevA.78.022106} {\bibfield  {journal} {\bibinfo  {journal} {Phys. Rev. A}\ }\textbf {\bibinfo {volume} {78}},\ \bibinfo {pages} {022106} (\bibinfo {year} {2008})}\BibitemShut {NoStop}%
\bibitem [{\citenamefont {Majenz}\ \emph {et~al.}(2013)\citenamefont {Majenz}, \citenamefont {Albash}, \citenamefont {Breuer},\ and\ \citenamefont {Lidar}}]{majenz_coarse_2013}%
  \BibitemOpen
  \bibfield  {author} {\bibinfo {author} {\bibfnamefont {C.}~\bibnamefont {Majenz}}, \bibinfo {author} {\bibfnamefont {T.}~\bibnamefont {Albash}}, \bibinfo {author} {\bibfnamefont {H.-P.}\ \bibnamefont {Breuer}},\ and\ \bibinfo {author} {\bibfnamefont {D.~A.}\ \bibnamefont {Lidar}},\ }\bibfield  {title} {\bibinfo {title} {Coarse graining can beat the rotating-wave approximation in quantum {Markovian} master equations},\ }\href {https://doi.org/10.1103/PhysRevA.88.012103} {\bibfield  {journal} {\bibinfo  {journal} {Physical Review A}\ }\textbf {\bibinfo {volume} {88}},\ \bibinfo {pages} {012103} (\bibinfo {year} {2013})},\ \bibinfo {note} {publisher: American Physical Society}\BibitemShut {NoStop}%
\bibitem [{\citenamefont {Lidar}(2020)}]{lidar2020lecture}%
  \BibitemOpen
  \bibfield  {author} {\bibinfo {author} {\bibfnamefont {D.~A.}\ \bibnamefont {Lidar}},\ }\href@noop {} {\bibinfo {title} {Lecture notes on the theory of open quantum systems}} (\bibinfo {year} {2020}),\ \Eprint {https://arxiv.org/abs/1902.00967} {arXiv:1902.00967 [quant-ph]} \BibitemShut {NoStop}%
\bibitem [{\citenamefont {Hartmann}\ and\ \citenamefont {Strunz}(2020)}]{hartmann_accuracy_2020}%
  \BibitemOpen
  \bibfield  {author} {\bibinfo {author} {\bibfnamefont {R.}~\bibnamefont {Hartmann}}\ and\ \bibinfo {author} {\bibfnamefont {W.~T.}\ \bibnamefont {Strunz}},\ }\bibfield  {title} {\bibinfo {title} {Accuracy {Assessment} of {Perturbative} {Master} {Equations} -- {Embracing} {Non}-{Positivity}},\ }\href {https://doi.org/10.1103/PhysRevA.101.012103} {\bibfield  {journal} {\bibinfo  {journal} {Physical Review A}\ }\textbf {\bibinfo {volume} {101}},\ \bibinfo {pages} {012103} (\bibinfo {year} {2020})},\ \bibinfo {note} {arXiv:1906.02583 [quant-ph]}\BibitemShut {NoStop}%
\bibitem [{\citenamefont {Winczewski}\ \emph {et~al.}(2024)\citenamefont {Winczewski}, \citenamefont {Mandarino}, \citenamefont {Suarez}, \citenamefont {Alicki},\ and\ \citenamefont {Horodecki}}]{winczewski2021bypassing}%
  \BibitemOpen
  \bibfield  {author} {\bibinfo {author} {\bibfnamefont {M.}~\bibnamefont {Winczewski}}, \bibinfo {author} {\bibfnamefont {A.}~\bibnamefont {Mandarino}}, \bibinfo {author} {\bibfnamefont {G.}~\bibnamefont {Suarez}}, \bibinfo {author} {\bibfnamefont {R.}~\bibnamefont {Alicki}},\ and\ \bibinfo {author} {\bibfnamefont {M.}~\bibnamefont {Horodecki}},\ }\bibfield  {title} {\bibinfo {title} {Intermediate-times dilemma for open quantum system: Filtered approximation to the refined weak-coupling limit},\ }\href {https://doi.org/10.1103/PhysRevE.110.024110} {\bibfield  {journal} {\bibinfo  {journal} {Phys. Rev. E}\ }\textbf {\bibinfo {volume} {110}},\ \bibinfo {pages} {024110} (\bibinfo {year} {2024})}\BibitemShut {NoStop}%
\bibitem [{\citenamefont {Tanimura}\ and\ \citenamefont {Kubo}(1989)}]{HEOM}%
  \BibitemOpen
  \bibfield  {author} {\bibinfo {author} {\bibfnamefont {Y.}~\bibnamefont {Tanimura}}\ and\ \bibinfo {author} {\bibfnamefont {R.}~\bibnamefont {Kubo}},\ }\bibfield  {title} {\bibinfo {title} {Time evolution of a quantum system in contact with a nearly gaussian-markoffian noise bath},\ }\href {https://doi.org/10.1143/JPSJ.58.101} {\bibfield  {journal} {\bibinfo  {journal} {Journal of the Physical Society of Japan}\ }\textbf {\bibinfo {volume} {58}},\ \bibinfo {pages} {101} (\bibinfo {year} {1989})},\ \Eprint {https://arxiv.org/abs/https://doi.org/10.1143/JPSJ.58.101} {https://doi.org/10.1143/JPSJ.58.101} \BibitemShut {NoStop}%
\bibitem [{\citenamefont {Lambert}\ \emph {et~al.}(2023)\citenamefont {Lambert}, \citenamefont {Raheja}, \citenamefont {Cross}, \citenamefont {Menczel}, \citenamefont {Ahmed}, \citenamefont {Pitchford}, \citenamefont {Burgarth},\ and\ \citenamefont {Nori}}]{qutip}%
  \BibitemOpen
  \bibfield  {author} {\bibinfo {author} {\bibfnamefont {N.}~\bibnamefont {Lambert}}, \bibinfo {author} {\bibfnamefont {T.}~\bibnamefont {Raheja}}, \bibinfo {author} {\bibfnamefont {S.}~\bibnamefont {Cross}}, \bibinfo {author} {\bibfnamefont {P.}~\bibnamefont {Menczel}}, \bibinfo {author} {\bibfnamefont {S.}~\bibnamefont {Ahmed}}, \bibinfo {author} {\bibfnamefont {A.}~\bibnamefont {Pitchford}}, \bibinfo {author} {\bibfnamefont {D.}~\bibnamefont {Burgarth}},\ and\ \bibinfo {author} {\bibfnamefont {F.}~\bibnamefont {Nori}},\ }\bibfield  {title} {\bibinfo {title} {Qutip-bofin: A bosonic and fermionic numerical hierarchical-equations-of-motion library with applications in light-harvesting, quantum control, and single-molecule electronics},\ }\href {https://doi.org/10.1103/PhysRevResearch.5.013181} {\bibfield  {journal} {\bibinfo  {journal} {Phys. Rev. Res.}\ }\textbf {\bibinfo {volume} {5}},\ \bibinfo {pages} {013181} (\bibinfo {year} {2023})}\BibitemShut {NoStop}%
\bibitem [{\citenamefont {Nazir}\ and\ \citenamefont {Schaller}(2018)}]{Nazir2018}%
  \BibitemOpen
  \bibfield  {author} {\bibinfo {author} {\bibfnamefont {A.}~\bibnamefont {Nazir}}\ and\ \bibinfo {author} {\bibfnamefont {G.}~\bibnamefont {Schaller}},\ }\bibinfo {title} {The reaction coordinate mapping in quantum thermodynamics},\ in\ \href {https://doi.org/10.1007/978-3-319-99046-0_23} {\emph {\bibinfo {booktitle} {Thermodynamics in the Quantum Regime: Fundamental Aspects and New Directions}}},\ \bibinfo {editor} {edited by\ \bibinfo {editor} {\bibfnamefont {F.}~\bibnamefont {Binder}}, \bibinfo {editor} {\bibfnamefont {L.~A.}\ \bibnamefont {Correa}}, \bibinfo {editor} {\bibfnamefont {C.}~\bibnamefont {Gogolin}}, \bibinfo {editor} {\bibfnamefont {J.}~\bibnamefont {Anders}},\ and\ \bibinfo {editor} {\bibfnamefont {G.}~\bibnamefont {Adesso}}}\ (\bibinfo  {publisher} {Springer International Publishing},\ \bibinfo {address} {Cham},\ \bibinfo {year} {2018})\ pp.\ \bibinfo {pages} {551--577}\BibitemShut {NoStop}%
\bibitem [{\citenamefont {Doll}\ \emph {et~al.}(2008)\citenamefont {Doll}, \citenamefont {Zueco}, \citenamefont {Wubs}, \citenamefont {Kohler},\ and\ \citenamefont {Hänggi}}]{Doll_2008}%
  \BibitemOpen
  \bibfield  {author} {\bibinfo {author} {\bibfnamefont {R.}~\bibnamefont {Doll}}, \bibinfo {author} {\bibfnamefont {D.}~\bibnamefont {Zueco}}, \bibinfo {author} {\bibfnamefont {M.}~\bibnamefont {Wubs}}, \bibinfo {author} {\bibfnamefont {S.}~\bibnamefont {Kohler}},\ and\ \bibinfo {author} {\bibfnamefont {P.}~\bibnamefont {Hänggi}},\ }\bibfield  {title} {\bibinfo {title} {On the conundrum of deriving exact solutions from approximate master equations},\ }\href {https://doi.org/10.1016/j.chemphys.2007.09.003} {\bibfield  {journal} {\bibinfo  {journal} {Chemical Physics}\ }\textbf {\bibinfo {volume} {347}},\ \bibinfo {pages} {243} (\bibinfo {year} {2008})}\BibitemShut {NoStop}%
\bibitem [{\citenamefont {Suarez}()}]{repo}%
  \BibitemOpen
  \bibfield  {author} {\bibinfo {author} {\bibfnamefont {G.}~\bibnamefont {Suarez}},\ }\href {https://github.com/gsuarezr/NonMarkovianMethods} {\bibinfo {title} {Non markovian methods https://github.com/gsuarezr/nonmarkovianmethods}}\BibitemShut {NoStop}%
\bibitem [{\citenamefont {Alicki}\ \emph {et~al.}(2004)\citenamefont {Alicki}, \citenamefont {Horodecki}, \citenamefont {Horodecki}, \citenamefont {Horodecki}, \citenamefont {Jacak},\ and\ \citenamefont {Machnikowski}}]{preparation}%
  \BibitemOpen
  \bibfield  {author} {\bibinfo {author} {\bibfnamefont {R.}~\bibnamefont {Alicki}}, \bibinfo {author} {\bibfnamefont {M.}~\bibnamefont {Horodecki}}, \bibinfo {author} {\bibfnamefont {P.}~\bibnamefont {Horodecki}}, \bibinfo {author} {\bibfnamefont {R.}~\bibnamefont {Horodecki}}, \bibinfo {author} {\bibfnamefont {L.}~\bibnamefont {Jacak}},\ and\ \bibinfo {author} {\bibfnamefont {P.}~\bibnamefont {Machnikowski}},\ }\bibfield  {title} {\bibinfo {title} {Optimal strategy for a single-qubit gate and the trade-off between opposite types of decoherence},\ }\href {https://doi.org/10.1103/PhysRevA.70.010501} {\bibfield  {journal} {\bibinfo  {journal} {Phys. Rev. A}\ }\textbf {\bibinfo {volume} {70}},\ \bibinfo {pages} {010501} (\bibinfo {year} {2004})}\BibitemShut {NoStop}%
\bibitem [{Note1()}]{Note1}%
  \BibitemOpen
  \bibinfo {note} {In this context we say an an operator is parallel to another if they commute and orthogonal if the commutator generates a different member of the SU(N) group.}\BibitemShut {Stop}%
\bibitem [{\citenamefont {Schlienz}\ and\ \citenamefont {Mahler}(1995)}]{SUN}%
  \BibitemOpen
  \bibfield  {author} {\bibinfo {author} {\bibfnamefont {J.}~\bibnamefont {Schlienz}}\ and\ \bibinfo {author} {\bibfnamefont {G.}~\bibnamefont {Mahler}},\ }\bibfield  {title} {\bibinfo {title} {Description of entanglement},\ }\href {https://doi.org/10.1103/PhysRevA.52.4396} {\bibfield  {journal} {\bibinfo  {journal} {Phys. Rev. A}\ }\textbf {\bibinfo {volume} {52}},\ \bibinfo {pages} {4396} (\bibinfo {year} {1995})}\BibitemShut {NoStop}%
\bibitem [{\citenamefont {Hioe}\ and\ \citenamefont {Eberly}(1981)}]{hioe}%
  \BibitemOpen
  \bibfield  {author} {\bibinfo {author} {\bibfnamefont {F.~T.}\ \bibnamefont {Hioe}}\ and\ \bibinfo {author} {\bibfnamefont {J.~H.}\ \bibnamefont {Eberly}},\ }\bibfield  {title} {\bibinfo {title} {$n$-level coherence vector and higher conservation laws in quantum optics and quantum mechanics},\ }\href {https://doi.org/10.1103/PhysRevLett.47.838} {\bibfield  {journal} {\bibinfo  {journal} {Phys. Rev. Lett.}\ }\textbf {\bibinfo {volume} {47}},\ \bibinfo {pages} {838} (\bibinfo {year} {1981})}\BibitemShut {NoStop}%
\bibitem [{\citenamefont {Cattaneo}\ \emph {et~al.}(2019{\natexlab{a}})\citenamefont {Cattaneo}, \citenamefont {Giorgi}, \citenamefont {Maniscalco},\ and\ \citenamefont {Zambrini}}]{Cattaneo_2019}%
  \BibitemOpen
  \bibfield  {author} {\bibinfo {author} {\bibfnamefont {M.}~\bibnamefont {Cattaneo}}, \bibinfo {author} {\bibfnamefont {G.~L.}\ \bibnamefont {Giorgi}}, \bibinfo {author} {\bibfnamefont {S.}~\bibnamefont {Maniscalco}},\ and\ \bibinfo {author} {\bibfnamefont {R.}~\bibnamefont {Zambrini}},\ }\bibfield  {title} {\bibinfo {title} {Local versus global master equation with common and separate baths: superiority of the global approach in partial secular approximation},\ }\href {https://doi.org/10.1088/1367-2630/ab54ac} {\bibfield  {journal} {\bibinfo  {journal} {New Journal of Physics}\ }\textbf {\bibinfo {volume} {21}},\ \bibinfo {pages} {113045} (\bibinfo {year} {2019}{\natexlab{a}})}\BibitemShut {NoStop}%
\bibitem [{\citenamefont {Potts}\ \emph {et~al.}(2021)\citenamefont {Potts}, \citenamefont {Kalaee},\ and\ \citenamefont {Wacker}}]{Potts_2021}%
  \BibitemOpen
  \bibfield  {author} {\bibinfo {author} {\bibfnamefont {P.~P.}\ \bibnamefont {Potts}}, \bibinfo {author} {\bibfnamefont {A.~A.~S.}\ \bibnamefont {Kalaee}},\ and\ \bibinfo {author} {\bibfnamefont {A.}~\bibnamefont {Wacker}},\ }\bibfield  {title} {\bibinfo {title} {A thermodynamically consistent markovian master equation beyond the secular approximation},\ }\href {https://doi.org/10.1088/1367-2630/ac3b2f} {\bibfield  {journal} {\bibinfo  {journal} {New Journal of Physics}\ }\textbf {\bibinfo {volume} {23}},\ \bibinfo {pages} {123013} (\bibinfo {year} {2021})}\BibitemShut {NoStop}%
\bibitem [{\citenamefont {de~la Pradilla}\ \emph {et~al.}(2024)\citenamefont {de~la Pradilla}, \citenamefont {Moreno},\ and\ \citenamefont {Feist}}]{delapradilla2024taming}%
  \BibitemOpen
  \bibfield  {author} {\bibinfo {author} {\bibfnamefont {D.~F.}\ \bibnamefont {de~la Pradilla}}, \bibinfo {author} {\bibfnamefont {E.}~\bibnamefont {Moreno}},\ and\ \bibinfo {author} {\bibfnamefont {J.}~\bibnamefont {Feist}},\ }\href@noop {} {\bibinfo {title} {Taming the bloch-redfield equation: Recovering an accurate lindblad equation for general open quantum systems}} (\bibinfo {year} {2024}),\ \Eprint {https://arxiv.org/abs/2402.06354} {arXiv:2402.06354 [quant-ph]} \BibitemShut {NoStop}%
\bibitem [{\citenamefont {Nielsen}\ and\ \citenamefont {Chuang}(2011)}]{Nielsen}%
  \BibitemOpen
  \bibfield  {author} {\bibinfo {author} {\bibfnamefont {M.~A.}\ \bibnamefont {Nielsen}}\ and\ \bibinfo {author} {\bibfnamefont {I.~L.}\ \bibnamefont {Chuang}},\ }\href {https://www.amazon.com/Quantum-Computation-Information-10th-Anniversary/dp/1107002176?SubscriptionId=AKIAIOBINVZYXZQZ2U3A&tag=chimbori05-20&linkCode=xm2&camp=2025&creative=165953&creativeASIN=1107002176} {\emph {\bibinfo {title} {Quantum Computation and Quantum Information: 10th Anniversary Edition}}}\ (\bibinfo  {publisher} {Cambridge University Press},\ \bibinfo {year} {2011})\BibitemShut {NoStop}%
\bibitem [{Eke(1996)}]{Ekert}%
  \BibitemOpen
  \bibfield  {title} {\bibinfo {title} {Quantum computers and dissipation},\ }\href {https://doi.org/10.1098/rspa.1996.0029} {\bibfield  {journal} {\bibinfo  {journal} {Proceedings of the Royal Society of London. Series A: Mathematical, Physical and Engineering Sciences}\ }\textbf {\bibinfo {volume} {452}},\ \bibinfo {pages} {567} (\bibinfo {year} {1996})}\BibitemShut {NoStop}%
\bibitem [{\citenamefont {Schaller}()}]{Schaller_notes}%
  \BibitemOpen
  \bibfield  {author} {\bibinfo {author} {\bibfnamefont {G.}~\bibnamefont {Schaller}},\ }\href@noop {} {\bibinfo {title} {{O}pen {Q}uantum {S}ystems {F}ar from {E}quilibrium --- link.springer.com}},\ \bibinfo {howpublished} {\url{https://link.springer.com/book/10.1007/978-3-319-03877-3}},\ \bibinfo {note} {[Accessed 07-03-2024]}\BibitemShut {NoStop}%
\bibitem [{\citenamefont {REDFIELD}(1965)}]{REDFIELD19651}%
  \BibitemOpen
  \bibfield  {author} {\bibinfo {author} {\bibfnamefont {A.}~\bibnamefont {REDFIELD}},\ }\bibfield  {title} {\bibinfo {title} {The theory of relaxation processes* *this work was started while the author was at harvard university, and was then partially supported by joint services contract n5ori-76, project order i.},\ }in\ \href {https://doi.org/https://doi.org/10.1016/B978-1-4832-3114-3.50007-6} {\emph {\bibinfo {booktitle} {Advances in Magnetic Resonance}}},\ \bibinfo {series} {Advances in Magnetic and Optical Resonance}, Vol.~\bibinfo {volume} {1},\ \bibinfo {editor} {edited by\ \bibinfo {editor} {\bibfnamefont {J.~S.}\ \bibnamefont {Waugh}}}\ (\bibinfo  {publisher} {Academic Press},\ \bibinfo {year} {1965})\ pp.\ \bibinfo {pages} {1--32}\BibitemShut {NoStop}%
\bibitem [{\citenamefont {Bloch}(1946)}]{bloch}%
  \BibitemOpen
  \bibfield  {author} {\bibinfo {author} {\bibfnamefont {F.}~\bibnamefont {Bloch}},\ }\bibfield  {title} {\bibinfo {title} {Nuclear induction},\ }\href {https://doi.org/10.1103/PhysRev.70.460} {\bibfield  {journal} {\bibinfo  {journal} {Phys. Rev.}\ }\textbf {\bibinfo {volume} {70}},\ \bibinfo {pages} {460} (\bibinfo {year} {1946})}\BibitemShut {NoStop}%
\bibitem [{\citenamefont {Winczewski}\ and\ \citenamefont {Alicki}(2023)}]{winczewski2023renormalization}%
  \BibitemOpen
  \bibfield  {author} {\bibinfo {author} {\bibfnamefont {M.}~\bibnamefont {Winczewski}}\ and\ \bibinfo {author} {\bibfnamefont {R.}~\bibnamefont {Alicki}},\ }\href@noop {} {\bibinfo {title} {Renormalization in the theory of open quantum systems via the self-consistency condition}} (\bibinfo {year} {2023}),\ \Eprint {https://arxiv.org/abs/2112.11962} {arXiv:2112.11962 [quant-ph]} \BibitemShut {NoStop}%
\bibitem [{\citenamefont {Hegerfeldt}\ and\ \citenamefont {Plenio}(1993)}]{approx_plenio}%
  \BibitemOpen
  \bibfield  {author} {\bibinfo {author} {\bibfnamefont {G.~C.}\ \bibnamefont {Hegerfeldt}}\ and\ \bibinfo {author} {\bibfnamefont {M.~B.}\ \bibnamefont {Plenio}},\ }\bibfield  {title} {\bibinfo {title} {Coherence with incoherent light: A new type of quantum beat for a single atom},\ }\href {https://doi.org/10.1103/PhysRevA.47.2186} {\bibfield  {journal} {\bibinfo  {journal} {Phys. Rev. A}\ }\textbf {\bibinfo {volume} {47}},\ \bibinfo {pages} {2186} (\bibinfo {year} {1993})}\BibitemShut {NoStop}%
\bibitem [{\citenamefont {Dodin}\ \emph {et~al.}(2018)\citenamefont {Dodin}, \citenamefont {Tscherbul}, \citenamefont {Alicki}, \citenamefont {Vutha},\ and\ \citenamefont {Brumer}}]{3lvl}%
  \BibitemOpen
  \bibfield  {author} {\bibinfo {author} {\bibfnamefont {A.}~\bibnamefont {Dodin}}, \bibinfo {author} {\bibfnamefont {T.}~\bibnamefont {Tscherbul}}, \bibinfo {author} {\bibfnamefont {R.}~\bibnamefont {Alicki}}, \bibinfo {author} {\bibfnamefont {A.}~\bibnamefont {Vutha}},\ and\ \bibinfo {author} {\bibfnamefont {P.}~\bibnamefont {Brumer}},\ }\bibfield  {title} {\bibinfo {title} {Secular versus nonsecular redfield dynamics and fano coherences in incoherent excitation: An experimental proposal},\ }\href {https://doi.org/10.1103/PhysRevA.97.013421} {\bibfield  {journal} {\bibinfo  {journal} {Phys. Rev. A}\ }\textbf {\bibinfo {volume} {97}},\ \bibinfo {pages} {013421} (\bibinfo {year} {2018})}\BibitemShut {NoStop}%
\bibitem [{\citenamefont {Xu}\ and\ \citenamefont {Cao}(2016)}]{Xu2016}%
  \BibitemOpen
  \bibfield  {author} {\bibinfo {author} {\bibfnamefont {D.}~\bibnamefont {Xu}}\ and\ \bibinfo {author} {\bibfnamefont {J.}~\bibnamefont {Cao}},\ }\bibfield  {title} {\bibinfo {title} {Non-canonical distribution and non-equilibrium transport beyond weak system-bath coupling regime: A polaron transformation approach},\ }\href {https://doi.org/10.1007/s11467-016-0540-2} {\bibfield  {journal} {\bibinfo  {journal} {Frontiers of Physics}\ }\textbf {\bibinfo {volume} {11}},\ \bibinfo {pages} {110308} (\bibinfo {year} {2016})}\BibitemShut {NoStop}%
\bibitem [{\citenamefont {Albert}(2018)}]{albert2018lindbladians}%
  \BibitemOpen
  \bibfield  {author} {\bibinfo {author} {\bibfnamefont {V.~V.}\ \bibnamefont {Albert}},\ }\href@noop {} {\bibinfo {title} {Lindbladians with multiple steady states: theory and applications}} (\bibinfo {year} {2018}),\ \Eprint {https://arxiv.org/abs/1802.00010} {arXiv:1802.00010 [quant-ph]} \BibitemShut {NoStop}%
\bibitem [{\citenamefont {Jeske}\ \emph {et~al.}(2015)\citenamefont {Jeske}, \citenamefont {Ing}, \citenamefont {Plenio}, \citenamefont {Huelga},\ and\ \citenamefont {Cole}}]{Jeske}%
  \BibitemOpen
  \bibfield  {author} {\bibinfo {author} {\bibfnamefont {J.}~\bibnamefont {Jeske}}, \bibinfo {author} {\bibfnamefont {D.~J.}\ \bibnamefont {Ing}}, \bibinfo {author} {\bibfnamefont {M.~B.}\ \bibnamefont {Plenio}}, \bibinfo {author} {\bibfnamefont {S.~F.}\ \bibnamefont {Huelga}},\ and\ \bibinfo {author} {\bibfnamefont {J.~H.}\ \bibnamefont {Cole}},\ }\bibfield  {title} {\bibinfo {title} {{Bloch-Redfield equations for modeling light-harvesting complexes}},\ }\href {https://doi.org/10.1063/1.4907370} {\bibfield  {journal} {\bibinfo  {journal} {The Journal of Chemical Physics}\ }\textbf {\bibinfo {volume} {142}},\ \bibinfo {pages} {064104} (\bibinfo {year} {2015})},\ \Eprint {https://arxiv.org/abs/https://pubs.aip.org/aip/jcp/article-pdf/doi/10.1063/1.4907370/14030726/064104\_1\_online.pdf} {https://pubs.aip.org/aip/jcp/article-pdf/doi/10.1063/1.4907370/14030726/064104\_1\_online.pdf} \BibitemShut {NoStop}%
\bibitem [{\citenamefont {Cattaneo}\ \emph {et~al.}(2019{\natexlab{b}})\citenamefont {Cattaneo}, \citenamefont {Giorgi}, \citenamefont {Maniscalco},\ and\ \citenamefont {Zambrini}}]{cattaneo_local_2019}%
  \BibitemOpen
  \bibfield  {author} {\bibinfo {author} {\bibfnamefont {M.}~\bibnamefont {Cattaneo}}, \bibinfo {author} {\bibfnamefont {G.~L.}\ \bibnamefont {Giorgi}}, \bibinfo {author} {\bibfnamefont {S.}~\bibnamefont {Maniscalco}},\ and\ \bibinfo {author} {\bibfnamefont {R.}~\bibnamefont {Zambrini}},\ }\bibfield  {title} {{\selectlanguage {en}\bibinfo {title} {Local versus global master equation with common and separate baths: superiority of the global approach in partial secular approximation}},\ }\href {https://doi.org/10.1088/1367-2630/ab54ac} {\bibfield  {journal} {\bibinfo  {journal} {New Journal of Physics}\ }\textbf {\bibinfo {volume} {21}},\ \bibinfo {pages} {113045} (\bibinfo {year} {2019}{\natexlab{b}})},\ \bibinfo {note} {publisher: IOP Publishing}\BibitemShut {NoStop}%
\bibitem [{\citenamefont {Vinjanampathy}\ and\ \citenamefont {Anders}(2016)}]{vinjanampathy_quantum_2016}%
  \BibitemOpen
  \bibfield  {author} {\bibinfo {author} {\bibfnamefont {S.}~\bibnamefont {Vinjanampathy}}\ and\ \bibinfo {author} {\bibfnamefont {J.}~\bibnamefont {Anders}},\ }\bibfield  {title} {{\selectlanguage {en}\bibinfo {title} {Quantum thermodynamics}},\ }\href {https://doi.org/10.1080/00107514.2016.1201896} {\bibfield  {journal} {\bibinfo  {journal} {Contemporary Physics}\ }\textbf {\bibinfo {volume} {57}},\ \bibinfo {pages} {545} (\bibinfo {year} {2016})}\BibitemShut {NoStop}%
\bibitem [{\citenamefont {Ángel Rivas}\ \emph {et~al.}(2010)\citenamefont {Ángel Rivas}, \citenamefont {Plato}, \citenamefont {Huelga},\ and\ \citenamefont {Plenio}}]{rivascritical}%
  \BibitemOpen
  \bibfield  {author} {\bibinfo {author} {\bibnamefont {Ángel Rivas}}, \bibinfo {author} {\bibfnamefont {A.~D.~K.}\ \bibnamefont {Plato}}, \bibinfo {author} {\bibfnamefont {S.~F.}\ \bibnamefont {Huelga}},\ and\ \bibinfo {author} {\bibfnamefont {M.~B.}\ \bibnamefont {Plenio}},\ }\bibfield  {title} {\bibinfo {title} {Markovian master equations: a critical study},\ }\href {https://doi.org/10.1088/1367-2630/12/11/113032} {\bibfield  {journal} {\bibinfo  {journal} {New Journal of Physics}\ }\textbf {\bibinfo {volume} {12}},\ \bibinfo {pages} {113032} (\bibinfo {year} {2010})}\BibitemShut {NoStop}%
\bibitem [{Note2()}]{Note2}%
  \BibitemOpen
  \bibinfo {note} {This convention has been chosen to match existing literature, when the reduced model is considered \cite {Rivas}}\BibitemShut {NoStop}%
\bibitem [{\citenamefont {Blais}\ \emph {et~al.}(2021)\citenamefont {Blais}, \citenamefont {Grimsmo}, \citenamefont {Girvin},\ and\ \citenamefont {Wallraff}}]{Girvin}%
  \BibitemOpen
  \bibfield  {author} {\bibinfo {author} {\bibfnamefont {A.}~\bibnamefont {Blais}}, \bibinfo {author} {\bibfnamefont {A.~L.}\ \bibnamefont {Grimsmo}}, \bibinfo {author} {\bibfnamefont {S.~M.}\ \bibnamefont {Girvin}},\ and\ \bibinfo {author} {\bibfnamefont {A.}~\bibnamefont {Wallraff}},\ }\bibfield  {title} {\bibinfo {title} {Circuit quantum electrodynamics},\ }\href {https://doi.org/10.1103/RevModPhys.93.025005} {\bibfield  {journal} {\bibinfo  {journal} {Rev. Mod. Phys.}\ }\textbf {\bibinfo {volume} {93}},\ \bibinfo {pages} {025005} (\bibinfo {year} {2021})}\BibitemShut {NoStop}%
\bibitem [{\citenamefont {Guarnieri}\ \emph {et~al.}(2021)\citenamefont {Guarnieri}, \citenamefont {Kol\'a\ifmmode~\check{r}\else \v{r}\fi{}},\ and\ \citenamefont {Filip}}]{guarnieriErratum}%
  \BibitemOpen
  \bibfield  {author} {\bibinfo {author} {\bibfnamefont {G.}~\bibnamefont {Guarnieri}}, \bibinfo {author} {\bibfnamefont {M.}~\bibnamefont {Kol\'a\ifmmode~\check{r}\else \v{r}\fi{}}},\ and\ \bibinfo {author} {\bibfnamefont {R.}~\bibnamefont {Filip}},\ }\bibfield  {title} {\bibinfo {title} {Erratum: Steady-state coherences by composite system-bath interactions [phys. rev. lett. 121, 070401 (2018)]},\ }\href {https://doi.org/10.1103/PhysRevLett.127.129901} {\bibfield  {journal} {\bibinfo  {journal} {Phys. Rev. Lett.}\ }\textbf {\bibinfo {volume} {127}},\ \bibinfo {pages} {129901} (\bibinfo {year} {2021})}\BibitemShut {NoStop}%
\bibitem [{\citenamefont {Chru\ifmmode \acute{s}\else \'{s}\fi{}ci\ifmmode~\acute{n}\else \'{n}\fi{}ski}\ \emph {et~al.}(2011)\citenamefont {Chru\ifmmode \acute{s}\else \'{s}\fi{}ci\ifmmode~\acute{n}\else \'{n}\fi{}ski}, \citenamefont {Kossakowski},\ and\ \citenamefont {Rivas}}]{measures}%
  \BibitemOpen
  \bibfield  {author} {\bibinfo {author} {\bibfnamefont {D.}~\bibnamefont {Chru\ifmmode \acute{s}\else \'{s}\fi{}ci\ifmmode~\acute{n}\else \'{n}\fi{}ski}}, \bibinfo {author} {\bibfnamefont {A.}~\bibnamefont {Kossakowski}},\ and\ \bibinfo {author} {\bibfnamefont {A.}~\bibnamefont {Rivas}},\ }\bibfield  {title} {\bibinfo {title} {Measures of non-markovianity: Divisibility versus backflow of information},\ }\href {https://doi.org/10.1103/PhysRevA.83.052128} {\bibfield  {journal} {\bibinfo  {journal} {Phys. Rev. A}\ }\textbf {\bibinfo {volume} {83}},\ \bibinfo {pages} {052128} (\bibinfo {year} {2011})}\BibitemShut {NoStop}%
\bibitem [{\citenamefont {Krug}\ and\ \citenamefont {Stockburger}(2023)}]{instabilities}%
  \BibitemOpen
  \bibfield  {author} {\bibinfo {author} {\bibfnamefont {M.}~\bibnamefont {Krug}}\ and\ \bibinfo {author} {\bibfnamefont {J.}~\bibnamefont {Stockburger}},\ }\bibfield  {title} {\bibinfo {title} {On stability issues of the heom method},\ }\href {https://doi.org/10.1140/epjs/s11734-023-00972-9} {\bibfield  {journal} {\bibinfo  {journal} {The European Physical Journal Special Topics}\ }\textbf {\bibinfo {volume} {232}},\ \bibinfo {pages} {3219} (\bibinfo {year} {2023})}\BibitemShut {NoStop}%
\bibitem [{\citenamefont {Takahashi}\ \emph {et~al.}(2024)\citenamefont {Takahashi}, \citenamefont {Rudge}, \citenamefont {Kaspar}, \citenamefont {Thoss},\ and\ \citenamefont {Borrelli}}]{thoss_review}%
  \BibitemOpen
  \bibfield  {author} {\bibinfo {author} {\bibfnamefont {H.}~\bibnamefont {Takahashi}}, \bibinfo {author} {\bibfnamefont {S.}~\bibnamefont {Rudge}}, \bibinfo {author} {\bibfnamefont {C.}~\bibnamefont {Kaspar}}, \bibinfo {author} {\bibfnamefont {M.}~\bibnamefont {Thoss}},\ and\ \bibinfo {author} {\bibfnamefont {R.}~\bibnamefont {Borrelli}},\ }\bibfield  {title} {\bibinfo {title} {{High accuracy exponential decomposition of bath correlation functions for arbitrary and structured spectral densities: Emerging methodologies and new approaches}},\ }\href {https://doi.org/10.1063/5.0209348} {\bibfield  {journal} {\bibinfo  {journal} {The Journal of Chemical Physics}\ }\textbf {\bibinfo {volume} {160}},\ \bibinfo {pages} {204105} (\bibinfo {year} {2024})},\ \Eprint {https://arxiv.org/abs/https://pubs.aip.org/aip/jcp/article-pdf/doi/10.1063/5.0209348/19961639/204105\_1\_5.0209348.pdf} {https://pubs.aip.org/aip/jcp/article-pdf/doi/10.1063/5.0209348/19961639/204105\_1\_5.0209348.pdf} \BibitemShut {NoStop}%
\bibitem [{\citenamefont {Lambert}\ \emph {et~al.}(2019)\citenamefont {Lambert}, \citenamefont {Ahmed}, \citenamefont {Cirio},\ and\ \citenamefont {Nori}}]{Lambert2019}%
  \BibitemOpen
  \bibfield  {author} {\bibinfo {author} {\bibfnamefont {N.}~\bibnamefont {Lambert}}, \bibinfo {author} {\bibfnamefont {S.}~\bibnamefont {Ahmed}}, \bibinfo {author} {\bibfnamefont {M.}~\bibnamefont {Cirio}},\ and\ \bibinfo {author} {\bibfnamefont {F.}~\bibnamefont {Nori}},\ }\bibfield  {title} {\bibinfo {title} {Modelling the ultra-strongly coupled spin-boson model with unphysical modes},\ }\href {https://doi.org/10.1038/s41467-019-11656-1} {\bibfield  {journal} {\bibinfo  {journal} {Nature Communications}\ }\textbf {\bibinfo {volume} {10}},\ \bibinfo {pages} {3721} (\bibinfo {year} {2019})}\BibitemShut {NoStop}%
\bibitem [{\citenamefont {Menczel}\ \emph {et~al.}(2024)\citenamefont {Menczel}, \citenamefont {Funo}, \citenamefont {Cirio}, \citenamefont {Lambert},\ and\ \citenamefont {Nori}}]{menczel2024nonhermitian}%
  \BibitemOpen
  \bibfield  {author} {\bibinfo {author} {\bibfnamefont {P.}~\bibnamefont {Menczel}}, \bibinfo {author} {\bibfnamefont {K.}~\bibnamefont {Funo}}, \bibinfo {author} {\bibfnamefont {M.}~\bibnamefont {Cirio}}, \bibinfo {author} {\bibfnamefont {N.}~\bibnamefont {Lambert}},\ and\ \bibinfo {author} {\bibfnamefont {F.}~\bibnamefont {Nori}},\ }\href@noop {} {\bibinfo {title} {Non-hermitian pseudomodes for strongly coupled open quantum systems: Unravelings, correlations and thermodynamics}} (\bibinfo {year} {2024}),\ \Eprint {https://arxiv.org/abs/2401.11830} {arXiv:2401.11830 [quant-ph]} \BibitemShut {NoStop}%
\bibitem [{\citenamefont {Strathearn}\ \emph {et~al.}(2017)\citenamefont {Strathearn}, \citenamefont {Lovett},\ and\ \citenamefont {Kirton}}]{Strathearn_2017}%
  \BibitemOpen
  \bibfield  {author} {\bibinfo {author} {\bibfnamefont {A.}~\bibnamefont {Strathearn}}, \bibinfo {author} {\bibfnamefont {B.~W.}\ \bibnamefont {Lovett}},\ and\ \bibinfo {author} {\bibfnamefont {P.}~\bibnamefont {Kirton}},\ }\bibfield  {title} {\bibinfo {title} {Efficient real-time path integrals for non-markovian spin-boson models},\ }\href {https://doi.org/10.1088/1367-2630/aa8744} {\bibfield  {journal} {\bibinfo  {journal} {New Journal of Physics}\ }\textbf {\bibinfo {volume} {19}},\ \bibinfo {pages} {093009} (\bibinfo {year} {2017})}\BibitemShut {NoStop}%
\bibitem [{\citenamefont {Somoza}\ \emph {et~al.}(2019)\citenamefont {Somoza}, \citenamefont {Marty}, \citenamefont {Lim}, \citenamefont {Huelga},\ and\ \citenamefont {Plenio}}]{DAMPF}%
  \BibitemOpen
  \bibfield  {author} {\bibinfo {author} {\bibfnamefont {A.~D.}\ \bibnamefont {Somoza}}, \bibinfo {author} {\bibfnamefont {O.}~\bibnamefont {Marty}}, \bibinfo {author} {\bibfnamefont {J.}~\bibnamefont {Lim}}, \bibinfo {author} {\bibfnamefont {S.~F.}\ \bibnamefont {Huelga}},\ and\ \bibinfo {author} {\bibfnamefont {M.~B.}\ \bibnamefont {Plenio}},\ }\bibfield  {title} {\bibinfo {title} {Dissipation-assisted matrix product factorization},\ }\href {https://doi.org/10.1103/PhysRevLett.123.100502} {\bibfield  {journal} {\bibinfo  {journal} {Phys. Rev. Lett.}\ }\textbf {\bibinfo {volume} {123}},\ \bibinfo {pages} {100502} (\bibinfo {year} {2019})}\BibitemShut {NoStop}%
\bibitem [{\citenamefont {Becker}\ \emph {et~al.}(2022)\citenamefont {Becker}, \citenamefont {Schnell},\ and\ \citenamefont {Thingna}}]{becker}%
  \BibitemOpen
  \bibfield  {author} {\bibinfo {author} {\bibfnamefont {T.}~\bibnamefont {Becker}}, \bibinfo {author} {\bibfnamefont {A.}~\bibnamefont {Schnell}},\ and\ \bibinfo {author} {\bibfnamefont {J.}~\bibnamefont {Thingna}},\ }\bibfield  {title} {\bibinfo {title} {Canonically consistent quantum master equation},\ }\href {https://doi.org/10.1103/PhysRevLett.129.200403} {\bibfield  {journal} {\bibinfo  {journal} {Phys. Rev. Lett.}\ }\textbf {\bibinfo {volume} {129}},\ \bibinfo {pages} {200403} (\bibinfo {year} {2022})}\BibitemShut {NoStop}%
\end{thebibliography}%

\end{document}